\documentclass[conference]{IEEEtran}
\usepackage{cite}
\usepackage{array}
\usepackage{url}

\usepackage{breakurl}
\usepackage{hyperref}
\hypersetup{breaklinks=true}
\usepackage[colorinlistoftodos]{todonotes}
\usepackage{marginnote} 

\usepackage{url}
\usepackage[T1]{fontenc}
\usepackage[utf8]{inputenc}
\usepackage{graphicx}
\usepackage{listings}
 \usepackage{multirow}
\usepackage{xargs,paralist,booktabs,array,tabularx,tikz,color,colortbl}
\usepackage{nameref}
\usepackage{xspace}
\usepackage{tabularx}
\usepackage{hhline}
\usepackage{color}
\usepackage{subcaption}
\usepackage{amsmath}
\usepackage{mdframed}
\usepackage{booktabs}   
\usepackage{longtable}
\usepackage{array}
\usepackage{mathtools}
\usepackage[export]{adjustbox}
\usepackage{array}
\usepackage{fontawesome5}
\newcolumntype{R}[2]{%
    >{\adjustbox{angle=#1,lap=\width-(#2)}\bgroup}%
    l%
    <{\egroup}%
}
\usepackage{placeins}[below]
\usepackage{float}

\definecolor{scone}{HTML}{0060AF}
\definecolor{sctwo}{HTML}{F0E442}
\definecolor{scthree}{HTML}{9D2A32}

\definecolor{match}{HTML}{009E73}
\definecolor{quasi}{HTML}{E69F00}
\definecolor{nocondition}{HTML}{000000}

\renewcommand{\arraystretch}{1.5}

\usepackage{booktabs}

\usepackage[normalem]{ulem}
\useunder{\uline}{\ul}{}

\newcommand{\parahighlight}[1]{\textbf{#1}:\xspace}
\newcommand{\parahighlightnoindent}[1]{\noindent\textbf{#1}:\xspace}

\definecolor{Gray}{gray}{0.9}

\definecolor{lightgray}{gray}{0.93}

\DeclareTextFontCommand{\red}{\color{red}}
\usetikzlibrary{trees,shapes.geometric,shapes.symbols,chains,calc,positioning,arrows,decorations.pathmorphing,decorations.pathreplacing}

\newlength\bubblesize
\newcommand\yestab{\tikz[baseline=0.0ex] \fill[black] (\bubblesize,\bubblesize) circle (\bubblesize);}
\newcommand\yes{%
  \begin{tikzpicture}[baseline=0.0ex, line width=0.2ex]%
    \fill[match] (2.5\bubblesize, 2.5\bubblesize) rectangle (-.5\bubblesize,-.5\bubblesize);
    \fill[black] (\bubblesize,\bubblesize) circle (\bubblesize);
  \end{tikzpicture}%
}

\newcommand\kinda{%
  \begin{tikzpicture}[baseline=0.0ex, line width=0.2ex]%
    \fill[quasi] (2.5\bubblesize, 2.5\bubblesize) rectangle (-.5\bubblesize,-.5\bubblesize);
    \draw[clip]  (\bubblesize,\bubblesize) circle (\bubblesize-.5\pgflinewidth);%
    \fill[white] (2\bubblesize, 2\bubblesize) rectangle (0,0);%
    \draw[clip]  (\bubblesize,\bubblesize) circle (\bubblesize-.5\pgflinewidth);%
    \fill[black] (\bubblesize, 2\bubblesize) rectangle (0,0);%
  \end{tikzpicture}%
}

\newcommand\kindatab{%
  \begin{tikzpicture}[baseline=0.0ex, line width=0.2ex]%
    \draw[clip]  (\bubblesize,\bubblesize) circle (\bubblesize-.5\pgflinewidth);%
    \fill[white] (2\bubblesize, 2\bubblesize) rectangle (0,0);%
    \draw[clip]  (\bubblesize,\bubblesize) circle (\bubblesize-.5\pgflinewidth);%
    \fill[black] (\bubblesize, 2\bubblesize) rectangle (0,0);%
  \end{tikzpicture}%
}

\newcommand\kindaright{%
  \begin{tikzpicture}[baseline=0.0ex, line width=0.2ex]%
    \fill[quasi] (2.5\bubblesize, 2.5\bubblesize) rectangle (-.5\bubblesize,-.5\bubblesize);
    \draw[clip]  (\bubblesize,\bubblesize) circle (\bubblesize-.5\pgflinewidth);%
    \fill[white] (\bubblesize, 2\bubblesize) rectangle (0,0);%
    \draw[clip]  (\bubblesize,\bubblesize) circle (\bubblesize-.5\pgflinewidth);%
    \fill[black] (2\bubblesize, 2\bubblesize) rectangle (\bubblesize,0);%
  \end{tikzpicture}%
}

\newcommand\kindarighttab{%
  \begin{tikzpicture}[baseline=0.0ex, line width=0.2ex]%
    \draw[clip]  (\bubblesize,\bubblesize) circle (\bubblesize-.5\pgflinewidth);%
    \fill[white] (\bubblesize, 2\bubblesize) rectangle (0,0);%
    \draw[clip]  (\bubblesize,\bubblesize) circle (\bubblesize-.5\pgflinewidth);%
    \fill[black] (2\bubblesize, 2\bubblesize) rectangle (\bubblesize,0);%
  \end{tikzpicture}%
}

\newcommand\no{\tikz[baseline=0.0ex] \draw[black, fill=white, line width=0.2ex] (\squarsize,\squarsize) rectangle (0,0);}
\newcommand\notab{\tikz[baseline=0.0ex] \draw[white, fill=white, line width=0.2ex] (\bubblesize,\bubblesize) circle (\bubblesize-.5\pgflinewidth);}

\newcommand\nocondition{\tikz[baseline=0.0ex] \draw[black, fill=black, line width=0.2ex] (\squarsize,\squarsize) rectangle (0,0);}

\newlength\squarsize
\newlength\bgsize
\begin{document}
%-------------------------------------------------------------------------------

\title{A Systematic Study of the Consistency of Two-Factor Authentication User Journeys on Top-Ranked Websites (Extended Version)}

\author{\IEEEauthorblockN{Sanam Ghorbani Lyastani, Michael Backes, Sven Bugiel}
\IEEEauthorblockA{CISPA Helmholtz Center for Information Security}
}

\maketitle

\begin{abstract}
Heuristics for user experience state that users will transfer their expectations from one product to another. A lack of consistency between products can increase users' cognitive friction, leading to frustration and rejection. This paper presents the first systematic study of the external, functional consistency of two-factor authentication user journeys on top-ranked websites. We find that these websites implement only a minimal number of design aspects consistently (e.g., naming and location of settings) but exhibit mixed design patterns for setup and usage of a second factor. Moreover, we find that some of the more consistently realized aspects, such as descriptions of two-factor authentication, have been described in the literature as problematic and adverse to user experience. Our results advocate for more general UX guidelines for 2FA implementers and raise new research questions about the 2FA user journeys.
\end{abstract}

\section{Introduction}
\label{sec:introduction}

Would you buy a car where the gas and brake pedals are interchanged? You would probably be able to learn to drive this car safely after some acclimatization period. Still, it would be an experience that is very inconsistent with what you are used to, and you would most likely not continue using such an unpleasant car. Like this everyday example, a consistent user experience is crucial for websites to fit the mental models that users built and avoid unnecessarily increasing the users' cognitive load and friction by forcing them to learn something new. This important best practice has been captured in \textit{Jakob's Law of Internet User Experience}~\cite{Nielsen2000,Nielsen_Law_Online,krause_jakob_law} as one of several heuristics for user experience~\cite{9781492055310,lawsofux_website} and usability~\cite{nielsen_10heuristics} that guide website design.
Striving for consistent user experience has ruled website design for years, evident in the design of, e.g., online shopping, banking, forums, blogs, or streaming services.
The same best practices also apply to user authentication as part of the user experience.

When it comes to the incumbent authentication scheme on the web today, text-based passwords, the user experience of passwords is highly consistent across different websites, although recent work~\cite{281214} discovered inconsistent password policies for blocklists, strength meters, and composition when setting passwords on the top websites. Regardless of this inconsistency, text-based passwords are notorious for their security issues. Among the different solutions proposed to strengthen user authentication on the web, two-factor authentication~(2FA) has been shown to have a very tangible positive effect on account security~\cite{10.1007/978-3-662-54970-4_25,microsoft_compromise,10.1145/3359386}. Nowadays, 2FA is frequently recommended to end users to improve their security hygiene~\cite{255292}. Fortunately, many websites are starting to offer 2FA options to their users~\cite{2fawebsite,duo_state_of_auth}. However, previous work~\cite{238317,tale_of_two_studies} demonstrated that users struggled with 2FA when their 2FA journey did not match their expectations or previous experiences and advocated for more standardized procedures. In a survey with 2FA adopters (see Appendix~\ref{appendix:survey}), we found corroborating evidence that inconsistent implementations of the 2FA user journey caused friction for users that lowered the usability of 2FA and led users to refuse 2FA or abandon websites. Unfortunately, up to today, \textit{we have only very few insights about how consistent the user experience of 2FA is across different websites}.

To provide new insights about how websites offer 2FA to their users and how consistent this user experience is across websites, we systematically study the 2FA user journeys on 85 popular websites in this paper. More specifically, we want to determine whether these websites consistently follow the same design patterns and strategies to offer 2FA to their users. Or, in other words, we are interested in the external functional consistency of the 2FA user journeys across popular websites. 

To approach our research question systematically, we need concrete factors based on which we can compare the different user journeys. Unfortunately, such a list of factors does not exist for two-factor authentication, and there is no common guideline or best practice on how to implement the 2FA user flow on websites. Furthermore, 2FA is a technology that has only started gaining wider adoption among websites in the last couple of years and was hence not part of the initial website design. Additionally, the 2FA ecosystem is fragmented into various options for 2FA, such as TOTP, WebAuthn, push notifications, SMS, or custom solutions, each with its own setup process, dependencies (e.g., hardware token or app), and benefits/drawbacks in terms of usability and security~\cite{Bonneau:2012:QRP:2310656.2310722,238325}. For these reasons, it was not a priori obvious which exact comparison factors could describe potentially diverse user journeys on different websites.

To solve this challenge, we devised a methodology to derive a list of comparison factors from open and axial coding of existing user journeys on the 85 websites in our data set. As a result, we created a list of 22 comparison factors that describe the user journey from \textit{discovery} of an offered (promoted) 2FA support during sign-in/registration, to the \textit{education} of the user about the available second-factor options and their \textit{setup} processes, to \textit{usage} and \textit{deactivation} of the chosen 2FA option(s). Based on those factors, we then compared the 85 websites to identify common design patterns and differences and highlight beneficial or detrimental patterns for user experience.

Our results show that there is no overarching design pattern for the user journey that most websites follow. Instead, we found the design space to be clustered into groups of websites with very similar patterns, some of those favored by the top websites and others by less popular sites. The only design aspects that almost all websites agree on about 2FA are that it is an optional feature, how it should be called and described, and where it should be found in the account settings. In contrast, for the crucial steps of setting up and using 2FA, we found that websites implement mixed strategies, such as varying numbers of simultaneously supported 2FA technologies, inconsistent presentation of device remembrance options, or varying degrees of feedback to users. 

According to UX guidelines, this lack of consistency increases users' cognitive load and should be avoided. However, consistency alone does not guarantee a good user experience. We found that several of the more consistently used design patterns have been described in prior work as problematic for user experience, including non-encouraging descriptions or missing possibilities to personalize the 2FA. We also discovered that the journeys of top websites, like \path{icloud.com}, are outliers from the best practices in the academic literature. Therefore, our results create a call for action to reinvestigate what constitutes a good overall 2FA user experience, to study whether there is a ``gold standard'' for implementing 2FA user journeys, or to explore the motivations of website developers to implement specific design patterns.

\section{Background}
\label{sec:background}

\subsection{Two-Factor Authentication}
\label{sec:background:2fa}

With two-factor authentication enabled on a website, a user must successfully provide two authentication factors to verify their identity. Almost always, the first factor is a traditional text-based password. For the second factor, there are different technical realizations of knowledge, possession, and inherence factors. Most common~\cite{2fawebsite,elie_2fa_blog} are \textit{one-time codes} delivered via SMS text-message, phone call, or TOTP~\cite{RFC6238} apps, like Google Authenticator, Duo, or custom apps that the user registered with the website; \textit{push notifications} by sending an alert message to a dedicated app on the user's phone that asks the user to confirm a login attempt; and hardware tokens via the \textit{U2F} or \textit{FIDO2/WebAuthn}~\cite{webauthn_spec} standards that rely on public key cryptography and challenge-response protocols.

Each of these comes with its own set of usability and security benefits and drawbacks~\cite{238325}. Important for our work is that a website with 2FA support can offer one or multiple of those 2FA options, may even allow users to set one of those solutions up multiple times, or may enforce a particular order in which they can be set up or used.

A commonly acknowledged problem with two-factor authentication is account recovery when a user loses access to a factor (e.g., a mobile device with the TOTP app is unavailable). Often the strategy to avoid lockout from a 2FA-protected account is to set up a dedicated recovery option, such as printed-out one-time passwords that can replace another 2FA option, or to configure multiple 2FA options, when supported by the website, e.g., multiple hardware security keys.

\subsection{User Experience}
\label{sec:background:ux}

Unfortunately, providing an exact definition of ``user experience'' is very difficult, as there is no consensus on the exact definition~\cite{uxdefinition, uxdefinition2, uxdefinition3, ISO9241}. However, a common topic among the definitions is that UX encompasses the various aspects of user interaction with a product, such as a website. Cooper et al.~\cite{9781118766576} note that there exist three overlapping concerns for UX: form, content, and behavior. While form and content (e.g., UI design or phrasing) have an impact on usability, this work focuses on behavior (i.e., functionality) and only touches on some aspects of form and content.

To help designers provide the best possible user experience, various best practices and general guidelines have been developed (e.g., books~\cite{9781118766576,9780321965516,9781492055310,9780136746911,shneiderman2004designing} or online resources, such as \textit{Laws of UX}~\cite{lawsofux_website}, \textit{Nielsen Norman Group}~\cite{nngroup_webiste}, or \textit{Interaction Design Foundation}~\cite{interaction_design}). Among the earliest are Shneiderman's eight "Golden Rules" for interface design~\cite{shneiderman2004designing,shneiderman_website} and Nielsen's "10 Usability Heuristics for User Interface Design"~\cite{nielsen_10heuristics,nielsen94}. Shneiderman's rules state, for instance, that one should strive for consistency and provide informative feedback to users. Of Nielsen's heuristics, heuristic nr.~4, also known as \textit{Jakob's law of Internet user experience}~\cite{Nielsen_Law_Online}, is the most important for this work and provides the motivation to study the consistency of 2FA user journeys across websites. This heuristic states that \textit{``users spend most of their time on other sites''} and that \textit{``users prefer a site to work the same way as all the other sites they already know.''} As a consequence, one should \textit{``design for patterns for which users are accustomed.''} Having such conventions and consistency helps users build upon existing mental models and avoid cognitive friction by forcing them to learn something new~\cite{9781492055310}. If an unconventional website mismatches the user's mental model, the website will be difficult to learn, difficult to use, or even rejected~\cite{9780136746911}. One way to drive \textit{external} consistency is to make ample use of guidelines. For instance, for apps there exist Google's Material Design Guidelines~\cite{materialdesign} and Apple's Human Interaction Guidelines~\cite{iosdesign}. We are not aware of any general guidelines for implementers and designers of two-factor authentication on websites, although there are case-specific guidelines (for example, FIDO2~\cite{fido_ux}) or small collections of best practices (e.g.,~\cite{dias21,amiconsult}).

Although in this work we focus on external, \textit{functional consistency}, some of the comparison factors for 2FA user journeys that we identified (see Section~\ref{sec:comparisonfactors}) also touch on other UX guidelines and best practices. Tesler's law~\cite{9781492055310} states that for any system there is a certain amount of complexity that cannot be reduced, and it is recommended that the product design ensures that as much as possible of the burden on the user is lifted. Krug~\cite{9780321965516} recommends that if a difficulty for the user cannot be avoided, the design should provide brief and timely guidance, and Cooper et al.~\cite{9781118766576} recommend contextual help and assistive interfaces without the need to break the user's flow. If it cannot be avoided that the user has to learn something new, users learn best from examples (e.g., pictures, screenshots, or short tutorial videos)~\cite{9780136746911}. In addition, Hick's law~\cite{9781492055310} recommends breaking down complex tasks into smaller steps to decrease the cognitive load. Moreover, excise tasks, such as navigational excise, should be reduced, e.g., by reducing the number of places that a user must go and providing clear overviews~\cite{9781118766576}. Hereby, it is important to consider that users do not read but scan webpages~\cite{9780321965516} and that this scanning is based on the mental model they built from past experiences, which creates expectations of what they want to see and where~\cite{9780136746911}. Furthermore, part of Postel's law~\cite{9781492055310}, similar to Shneiderman's third golden rule~\cite{shneiderman2004designing,shneiderman_website}, recommends providing clear feedback to users, and the Peak-End Rule~\cite{9781492055310} recommends paying attention to the final moments of the user journey because people judge an experience largely based on how they felt at its peak and recall negative experiences more vividly than positive ones. Lastly, personalization can enhance the user experience. Although we did not explicitly investigate websites for their quality of those additional guidelines, some of our comparison factors indicate if 2FA settings are found in common places, if additional information and instructions are provided, if user notifications are present, or if users can set preferences.
\section{Related Work}
\label{sec:relatedwork}

Several works have studied two-factor authentication problems and focused on the usability component and user attitudes. Bonneau et al.~\cite{Bonneau:2012:QRP:2310656.2310722} conducted a systematic expert assessment of various authentication solutions, including many of the solutions used for 2FA. They concluded that the usability of these solutions falls very often short compared to text-based passwords. In contrast to Bonneau et al., most other works relied on user studies to investigate problems of 2FA.

A focal point of prior user studies was the setup and usage of different two-factor authentication solutions to understand users' attitudes toward 2FA, obstacles for its adoption, and how to improve the usability and user experience. Previous works studied two-factor authentication in settings such as online banking~\cite{WEIR200947, WEIR2010153, GUNSON2011208, journals/corr/KrolPCS15} or military~\cite{strouble_dod_cac} services. Like other studies of 2FA~\cite{DBLP:conf/hicss/DasWKC20, Braz:2006:SUC:1132736.1132768,fagan_soups16,das_haisa19} they found that users consider 2FA often burdensome and slow, that convenience trumps perceived security, and that users do not always understand the risks that 2FA tries to remedy. Several works have studied 2FA problems in organizational contexts~\cite{DBLP:journals/corr/abs-2011-02901,10.1145/3313831.3376457,Colnago2018,255254,Dutson2019,Weidman2017} where the use of MFA can be mandated. While those studies show that many of the problems overlap with non-organizational settings, they could also shed new light on the positive influence of features such as device remembrance~\cite{255254,Dutson2019} or better help and instructions. 

Several studies~\cite{238325,WEIR200947, WEIR2010153, journals/corr/KrolPCS15} compared different options for the second factor to identify option-specific differences in user attitudes and usability, while other works specifically studied security keys~\cite{johnny_two_factor, tale_of_two_studies,238317} or authenticator apps~\cite{das_haisa19_2}. An interesting aspect of those works~\cite{238325,238317,tale_of_two_studies} for our study is that they differentiated between 2FA setup and login, where users often struggled in the setup due to unclear instructions/workflows. Strong recommendations from those works were clearer instructions and guidance for the setup to avoid user frustration that often leads to non-adoption. 

Additionally, improved notification design patterns~\cite{driving_2fa} have been shown to encourage users to adopt 2FA.

Lastly, recent works~\cite{GhorbaniLyastani2020,Farke2020, Owens2021,274547} studied specifically FIDO2 \textit{single}-factor authentication. They found similar user concerns as for 2FA. However, the weighting of the concerns shifted (e.g., loss of the authenticator device is ranked very high) or new concerns were added (e.g., misunderstanding biometric WebAuthn). Relevant to our work, the FIDO Alliance has recently published UX guidelines for security keys~\cite{fido_securitykey_ux} and implementers of desktop authenticators~\cite{fido_ux} that, similar to our methodology, divide the user journey into different steps and provide recommendations for the design of each step; however, explicitly tailored to the technical details of FIDO2/WebAuthn with biometric authenticator devices or security keys. Nevertheless, those guidelines incorporate many of the UX guidelines explained in Section~\ref{sec:background:ux}.

The key difference of our work is that we do not study how concrete changes in form, content, or functionality affect the usability and concrete experience of 2FA, but that we are first to systematically study how \textit{consistent} the user experience is across existing popular websites. Our work, in contrast to previous works, strongly focuses on Jakob's law of Internet user experience which states that an inconsistent user experience across websites increases cognitive friction and can be detrimental to users' adoption. Providing first insights into how well the 2FA user journeys adhere to this law is the core contribution of this work.
Further, we are not aware of prior studies that measured Jakob's law across a larger number of websites but instead, to the best of our knowledge, qualitative and quantitative testing of websites focuses on single websites or comparative user studies between a small set of websites based on general UX best-practices and guidelines. Therefore, we had to devise a methodology to measure the consistency of the 2FA user journeys on different websites.

\section{Methodology}
\label{sec:methodology}

\begin{figure*}[t]
\centering
        \includegraphics[width=\linewidth]{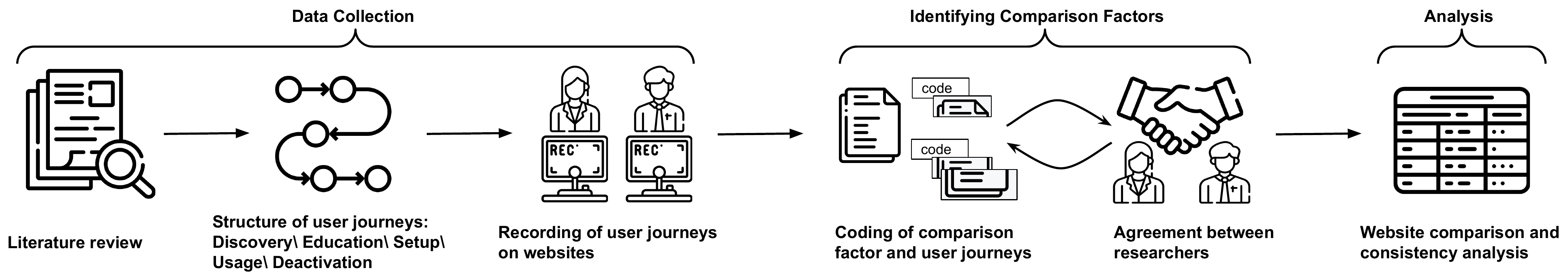}
        \caption{Overview of our methodology}
        \label{fig:methotology_overview}
\end{figure*} 

To compare the 2FA user journeys of different websites and measure their consistency, we require concrete comparison factors that describe these journeys. Unfortunately, there is no existing list of such comparison factors, of which we are aware, or general guidelines for implementing 2FA on websites from which we could extract such factors. Therefore, a crucial challenge for our study is to create a list of relevant and representative factors. 
We used inductive research methods (e.g., \cite[Chapter~11.4]{LazarFengHochheiser17}) to solve this challenge.
Figure~\ref{fig:methotology_overview} gives an overview of our methodology, whose data collection (Section~\ref{sec:methodology:recording}) and identification of comparison factors (Section~\ref{sec:methodology:idfactors}) we explain in the following. In a nutshell, we use open and axial coding from grounded theory on the screen-recorded 2FA user journeys of different websites to identify the list of comparison factors and to form an agreement about how each website matches each factor. Using the coding results, we then compare the different websites and study how consistently they implement the 2FA user journey and where they differ (Sections~\ref{sec:comparisonfactors} and~\ref{sec:results}).

\subsection{Data Collection}
\label{sec:methodology:recording}

The first part of our methodology is to collect a representative data set of user journeys recorded on different websites that we can analyze. Since we are building our knowledge about user journeys inductively, the screen recordings must have as high as possible coverage of all steps and choices along each journey. To this end, an automated tool, such as a Web crawler, could be used to explore various websites. Unfortunately, the need for a priori knowledge about how websites might implement their user journeys to guide the crawler and the need to use additional authentication devices (e.g., phone or security key) hamper an automated collection. Alternatively, we could use a crowd-sourced data collection, e.g., Amazon Mechanical Turk. Unfortunately, this was not possible in our setting for ethical reasons. We would need to ask our participants to use private accounts (or create fake accounts) on different websites and explore security settings for which they might need to provide a (personal) email address, phone number, or security key, and risk accidentally locking themselves out of an account as a result of a misconfiguration.

Instead, two researchers independently explored and screen-recorded the 2FA user journeys for our study. Their general instruction was to ``thoroughly explore all aspects'' of these journeys. However, this exploration could be informed beforehand from the literature, which discusses different aspects of 2FA user journeys (see also Section~\ref{sec:relatedwork}). For example, recent works~(e.g., \cite{driving_2fa,274547,Das2020}) and guidelines~\cite{fido_ux} identify discovery of 2FA options and user education, different works studied 2FA setup and login (e.g., \cite{238325,tale_of_two_studies}) or mandating 2FA (e.g., \cite{10.1145/3313831.3376457}), and account recovery is a commonly identified problem. Based on those insights, we structure the exploration of user journeys into five steps: The first step is \textit{Discovery} of 2FA support on the website. We explore the landing pages, FAQ, and account registration for information on 2FA and follow all linked information. To further encourage users, there might also be nudges and messages about securing the account with 2FA, for which we scan the websites' interfaces. To use 2FA, the user must find the corresponding settings in their account settings, which we explore for the locations and options for authentication. In the next step, \textit{Education}, we examine how a website introduces 2FA and if it gives further explanations, such as descriptions of how 2FA works and what it offers. Once the user has decided to use 2FA, they need to \textit{Setup} their second factor(s). We explore the workflow of setting up all supported 2FA options (e.g., TOTP or Security Key). This exploration includes examining the websites' instructions, exploring the different settings choices (e.g., personalization choices), and feedback from the website on successful setup. After setting up two-factor authentication, we examine the \textit{Usage} of 2FA on the website. We re-login and observe how the website prompts us to authenticate and whether it provides any options (e.g., device remembrance), which we explore. Finally, we explore the 2FA \textit{Deactivation} procedure in the website settings and how the website communicates those changes.

For data collection, we maintained identical study conditions. All recordings were made on MacBooks running mac\-OS~11 in the same network with the latest version of the Chrome browser when we started our data collection. Data collection was carried out between 06/2021 and 08/2021. This fixed setup should minimize the risk~\cite{article_ifipsec2019_wiefling} of external factors (e.g., varying geolocation) and possible risk-based authentication to distort the data.

It is important to note that we focus \textit{only} on the workflow for account creation, initial 2FA setup, and 2FA usage. We do not explore the workflows for account recovery or to change personal information relevant to 2FA after 2FA setup, such as a phone number or email address. We consider those follow-up problems to be studied after we have insights into the consistency of the fundamental steps that mint the users' first impressions about 2FA on a particular website.

\subsection{Identifying Comparison Factors}
\label{sec:methodology:idfactors}

Since there is no predefined set of factors to compare 2FA user journeys, we applied emergent coding~\cite[Chapter~11.4]{LazarFengHochheiser17}, in particular open and axial coding from grounded theory, to identify comparison factors from our recorded user journeys.
These coding techniques are commonly applied in qualitative data analysis for text content.
To still use those established methods, we treated the screen-recorded journeys like semi-structured interviews.
Semi-structured interviews follow a set of predetermined questions, but the remaining questions are made up during the interview based on the interviewee’s answers. 
We transferred this idea to our data collection (see Section~\ref{sec:methodology:recording}): The exploration of user journeys follows a set of predetermined questions for discovery over usage to deactivation but allows the researcher to divert to individually explore a website in more detail and discover new or unique aspects of 2FA user journeys.
Two researchers separately iterated through the set of recorded user journeys and segmented the observed journeys into meaningful parts to which they assigned concepts (i.e., codes).
This is followed by axial coding, where the two researchers combined those concepts via induction and deduction into categories.
For example, the codes ``2FA advertised on the landing page'' and ``2FA recommended during account creation'' can be combined into ``Promotion of 2FA.''
These combined concepts can be used as comparison factors on all websites. 
The researchers also noted whether there exists a functional dependency between factors. 
After agreeing on the list of comparison factors, the researchers discussed how each website matches each comparison factor (e.g., fully, partially, or not at all). 
Since the matching of comparison factors might reveal that the list of factors is too fine-grained, potentially weighting small differences too heavily, or too coarse-grained, potentially hiding important differences, the researchers repeated the axial coding process until a set of comparison factors and website matching was found to which all involved researchers agreed. 
The focus of coding was on the \textit{functional} aspects of the websites, and less on the elements of the content or user interface since this study focuses on the consistency between websites and \textit{not} rating the quality of each website's user journey.

\section{Data Set}
\label{sec:methodology:dataset}

To gather a set of websites for our study, we rely on the open source project \url{2fa.directory}~\cite{2fadirectory, 2fawebsite} that maintains a list of websites with 2FA support, which almost 1,000 contributors currently curate. The websites are assigned to different categories, such as social, communication, or retail. Since \url{2fa.directory} distinguishes websites at the level of subdomains, we merged subdomains into their domain when we were aware that they use the same account for authentication. For example, \url{drive.google.com}, \url{cloud.google.com}, and \url{mail.google.com} are in different categories but rely on the same Google account, while \url{amazon.com} and \url{aws.amazon.com} have separate accounts. For merged entries, we chose the category we thought end users most likely knew the domain for (e.g., \textit{mail} for \url{google.com}). Since we rely on a manual investigation of the user journeys of each website, we needed to reduce the set of all websites listed on \url{2fa.directory} to a feasible number. First, we excluded categories for which we cannot create an account, for example, almost all websites in the \textit{banking} and \textit{government} categories. Second, we used the Tranco~\cite{tranco_paper,tranco_website} data set to rank websites according to their popularity. We selected the top websites from each category, where we selected the number of websites from each category based on the category's weight in the 2fa.directory data set. For example, there were only four \textit{VPN provider} websites in the \url{2fa.directory} set but 45 \textit{Gaming} websites. This initially resulted in 120 websites. Unfortunately, we had to exclude 35 websites that we could not study for different reasons, such as language barriers, geo-restrictions, or the need for financial expenditures. In the end, we recorded the 2FA user journey on 85 websites with 2FA support from 26 categories. 

\section{Comparison Factors}
\label{sec:comparisonfactors}

In this section, we explain the comparison factors that we identified in our analysis of 85 popular websites following the methodology of Section~\ref{sec:methodology} and describe informally how we categorize websites according to these factors. We apply the methodology of Bonneau et al.~\cite{Bonneau:2012:QRP:2310656.2310722} by categorizing every website if it matches~(\yes), partially matches~(\kinda, \kindaright), or not matches~(\no) a factor. However, in our categorization, some factors are dependent on other factors, and we denote it explicitly when a conditional factor's prerequisite is not fulfilled (\nocondition) and this factor does not apply to a website. Further, in contrast to Bonneau et al., we do \textit{not} use the categorization as a ranking to determine if a website is better than another website, but we use the categorization to identify patterns in how websites realize their 2FA user journey and to study whether websites realize this journey in a consistent way. Although, for some of the factors described below, this categorization overlaps with a scale from known best practices to known poor practices from the literature.
We found 22 comparison factors; 8 are conditional and depend on other factors to be applicable. Appendix~\ref{appendix:examples} provides various examples of the different comparison factors.

\subsection{Factors for Discovery}

\parahighlight{Promotion} The website promotes its 2FA support in a clear and obvious way during account creation or immediately after login (e.g., through a banner, pop-up, or highlighted message) (\yes). If the website does not clearly promote but only mentions the 2FA support in a way that could be easily missed by the user (for example, only a quick link in the footer of the landing page), we categorize this as \textit{quasi-}promotion (\kinda). If the service does not promote its 2FA support and the user has to discover it themselves (e.g., browsing the settings pages), we categorize this as not matching (\no).

\parahighlight{Non-Optional} The website mandates setting up 2FA for user accounts (\yes). For instance, without setting a 2FA option up, the account registration cannot be completed; or after account registration, core functionality and features of the website are not available to the user until the user sets up 2FA for their account. Otherwise, using 2FA is optional and not mandatory for the website (\no).

\parahighlight{Common-Naming-and-Location} The website denotes its 2FA settings with a commonly used name, and the 2FA settings are in a commonly used location in the account settings (\yes). We identify commonly used names and locations in our analysis of our selected websites and summarize the results in Section~\ref{sec:results:overview}.

If either the name (\kinda) or the location (\kindaright) is uncommon, we categorize this as \textit{quasi-common-naming-and-location}.

If the naming and location are uncommon, we categorize the website as not matching this factor (\no).

\subsection{Factors for Education}

\parahighlight{Descriptive-Notification} The website briefly describes what 2FA is in general or why it is important to users. The description is provided to the user \textit{before} the user clicks to enable 2FA (\yes), e.g., located together with a notification about 2FA availability or within the settings page; or the description is only provided \textit{after} the user starts the 2FA setup process (\kinda) at which point the user can still abort the setup. If the website does not present a description of 2FA, we categorize this website as not matching (\no).

\parahighlight{Additional-Information} The website provides more detailed information through a link (e.g. ``learn more'') to help users understand 2FA (\yes). If no such information is provided or the link is broken, the factor does not match (\no).

\subsection{Factors for Setup}

\parahighlight{Option-Specific-Information} The website provides specific information about all 2FA options it supports (\yes). For instance, it informs the user that TOTP or Push-notifications require the installation of an app or that WebAuthn requires a hardware authenticator. If the website does not provide this information but directly starts the setup process (e.g., asking users to scan a QR code or to use a security key without further explanation), this factor does not match~(\no).

\parahighlight{Step-Wise-Instructions} The website gives an overview of the steps involved in setting up a specific 2FA option (e.g., linking a device or app, verifying the link, setting a recovery option) and/or details the instructions for each step for all 2FA options (\yes). Otherwise, this factor does not match (\no).

\parahighlight{Multiselection} The website offers multiple 2FA options (or setting up one method multiple times) and allows the user to set up multiple 2FA options (\yes), e.g., TOTP and Push-notification or multiple security keys. If the website supports multiple 2FA options but only allows the user to select one non-repeated option, we categorize this as \textit{quasi-multiselection} (\kinda). This factor does not match (\no) if the website only offers a single, one-time configurable option.

\parahighlight{Grouped-Setting} The website's user settings present the 2FA options grouped, and users have a single setting location to manage all their 2FA options (\yes), e.g., all under the same settings tab. If the 2FA settings are split between different sections of the settings, we consider this to be not matching this factor (\no). For instance, the management of security keys is organizationally separated from managing other 2FA options and, hence, might not be obvious to users. This factor depends on \textit{Multiselection} being (quasi-)matched.

\parahighlight{No-Enforced-Options} The website immediately presents all supported 2FA options to the user and allows them to choose their options themselves (\yes). If the website mandates the setup of specific 2FA options before the user can set up other options, we consider this not to match this factor (\no). For example, the user must configure SMS-based 2FA before having the possibility to configure TOTP or WebAuthn options. This factor depends on \textit{Multiselection} being (quasi-)matched.

\parahighlight{Selectable-Primary-Option} If the website allows the configuration of multiple 2FA options and allows the user to select a primary option, which is the first option requested during login before falling back to other configured options (or recovery), we consider this a match (\yes). If the website does not support setting a user-selected primary 2FA option, we consider this not matching (\no). This factor depends on \textit{Multiselection} being matched.

\parahighlight{Settings-Changed-Verification} The website requires the user to verify their identity before being able to change the 2FA settings (\yes).  Otherwise, this factor does not match (\no).

\parahighlight{Settings-Changed-Notification} The website notifies the user about the changed 2FA settings via an out-of-band channel, e.g., by email or push notification (\yes).  If there is no notification, this website does not match this factor (\no).

\parahighlight{Confirm-Successful-Setup} The website requires the user to confirm the 2FA authentication to complete the setup successfully and provides clear messaging about the successful setup for all options (\yes). For example, the user must enter the current TOTP or confirm a push notification to complete the setup, and the website shows a highlighted message in the settings. If the messaging is missing, but confirmation is required, we consider this as \textit{quasi-confirm-successful-setup} (\kinda). If the website does not require confirmation (for all options), this website does not match this factor (\no).

\parahighlight{Informed-2FA-Recovery-Options} The website offers dedicated recovery options (such as one-time codes or asking to set up multiple 2FA options) and explains to the user why configuring dedicated 2FA recovery options is important for preparing for cases where the default 2FA options are not available, e.g., to prevent account lockout due to a lost or broken authentication device (\yes). If the website offers such recovery options but does not explicitly inform the user about their benefits and importance, we consider this as \textit{quasi-informed-recovery-options} (\kinda). If the website does not offer explicit 2FA recovery options (e.g., it relies on a general account recovery or customer support), we consider this as not matching this factor~(\no).

\parahighlight{Enforced-2FA-Recovery-Setup} Setting up recovery options is a mandatory step in setting up 2FA for this website (\yes), and the user cannot finish or continue setting up 2FA unless they set up the recovery option first. For example, the user has to confirm that they printed one-time backup codes to finish the 2FA setup or the website enforces setting up multiple 2FA options with a clear hint at account recovery.
If setting up recovery options is not mandatory, but the website nudges users or strongly recommends them to set up a recovery option, we consider this \textit{quasi-mandatory-recovery-setup} (\kinda).
If setting up dedicated recovery options is at the user's discretion (without nudging or recommending), we consider this factor not matching (\no). This factor depends on \textit{Informed-2FA-recovery-options} being (quasi-)matched.

 \subsection{Factors for Usage}
 \parahighlight{Device-Remembrance} The website offers a device remembrance during login, such that the user does not have to use 2FA on subsequent logins on the same device (e.g., "remember this device" checkbox). If the website automatically sets device remembrance without involving the user, e.g., during the first login after 2FA setup or during 2FA setup, we categorize this as \yes. If device remembrance is at the discretion of the user and is stated as opt-\textit{out} (e.g., an unchecked checkbox described as "ask me again on this device" or a pre-ticked checkbox "trust this device"), we categorize this as \kindaright. If device remembrance is stated as opt-\textit{in} (e.g., "trust this device" checkbox that was not pre-checked), we categorize this as \kinda. If device remembrance is not offered, we categorize as~\no.
 
\parahighlight{No-Preselected-Option} If the website supports more than one active 2FA option at a time and no primary method is set (or could be set), how does the website present the configured 2FA options to their end users: the website shows all configured 2FA options at the same time during login (\yes), e.g., as a drop-down list. Alternatively, the website selected the primary option based on internal metrics (\no), e.g., a security policy or the user's usage history. This preselection is usually intransparent to the user. This factor depends on \textit{Multiselection} being matched and \textit{Selectable-primary-option} not being matched.

\subsection{Factors for Deactivation}

\parahighlight{Informed-Deactivation} The website allows the user to deactivate 2FA options and also explains to the user the potential risks associated with this (\yes) or does not provide any explanation or warning (\kinda). If the website does not allow the user to deactivate two-factor authentication, we consider this a mismatch for this factor (\no).

\parahighlight{Deactivation-Verification} The website requires the user to verify their identity before being able to deactivate a 2FA option (\yes). If a 2FA option can be disabled by the user without further authorization, we consider this factor to not match the website (\no). This factor depends on \textit{Informed-deactivation} being (quasi-)matched.

\parahighlight{Deactivation-Notification} The website notifies the user about the deactivated 2FA option via an out-of-band channel, e.g., by email (\yes).  If there is no notification, this website does not match this factor (\no). This factor depends on \textit{Informed-deactivation} being (quasi-)matched.

\parahighlight{Communicate-Successful-Deactivation} The website communicates successful deactivation to the user as part of its user interface (\yes), e.g., highlighted message or pop-up. 

Otherwise, we consider this website not to match this factor~(\no). 
This factor depends on \textit{Informed-deactivation} being (quasi-)matched.

\section{Results}
\label{sec:results}

We first provide an overview of the collected data (Section~\ref{sec:results:overview}), followed by exploratory data analysis of the comparison factors (Section~\ref{sec:results:comparison}). Lastly, we discuss the results of qualitative data analysis of our observations (Section~\ref{sec:results:qualitative}).

\subsection{Overview of Website Data}
\label{sec:results:overview}

\begin{table*}[!htbp]
  \centering
 
  \caption{Comparison of popular websites based on the factors introduced in Section~\ref{sec:comparisonfactors} and clustering described in Section~\ref{sec:results}.}
  \label{tab:comparisonfactors}

  \footnotesize
  \setlength{\tabcolsep}{2 pt} % width of column
  \def\arraystretch{1.0} % height of row

\begin{tabular}{@{}p{2cm} | *{1}{c} | *{1}{l} | *{3}{c} | *{2}{c} | *{11}{c} | *{2}{c} | *{4}{c} |}

    \multicolumn{1}{c}{} &
    \multicolumn{1}{c}{} & 
    \multicolumn{1}{c|}{} & 
    \multicolumn{3}{c|}{\textbf{Discovery}} &
    \multicolumn{2}{c|}{\textbf{Education}} &
    \multicolumn{11}{c|}{\textbf{Setup}} &
    \multicolumn{2}{c|}{\textbf{Usage}} &
    \multicolumn{4}{c|}{\textbf{Deactivation}}

    \\
    \hhline{*{3}{~}*{22}{-}}
    \textbf{Website} &
    \tikz \node[rotate=90] {\textbf{Subcluster}}; &
    \tikz \node[rotate=90] {\textbf{Category}}; &
 
    \tikz \node[rotate=90] {Promotion}; &
    \tikz \node[rotate=90] {Non-optional}; &
    \tikz \node[rotate=90] {Common-Naming-and-Location}; &
    \tikz \node[rotate=90] {Descriptive-notification}; &
    \tikz \node[rotate=90] {Additional-information}; &
    \tikz \node[rotate=90] {Option-specific-information}; &
    \tikz \node[rotate=90] {Stepwise-instructions}; &
    \tikz \node[rotate=90] {Multiselection}; &
    \tikz \node[rotate=90] {\cellcolor{gray!20}Grouped-setting}; &
    \tikz \node[rotate=90] {\cellcolor{gray!20}No-enforced-option}; &
    \tikz \node[rotate=90] {\cellcolor{gray!20}Selectable-primary-option}; &
    \tikz \node[rotate=90] {Settings-changed-verification}; &
    \tikz \node[rotate=90] {Settings-changed-notification}; &
    \tikz \node[rotate=90] {Confirm-successful-setup}; &
    \tikz \node[rotate=90] {Informed-2FA-recovery-options}; &
    \tikz \node[rotate=90] {\cellcolor{gray!20}Enforced-2FA-recovery-setup}; &
    \tikz \node[rotate=90] {Device-remembrance}; &
    \tikz \node[rotate=90] {\cellcolor{gray!20}No-preselected-option}; &
    \tikz \node[rotate=90] {Informed-deactivation}; &
    \tikz \node[rotate=90] {\cellcolor{gray!20}Deactivation-verification}; &
    \tikz \node[rotate=90] {\cellcolor{gray!20}Deactivation-notification}; &
    \tikz \node[rotate=90] {\cellcolor{gray!20}Communicate-successful-deact.};
 
    \\\midrule
    \multicolumn{25}{c}{\textbf{Cluster 1 ($n=30$)}}\\
\end{tabular}

\end{table*}

\begin{table*}[!htbp]
  \centering

  \footnotesize
  \setlength{\tabcolsep}{2 pt} % width of column
  \def\arraystretch{1.0} % height of row

  \begin{tabular}{@{}p{2cm} | *{1}{c} | *{1}{l} | *{3}{c} | *{2}{c} | *{11}{c} | *{2}{c} | *{4}{c} |}

    \multicolumn{1}{c}{} &
    \multicolumn{1}{c}{} & 
    \multicolumn{1}{c|}{} & 
    \multicolumn{3}{c|}{\textbf{Discovery}} &
    \multicolumn{2}{c|}{\textbf{Education}} &
    \multicolumn{11}{c|}{\textbf{Setup}} &
    \multicolumn{2}{c|}{\textbf{Usage}} &
    \multicolumn{4}{c|}{\textbf{Deactivation}}

    \\
    \hhline{*{3}{~}*{22}{-}}
    \textbf{Website} &
    \tikz \node[rotate=90] {\textbf{Subcluster}}; &
    \tikz \node[rotate=90] {\textbf{Category}}; &
 
    \tikz \node[rotate=90] {Promotion}; &
    \tikz \node[rotate=90] {Non-optional}; &
    \tikz \node[rotate=90] {Common-Naming-and-Location}; &
    \tikz \node[rotate=90] {Descriptive-notification}; &
    \tikz \node[rotate=90] {Additional-information}; &
    \tikz \node[rotate=90] {Option-specific-information}; &
    \tikz \node[rotate=90] {Stepwise-instructions}; &
    \tikz \node[rotate=90] {Multiselection}; &
    \tikz \node[rotate=90] {\cellcolor{gray!20}Grouped-setting}; &
    \tikz \node[rotate=90] {\cellcolor{gray!20}No-enforced-option}; &
    \tikz \node[rotate=90] {\cellcolor{gray!20}Selectable-primary-option}; &
    \tikz \node[rotate=90] {Settings-changed-verification}; &
    \tikz \node[rotate=90] {Settings-changed-notification}; &
    \tikz \node[rotate=90] {Confirm-successful-setup}; &
    \tikz \node[rotate=90] {Informed-2FA-recovery-options}; &
    \tikz \node[rotate=90] {\cellcolor{gray!20}Enforced-2FA-recovery-setup}; &
    \tikz \node[rotate=90] {Device-remembrance}; &
    \tikz \node[rotate=90] {\cellcolor{gray!20}No-preselected-option}; &
    \tikz \node[rotate=90] {Informed-deactivation}; &
    \tikz \node[rotate=90] {\cellcolor{gray!20}Deactivation-verification}; &
    \tikz \node[rotate=90] {\cellcolor{gray!20}Deactivation-notification}; &
    \tikz \node[rotate=90] {\cellcolor{gray!20}Communicate-successful-deact.};
 
    \\\midrule
    \multicolumn{25}{c}{\textbf{Cluster 2} (continued)}\\
\end{tabular}

\end{table*}
\begin{table}[t]
    \caption{Naming, location, descriptions of 2FA, and description of remembrance (where applicable).}
    \label{tab:overview_nld}
    \centering
    % \footnotesize
    \def\arraystretch{1.0} % height of row
    
\begin{tabular}{lr@{ }r}
\toprule
\multicolumn{3}{c}{\textbf{Name of 2FA in website settings}}\\
\midrule
  Two-Factor Authentication (2FA) & 42 & (49.41\%) \\
      Two-Step Verification (2SV) & 24 & (28.24\%) \\
                      Other & 12 & (14.12\%) \\
Multi-Factor Authentication (MFA) &   4 & (4.71\%) \\
    Two-Step Authentication (2SA) &   3 & (3.53\%) \\
\midrule
\multicolumn{3}{c}{\textbf{Location of 2FA settings}}\\
\midrule
Security / Account & 78 & (91.76\%) \\
             Other &   7 & (8.24\%) \\
\midrule
\multicolumn{3}{c}{\textbf{Focus of 2FA description}}\\
\midrule
Security & 69 & (92.00\%) \\
  Device &   6 & (8.00\%) \\
 \midrule
\multicolumn{3}{c}{\textbf{Description of device remembrance}}\\
\midrule
Remember & 16 & (51.61\%) \\
   Trust &  9 & (29.03\%) \\
    Skip &  4 & (12.90\%) \\
   Other &  2 & ( 6.45\%) \\
\bottomrule
\end{tabular}
\end{table}

Table~\ref{tab:comparisonfactors} summarizes how each of the 85 websites in our data set matches the 22 comparison factors that we identified. We will explore these data further in the following sections. 
Table~\ref{tab:overview_nld} summarizes the naming and location of the 2FA settings, the type of 2FA description, and the forms of device remembrance. 
Table~\ref{tab:device_remembrance_codebook} in Appendix~\ref{appendix:codebook_dr} provides the codebook for device remembrance descriptions. 
And Table~\ref{tab:websitedetails} in Appendix~\ref{appendix:codebook} gives further details per website, including Tranco~\cite{tranco_paper,tranco_website} rank and website category according to 2fa.directory~\cite{2fadirectory, 2fawebsite}.
In summary, we found that 73 (86\%) of the websites use a combination of ``two-factor''/``two-step''/``multiple-factor'' with ``authentication''/``verification'' for the naming, and on 78 (92\%) websites, the 2FA settings are located in the security settings of the account settings under similar paths (e.g., ``Security,'' ``Login security,'' or ``Authentication''). We considered those names and locations the common naming and location during our evaluation of the \textit{Common-Naming-and-Location} factor. Of the 75 websites that describe 2FA in their settings, 69 (92\%) describe 2FA in the form of ``an additional layer of security,'' while 6 websites describe the 2FA mechanism with a focus on the user device (e.g., ``we ask for additional authentication when logging in from a device that we do not know''). Only 31 websites in our data set offered a device remembrance feature, and half of those (16; 52\%) describe this feature in terms of remembering the device or client (e.g., ``Do not require OTP on this browser'' or ``Do not ask again on this device''). Almost a third (9; 29\%) describe it in terms of trust (e.g., ``Trust this device for \{duration\}''), and only four websites (13\%) phrase it as skipping the additional step (e.g., ``We won’t ask for the next \{duration\}''). We also noticed that websites have a mixed strategy for phrasing the user's choice (i.e., opt-\textit{in} versus opt-\textit{out}), which we encoded in our factor \textit{Device-remembrance} in Table~\ref{tab:comparisonfactors} (i.e., \kinda\ vs.~\kindaright).

\subsection{Exploratory Data Analysis}
\label{sec:results:comparison}

Our factors allow us to compare the 2FA user journeys of different websites. We first explore our collected data (in Table~\ref{tab:comparisonfactors}) through similarity analysis and clustering to gain insight into the overall consistency of those journeys on different websites and to identify potential clusters of websites that follow similar design patterns for their 2FA user experience.

\subsubsection{Website similarity and factor consistency}
\label{sec:websitesimilarity}

To get a general impression of how similar the journeys are on the 85 websites, we compared them pairwise. Since our comparison factors are feature vectors of nominal (i.e., categorical) variables for each website, there is no intrinsic ordering and no equal space between variable values to measure the distance between values.
We used the Hamming distance between pairs of websites as a measure of similarity.
Since our variables have only values between 2--4, Hamming distance (i.e., ``overlap without weights'') is the most efficient measure of similarity for our data to obtain an overall impression of consistency between websites instead of measures that consider the number and/or frequency of values per variable, such as (Inverse) Occurrence Frequency, Goodall~\cite{Goodall1966ANS}, or Eskin et al.~\cite{Eskin2002}.
To avoid artificial inflation of similarity from unfulfilled conditional factors, we calculate the Hamming distance only for the 14 non-conditional factors. 
We find that the average website in our data set differs in 6--7 of those 14 factors from the other websites, indicating that from a bird's-eye view, the user journeys are not very consistent across those websites. Further details about the frequency distribution of the pairwise Hamming distances are provided in Appendix~\ref{appendix:websitesimilarity}.

\begin{table}[t]
\caption{Shannon entropy of each non-conditional factor}
\label{tab:shannon}
\centering

\def\arraystretch{1.1} % height of row
\scriptsize

\begin{tabular}{llrr}
\toprule
&\textbf{Comparison Factor} & $H(X)$ & Max ent. \\
\midrule

\multirow{6}{2.0cm}{\textbf{Two-point scale}} & Non-optional & {\cellcolor[HTML]{B6CEFA}} \color[HTML]{000000} 0.37 & \multirow{6}{*}{1.0}\\
& Additional-information & {\cellcolor[HTML]{D65244}} \color[HTML]{F1F1F1} 0.90 & \\
& Option-specific-information & {\cellcolor[HTML]{B70D28}} \color[HTML]{F1F1F1} 0.99 & \\
& Stepwise-instructions & {\cellcolor[HTML]{DE614D}} \color[HTML]{F1F1F1} 0.87 & \\
& Settings-changed-verification & {\cellcolor[HTML]{B70D28}} \color[HTML]{F1F1F1} 0.99 & \\
& Settings-changed-notification & {\cellcolor[HTML]{B50927}} \color[HTML]{F1F1F1} 1.00 & \\
\midrule

\multirow{6}{2.0cm}{\textbf{Three-point scale}} & Promotion & {\cellcolor[HTML]{F7AA8C}} \color[HTML]{000000} 1.12 & \multirow{6}{*}{1.57}\\
& Descriptive-notification & {\cellcolor[HTML]{F7AC8E}} \color[HTML]{000000} 1.11 & \\
& Multiselection & {\cellcolor[HTML]{B70D28}} \color[HTML]{F1F1F1} 1.57 & \\
& Confirm-successful-setup & {\cellcolor[HTML]{F08B6E}} \color[HTML]{F1F1F1} 1.24 & \\
& Informed-2FA-recovery-options & {\cellcolor[HTML]{EE8669}} \color[HTML]{F1F1F1} 1.26 & \\
& Informed-deactivation & {\cellcolor[HTML]{F7B99E}} \color[HTML]{000000} 1.05 & \\
\midrule

\multirow{2}{2.0cm}{\textbf{Four-point scale}} & Common-Naming-and-Location & {\cellcolor[HTML]{DCDDDD}} \color[HTML]{000000} 1.00 & \multirow{2}{*}{2.0}\\
& Device-remembrance & {\cellcolor[HTML]{EE8468}} \color[HTML]{F1F1F1} 1.60 & \\
\bottomrule
\end{tabular}
\end{table}

Furthermore, we measured the consistency of individual, non-conditional factors across all websites using Shannon entropy.
High entropy means high inconsistency, whereas low entropy implies high consistency.
The results are summarized in Table~\ref{tab:shannon}.
Since some factors can also \textit{quasi}-match (\kinda/\kindaright) and, thus, have a different maximum entropy from binary (two-point scale) factors with only \yes\xspace and \no, we distinguish between the point scales for each factor.
The maximum possible entropy for each scale is indicated in column \textit{Max ent}.
Noticeable outliers with high entropy, i.e., low consistency, are \textit{Multiselection} and most of the two-scale factors, which are close to the highest possible entropy.
For example, \textit{Multiselection} is almost evenly split ($34\times$\yes, $28\times$\kinda, $23\times$\no). 
In contrast, \textit{Non-optional} is very consistent ($6\times$\yes, $79\times$\no) and \textit{Common-Naming-and-Location} shows a strong tendency ($67\times$\yes, $6\times$\kindaright, $11\times$\kinda, $1\times$\no).
In summary, we found that none of the factors exhibit high consistency, except for \textit{Non-optional} 2FA and \textit{Common-Naming-and-Location}.

\subsubsection{Factor clusters}
\label{sec:clustering}
Since our data do \textit{not} indicate a ``global consistency,'' we explore further whether there exist clusters of websites that have close similarities to each other but are more dissimilar from others. 
We applied a two-stage clustering process: first, we cluster websites based on their non-conditional factors (\textit{inter-cluster}) and, additionally, assign each website to a subcluster based on the conditional factors (\textit{intra-cluster}).
Our intention for this two-stage process was that inter-clusters based on non-conditional factors provide the primary view of the different strategies for the 2FA user journeys across all websites, while additional intra-clusters based on conditional factors could support a more differentiated discussion of the overall strategies.
Since our comparison factors are nominal variables, we apply k-modes~\cite{chaturvedi2001k} clustering in both stages. For inter-clustering, Silhouette testing~\cite{ROUSSEEUW198753} indicated that 2, 5, or 6 clusters fit the data best, and we decided on 6 clusters due to the best descriptive performance of those clusters. For the intra-clustering of the conditional factors, we found 3 clusters to best describe the data. The result of the final clustering is noted in Table~\ref{tab:comparisonfactors}. Appendix~\ref{sec:appendix:clusters} provides less noisy views of the cluster structures.

When comparing the characteristics of the \textit{inter-clusters}, we find three aspects that differentiate the clusters the most: how they inform and instruct their users, how they offer support for multiple 2FA options, and whether they offer device remembrance. In terms of informing and instructing users, the six clusters can be combined into two larger clusters. Websites in \textit{\textbf{Clusters~1, 2}} and \textit{\textbf{3}} generally do not verify or notify about changes in 2FA settings (except for \textit{\textbf{Cluster~2}}), omit additional information, do not give step-wise instructions (with the exception of \textbf{\textit{Cluster~3}}), and often do not provide specific information about 2FA options. In contrast, \textit{\textbf{Clusters~4, 5}} and \textit{\textbf{6}} provide this information and instructions more regularly and, in addition, the websites in \textit{\textbf{Cluster~4}} warn users about the deactivation of 2FA. Alternatively, the six \textit{inter-clusters} could be combined into two groups based on their strategy to support multiple 2FA options. Here, websites in \textit{\textbf{Clusters~1, 5}} and \textit{\textbf{6}} usually allow only one option to be activated simultaneously, although they usually offer multiple options. In contrast, websites in \textit{\textbf{Clusters~2, 3}} and \textit{\textbf{4}}, when supporting multiple 2FA options, usually allow users to choose between multiple activated 2FA options for login. Lastly, regarding device remembrance, the websites in \textit{\textbf{Clusters~1, 3, 4}} and \textit{\textbf{5}} have in common that they mostly do not offer device remembrance for future logins. \textit{\textbf{Clusters~2}} and \textit{\textbf{6}} usually offer this.

Regarding \textit{intra-clusters}, \textit{\textbf{Subcluster~1}} websites do not usually provide a selection of multiple 2FA options, and when they do, they enforce certain 2FA options. The websites in \textit{\textbf{Subclusters~2}} and \textit{\textbf{3}} support multiple 2FA options but differ in their strategy to enforce certain 2FA options and verify 2FA deactivation. Websites in \textit{\textbf{Subcluster~2}} almost always verify 2FA deactivation, while websites in \textit{\textbf{Subcluster~3}} do not enforce certain 2FA options. Unfortunately, \textit{\textbf{Clusters~3}} to \textit{\textbf{6}} are too small to reliably comment on the relationship between \textit{inter-clusters} and \textit{intra-cluster}.

\subsubsection{Clusters vs.\ Website Ranks}\label{sec:results:clustervsrank}

We divide the websites into our data set into three roughly equal-sized groups through the 36th and 71st percentiles of the websites' Tranco ranks.
Figure~\ref{fig:cfd_ranking} in Appendix~\ref{appendix:codebook} illustrates the CFD of the Tranco ranking in our data set.
Based on this CFD, the first group of websites ($n=31$) is in the \textit{Top-500} of Tranco, the second group of websites ($n=29$) ranks between 501 and 4,000 (denoted as \textit{Top-4000}), and the third group ($n=25$) is the \textit{``long tail''} with a rank greater than 4,000. 
Since we initially selected the most popular websites in each category, this distribution is naturally heavily skewed toward the top ranks. 
We then used the \textit{inter}-cluster to describe each website's underlying 2FA user flow, which we analyzed for an association with the website Tranco rank group.
Fisher's exact test ($p=0.04388$) shows that this association is statistically significant.
The contingency table for cluster vs.~rank can be found in Appendix~\ref{appendix:codebook}.

We also considered the association between website categories and clusters, but unfortunately, the website categories are too diverse, and the number of websites per category is too small to derive a meaningful connection between cluster and category. We are also unaware of any reliable, more coarse-grained website categorization that could be used. 

\subsubsection{Opinionated Separation of Comparison Factors}
\label{sec:results:separated_factors}

Our analysis considered all the comparison factors at once and did not differentiate between different categories of factors.
To provide a different view on the consistency of 2FA user journeys, we conducted an expert evaluation of our factors to create an opinionated separation of factors by their impact on security, user experience, both, or neither.
The evaluation process is described in Appendix~\ref{appendix:separation_factors}.
As a result, we split our comparison factors into four disjoint sets: \textit{Non-conditional-UX} (7 factors), \textit{Non-conditional-Security} (6 factors), \textit{Conditional-UX} (5 factors), and \textit{Conditional-Security} (3 factors).
Only the factor \textit{Additional-information} was considered irrelevant for UX or security.
Based on the four sets of factors, we repeated the data analysis of Sections~\ref{sec:websitesimilarity} and \ref{sec:clustering}.

\paragraph*{Pairwise Hamming distance}
Considering only \textit{non-conditional-UX} factors, the average website differs in 3--4 of the 7 factors from other websites, and considering only \textit{non-conditional-security} factors, the average website differs in 2--3 of the 6 factors from other websites.
Thus, with this distance metric, the websites in our data set do not show better consistency when considering separated sets of factors.

\paragraph*{Factor clusters}
For each set of factors, we calculate the mean Silhouette coefficient for different numbers of clusters with KModes to determine the best number of clusters to describe our data set.
Compared to clustering with all factors, we found that the best-fitting number of clusters is larger when considering our separated factors.
For \textit{Non-conditional-UX} comparison factors, we found 5 clusters, and for \textit{Conditional-UX} comparison factors, 10 clusters.
For \textit{Non-conditional-security} comparison factors, we calculated 9 clusters as the best number of clusters. For \textit{Conditional-security} comparison factors, Silhouette testing showed 8 to be the best number of clusters.
As a result, considering sets of separated factors, we found more diverse strategies for how websites in our data set implement their 2FA user journeys with regard to purely UX or security.
Appendix~\ref{appendix:separation_factors:clusters} illustrates the corresponding clusters.

\subsection{Qualitative Data Analysis}
\label{sec:results:qualitative}

We discuss the consistencies and inconsistencies we observed during our analysis of the 2FA user journeys.

\subsubsection{Consistent Discovery for Self-Motivated Users}

Our analysis shows that the vast majority of websites in our data set did not \textit{immediately} promote 2FA to their end users in any form before/during sign-up and login---a website might promote 2FA only at a later point (e.g., an account existed for some time or the user takes actions that increase the severity of an account compromise), which our recording of journeys does not cover. The few websites that immediately promoted their 2FA support did this with mixed strategies, where most of them promoted 2FA during or immediately after account creation. In contrast, the remaining websites mentioned it only on their landing page, where users could easily miss it. However, we discovered that some websites' nudging to 2FA merely redirected the user to the account settings' security section, where the user has to pick up the journey themselves. Furthermore, six websites in our data set mandated 2FA, most of those sites in the cryptocurrency category. However, for two websites that mandate 2FA, we found that the intention to use the phone number or verified email address as a second factor was not clearly communicated to the user during account creation (e.g., Fig.~\ref{fig:icloud} in Appendix~\ref{appendix:examples}).

Our analysis showed that users looking for 2FA settings have a consistent experience across websites. Almost all websites used a common location for their 2FA settings. Therefore, users who once went through the 2FA workflow can find the 2FA settings more easily on other websites. Most websites also use similar descriptions of 2FA (e.g., ``second layer of security,'' ``prevent unauthorized access,'' or ``ask for authentication on new devices''), which helps the user to recognize the 2FA settings despite variations in the naming. Examples of clear exceptions to this pattern are illustrated in Figure~\ref{fig:Tumblr} and Figure~\ref{fig:callcentric} in Appendix~\ref{appendix:examples}.

\subsubsection{Consistent Lack of Informing and Educating Users}

We found that only a minority of the websites provided additional information (e.g., ``learn more'' link to detailed information including pictures and tutorials), and even fewer websites educate the user about the benefits and drawbacks of the 2FA options that they support, but instead immediately start the setup process. During this setup, only about a third of all websites guided the user with step-by-step instructions for setting up a chosen 2FA option. 
Most websites require the user to verify their identity to change their 2FA settings and inform them about such changes (e.g., by email). Very noticeable exceptions are the websites in Cluster~1, which almost entirely omit both verification and notification of settings changes.

The most consistent behavior we have observed to inform users is the confirmation of a successful setup. More than four-fifths of the websites required a positive confirmation from the user (e.g., the user had to enter the current TOTP code to complete the setup) and, in most cases, also provided some visual feedback to the user to inform them about the successfully concluded 2FA setup.

\subsubsection{Mixed Strategies for 2FA Setup and Configuration}

We observed the most inconsistent behavior when it comes to the setup of 2FA options and their possible configurations by the user. First, there is an almost even split between three basic strategies: ``offering only one 2FA option,'' ``offering multiple 2FA options but only one can be active at a time,'' and ``offering multiple 2FA options and multiple can be active at the same time.'' Unfortunately, we could not find an explanation on any of the websites that let their users select only one 2FA option about why they implement 2FA this way.
Second, among the websites that support multiple 2FA options, all but six websites show the 2FA options grouped in the same settings location, while those six exceptions, for instance, differentiate between 2FA and security keys in their security settings. However, thirdly, half of the websites with support for multiple 2FA options enforce a particular 2FA option to be set up before they offer the other options to the user. For example, only after providing their phone number can the user set up security keys or TOTP as an alternative. Fourth and last, websites are very consistent in proposing the 2FA option that should be used to login. Very few websites allow the user to select the 2FA option that should be used primarily for the login. Only a single website asked the user upfront during login which 2FA option they would like to use for the current login (see Figure~\ref{fig:idme} in Appendix~\ref{appendix:examples}). The vast majority of websites used internal metrics to determine which 2FA option should be used for login, and the user could only navigate through the ``use a different method'' or ``do you have difficulties'' menus to select another 2FA option.

\subsubsection{Mainly Optional Recovery}

Three-quarters of all websites offer recovery options, and most of those websites also explain to the user the importance of setting up recovery options or the risks of neglecting to set up a recovery option. The preferred recovery option among these websites was printable one-time codes. Also, websites are very consistent in enforcing the setup of a recovery option. Almost three-quarters of the websites with a recovery option nudge the user to set up the recovery, and only six websites enforce this during the 2FA setup. This low number of websites with mandatory recovery also means that there is no fail-safe account recovery strategy by websites that support at most one active 2FA option. It would be intuitive that such websites would enforce a recovery option to prevent account lockout in case this single option is unavailable. Still, our data set does not support this. For the usage of recovery options, we found only one website that, although supporting one-time codes, does not offer an obvious way to use them (i.e., there was no link to a recovery page and no instructions to use the recovery codes as input to the regular OTP form field).

\subsubsection{Mixed Strategies for Device Remembrance}

More than half of the websites in our data set do not support device remembrance, i.e., the user cannot explicitly select to skip the second-factor authentication during future logins. For the websites that support this feature, we found that not only do they describe it in different ways but also that their remembrance logic differs. Almost two-thirds of the websites need the user to opt-\textit{in} to this feature, a fifth of the websites needs the user to explicitly opt-\textit{out} from remembering the device, and another fifth of the websites automatically places a device remembrance cookie without asking the user.

\subsubsection{Consistent Support for Deactivation}

All but five websites in our data set support the deactivation of 2FA (and only two of those exceptions mandate the setup of 2FA). However, only a minority of websites communicate to the user the risk of deactivating 2FA (e.g., easier account hijacking). Furthermore, similar to the previously mentioned lack of consistently informing and educating users about 2FA, we find that only about half of all websites verify the user identity before deactivation, notify the user about deactivation, or communicate a successful deactivation in the website settings.

\section{Discussion and Future Work}
\label{sec:discussion}

\textit{Are the websites in our data set consistently following the same design patterns and strategies in offering 2FA?}
% No.
Although some factors individually show high consistency, we did not find a single start-to-end design pattern for the user journey that is consistently followed by the majority of websites in our data set.
Instead, we found that 2FA user journeys are clustered into smaller groups of websites with similar journeys.
Separating the comparison factors by their impact on UX and security did not indicate a more consistent strategy for pure usability or security-related steps along the user journey.
In fact, we found that the separated sets of factors were more clustered.
Taking into account the rankings of websites, our results indicate that the design represented by clusters~2, 3, and 6 is more popular among the top-ranked websites, while more than 50\% of the websites from the long-tail ranks are in cluster~1.

\textit{Implications for developers and users:}
UX guidelines state that users prefer a site to work in the same way as all other sites they already know.
This heuristic has been shown to be successful on the Web, for example, when it comes to online shopping or banking experiences.
To follow this heuristic vis-a-vis two-factor authentication as a factor for the overall user experience on the Web, developers could follow the 2FA experience on the majority of websites or on the most popular websites, which likely minted the users' mental models.
Our results show that such a majority does not exist among the popular websites and that even the leading websites (e.g., Google and Apple) do not agree in their user journeys.
Therefore, a recommendation for influential industry associations and consortia would be to draft recommendations for website developers on how to achieve a consistent strategy for 2FA user journeys.
Possible avenues for the community and future work to contribute to this endeavor could be to create guidelines that foster consistent strategies for implementing the best possible 2FA UX on the Web.

A crucial consideration when striving for consistency is that consistency in itself does not guarantee a good UX; a bad design could be consistently implemented, but users have learned to live with it. 
As a concrete example from our data set, Apple's \url{icloud.com} is an outlier in various comparison factors: it mandates a phone number as the only 2FA option without clearly informing the user during account creation and without the option to add other options later or to deactivate it (see Appendix~\ref{appendix:icloudcom}).
But do users conceive \url{icloud.com}'s 2FA user journey as a bad or good experience?
Our study design does not attempt to assign a quality measurement to individual factors, and it does not measure the quality of user experiences attached to the different clusters of user journeys.
But clearly, our results motivate that the impact of the different strategies for the 2FA user journey on the perceived usability by users has to be thoroughly investigated in an effort to make the best strategy consistent across websites.
Our data indicated a connection between the rank of a website and the site's strategy, but it is unclear to what extent the guidelines for 2FA journeys need to be contextualized.
For example, certain comparison factors may be dictated or recommended by regulations, such as PCI-DSS or the EU Revised Payment Services Directive (PSD2) with strong customer authentication (SCA), different types of websites may have different security policies, or specific user groups~\cite{das_hisc10,Napoli2021}) require different support.
Thus, it is unclear whether consistency between \textit{all} comparison factors is required or even desirable.

\textit{Indications for the external validity of prior works:}

Although we did not study the usability of individual instruments, comparison factors, or steps in the user journey, we can provide a new perspective on some aspects of previous work and UX guidelines we observed during our data analysis.
We noted that the discovery of 2FA and the initial education of users are very consistent and that there is a common naming and description of 2FA in place.
But neither of the two types of 2FA descriptions that we have noted in our analysis (see Table~\ref{tab:overview_nld}) complies with recent results by Golla et al.~\cite{driving_2fa} and Lassak et al.~\cite{274547} on how users should be nudged and educated to encourage the adoption of 2FA.
Furthermore, Ciolino et al.~\cite{238317} conducted a user study of 2FA setup and login ``in the wild.'' 
Their participants encountered some of the patterns we identified in our work and described them as problematic.
For instance, enforcing the SMS 2FA option while not communicating that additional 2FA options become available after registering the phone number confused participants that were explicitly looking for registration of security keys; an opt-out device remembrance, which we found on several websites ($7\times$\yes, $8\times$\kindaright), frustrated participants that were expecting to be prompted for a second factor on login but missed that they had to take explicit action for that; and their participants expressed the desire for personalization by being able to select the preferred 2FA option for logins, which we found is not a widespread feature but, on the contrary, the 2FA option is in most cases chosen by the website (only 8 out of 34 websites with fully matched \textit{Multiselection} allowed setting a primary option, and only one website of the remaining 26 sites did \textit{not} pre-select the option).
Lastly, from our clustering, we noticed that recommendations by UX guidelines to provide adequate contextual help and break down complex tasks, in this case for setting up 2FA, were ignored on many websites that did not offer additional or option-specific information or simply step-wise instructions.
Also, the recommendation to provide clear feedback to users was not realized on many websites that did not notify users or communicate a successful 2FA setup or deactivation.
Thus, our study provides indications for the external validity of prior results.
In our opinion, measuring to what extent each pattern we detected matches the recommendations and settings of related work would be an interesting follow-up study to provide better insights into the external validity of previous studies (e.g., taking textual content and UI designs into more consideration).
Those indications also emphasize the need to establish more general UX guidelines for implementers of 2FA user journeys to improve the usability of 2FA.
The first option-specific guidelines~\cite{fido_ux,fido_securitykey_ux} or collections of best-practices~\cite{dias21,amiconsult} are a good starting point.

\textit{FIDO UX Guidelines~\cite{fido_ux,fido_securitykey_ux}:} The FIDO Alliance UX guidelines also consider similar steps in the user journey (promotion, invitation, registration, and login). They recommend the promotion of biometric awareness or security keys at sign-in and registration, educating users about the FIDO value proposition of a ``simple and secure sign-in without password'' or about authentication with security keys, providing a ``learn more'' link and giving concrete statements based on user studies, confirming successful registration with a clear indication to users, encouraging users to register mulitple keys for recovery and backup~\cite{fido_securitykey_ux}, and explicitly promoting ``Security and Privacy settings'' to manage 2FA options. Unfortunately, these guidelines are not suitable as a general guideline and, in some points, conflict with recent recommendations from research (e.g., the promotion message~\cite{274547,driving_2fa} or automatically setting FIDO2 as the default sign-in option~\cite{238317}). The guidelines~\cite{fido_ux} are strongly tailored to promote biometric authentication as a convenient alternative to passwords or to promote 2FA with security keys to consumers of regulated industry websites~\cite{fido_securitykey_ux}, such as banking or healthcare. 
They do not target a 2FA setting~\cite{fido_ux}, and the guidelines do not address the setup and UX of multiple authentication options, or they limit themselves to only security keys as the second factor~\cite{fido_securitykey_ux}.
For desktop authenticators~\cite{fido_ux}, the password is considered a fallback option; therefore, these guidelines omit explicit recovery steps.

\textit{Limitations:}
Like any other qualitative study, our work also has some limitations.
Despite our best efforts, we cannot exclude a subjective bias by the involved researchers, e.g., in identifying the comparison factors or selecting a clustering with the best descriptive power.
We aimed to study 120 popular websites, but only 85 were possible due to various restrictions and obstacles. Thus, our study is skewed toward top websites in English and from specific categories.
We fixed the conditions for data collection to increase the internal validity of our data, but we cannot exclude that our setting is considered high-risk or low-risk by a website and that we experienced a different user journey than other users of the same site.
Moreover, we collected our data only from desktop computers; thus, our comparison factors may differ on mobile devices.
Furthermore, with the adoption of new technologies (e.g., WebAuthn) and changes in website policies (e.g., Google plans to mandate 2FA for an increasing number of its users~\cite{google_mandatory}), our comparison might not capture the most recent picture. However, we believe that our general results remain valid.
Lastly, we did not continue to monitor the websites, nor did we explore the user flow for going through account recovery or changing 2FA-relevant information (e.g., phone number or email address), since we focused on the steps of the user journey that mint the users' initial impressions of 2FA.

\textit{Future work:} Conducting user and developer studies is an obvious path to follow up on our results. While we detected a lack of consistency in the 2FA user journeys that can increase users' cognitive friction, it is unclear if this contributes to the notoriously low adoption rate of 2FA among end-users. Our survey indicated that several users did indeed refrain from setting up 2FA or deactivated it due to differences in the user experience between websites. Furthermore, it is unclear whether a ``gold standard'' for journeys exists or to what extent journeys need to be contextualized (e.g., website category, regulations, or specific user groups). Comparative studies of different design patterns could answer those questions and others, like a weighting of comparison factors by their impact on, e.g., the UX or 2FA security. Moreover, we consider it worthwhile to explore developers' reasons for choosing a particular design pattern to understand the reasons behind those inconsistent journeys. In addition to human-centered studies, extending our methodology to user journeys for account recovery, to other device form factors, such as mobile devices, or to entirely new solutions, such as Passkey, would complement our results. Lastly, we think that studying the 2FA user journeys can provide insights into the external validity of (prior) studies of individual aspects of 2FA and shed new light on what constitutes a good 2FA user experience.

\section{Conclusion}

This work contributes a methodology for comparing 2FA user journeys on websites and presents the first systematic study of the consistency of those journeys on top-ranked websites. Our results show a lack of consistency for the various steps along those journeys. We find that even the more consistent design patterns were described as problematic for usability in the literature. We strongly believe that our results motivate different future works that can lead to the creation of more general user experience guidelines for implementers of two-factor authentication.

\bibliographystyle{IEEEtranS}
\bibliography{references}

\appendices
\section{Sampling Survey about 2FA User Experience}
\label{appendix:survey}

We conducted a survey among 2FA users to gather users' experiences with different 2FA journeys and gain insights into whether users had negative experiences transferring their 2FA experiences between websites and whether this has stopped them from enabling or continuing to use 2FA.
In the following, we first explain the structure of our survey (Appendix~\ref{appendix:survey:methodology}), followed by an overview of the results (Appendix~\ref{appendix:survey:results}, and lastly we provide a brief conclusion (Appendix~\ref{appendix:survey:conclusion}).

\subsection{Sampling Survey Questions}
\label{appendix:survey:methodology}

The structure of our survey was as follows: First, we presented a general welcome message that explained 1)~that this survey is about 2FA user experiences, 2)~which types of personal data are collected and how the collected data is handled, 3)~that participation is voluntary and participants have the right to withdraw their consent at any time during the study, 4)~that are bonus tasks in this survey but not when and how, and 5)~who is conducting the study and which contact points exist.
Afterward, participants followed the survey structure below, on which we converged after three rounds of pre-testing with 10 randomly selected Prolific participants in each round (cf.~Appendix~\ref{appendix:survey:results:demographics}):

{\small
\noindent\textbf{[Q1: Same notion of 2FA]} 
As you mentioned in the first part of the study that you use two-factor authentication on different websites, we assume that we don't need to explain the concept to you anymore. However, to make sure that we all refer to exactly the same thing, here is a short definition from our side:
Two-factor authentication (short: 2FA, but also sometimes referred to as two-step verification) is a method to confirm a user has claimed online identity by using a combination of two different types of authentication methods. Typically a password is considered one type, and with 2FA, the password is combined with another type to increase login security (for example, code via SMS or from an app like Google Authenticator or Duo Security, or fingerprint). However, websites often differ in how they implement 2FA in detail and therefore also in the experience for the user.\\  
\textit{Does this definition match what you mean by 2FA? (i)~Yes, (ii)~No}

\noindent\textbf{[Q2: Current and past use of 2FA]} 
Please tell us on which websites you currently or in the past have used 2FA.
Additionally, we are also interested in websites where you have started to set up 2FA but have aborted the setup (please scroll further down to enter those).\\
\noindent\textbf{[Q2a]} \textit{I currently use 2FA on the following websites:
(Up to 20, please name the site and not use generic terms such as "banking websites"):} \textbf{[Form field]} \\
\noindent\textbf{[Q2b]} \textit{I have used 2FA on the following websites in the past but stopped using it, or I started to set up 2FA on these websites but aborted the setup (up to 10, please name the site and not use generic terms such as "banking websites"):} \textbf{[Form field]}

\noindent\textbf{[Q3: Recognized differences between websites]} Please read the following instructions really carefully and take your time to think about your answer! Since this question is really important to us, we have set a short timer so that you can continue the survey only after 45 seconds (the "Next" button will appear automatically). Now, please take a moment to compare your 2FA experiences with all websites you listed on the previous page \textit{(they are also listed below on this page)}. Those experiences include how the individual sites present 2FA, the options for the second factor each site offers, or the steps to set up 2FA and log in on each site. Please note that we are not asking about your general impression of 2FA but the differences between the 2FA on different websites that you noticed. If you did not notice such differences, please answer accordingly below. \\
\textit {Are there one or more websites where your experiences, based on the above definition, differed significantly positively or negatively from the other sites? If so, select them below. If you did not notice any differences, select none of the websites and click "Next".
[Website A], [Website B]...[Website N]}\\

\noindent\textbf{[If no website selected in Q3 go to Q6]}\\

\noindent\textbf{[Q4: Opt-in to bonus task]} You are almost through the survey, and only one page of questions about your experience with 2FA is missing. However, we would like to offer you a bonus task where you can earn some extra money. On the previous page, you stated that your experience with 2FA is very different on the following website(s): [website A] [Website B] ... [Website N]. We would be very interested in what exactly made the difference in your experiences with 2FA on these websites. For each of these sites, we have a few questions for you.
It usually takes no longer than 2 minutes to answer all the questions per website, and you will get paid a bonus of 50 cents per website. With this bonus task, you can earn another [number of websites in Q3 $\times$ 0.50] pounds with this bonus task. \\
\textit {Would you like to do the bonus task? \\
(i)~Yes, take me to the bonus task!, (ii)~No, take me to the last questions of your survey! } \\

\noindent\textbf{[If answer to Q4 is \textit{No} then go to Q6]} \\

\noindent\textbf {[For each website selected in Q3 ask:]}
\mdfsetup{%
    topline=false,
    rightline=false,
    bottomline=false,
    linecolor=black,linewidth=2pt
}

\begin{mdframed}
\noindent\textbf{[Q5]} Please tell us a little bit more about [Website]. \\
\noindent\textbf{[Q5a: Differences]} \textit{How exactly did your 2FA experience differ on [Website] from the other websites that you listed? (Those experiences include how [Website] presents 2FA, its options for the second factor, or the steps to set up 2FA and log in on [Website])[\textbf{Open-text question]}} \\
\noindent\textbf{[Q5b: Usability]} \textit{Did this difference make 2FA on [Website] easier or harder for you to use in comparison to other websites with 2FA? (i)~Easier, (ii)~Equally easy, (iii)~harder}\\
\noindent\textbf{[Q5c: Behavior]} \textit{How has this changed your usage behavior on [Website] compared to other websites on which you use 2FA? For example, do you log in to the website more/less? Or was that maybe even a reason why you don't use 2FA on the site in the end?} [\textbf{Open-text question]}\\
\noindent\textbf{[Q5d: Recommendation]} \textit{Based on your experience of 2FA on [Website] Is there something that the other websites should adopt for their 2FA or should they avoid?} [\textbf{Open-text question]}\\
\end{mdframed}

\noindent\textbf{[Q6: Concrete situation]} You might be used to a particular experience of 2FA, meaning how websites present 2FA, which second factors you can use, or how to set 2FA up and log in. \\
\noindent\textbf{[Q6a: Problematic inconsistency]} \textit{Do you remember a concrete situation where your 2FA experience was challenging to you because it differed from what you were used to? Please give us as many details as you can remember about this situation.  [\textbf{Open-text question].}}\\
\noindent\textbf{[Q6b: Solution]} \textit{How did you behave in this situation? For example, you aborted setting up 2FA, you learned a new way to use 2FA, or it didn't bother you. [\textbf{Open-text question]}}

\noindent\textbf{[Q7: General feedback]} Lastly, here is some space to leave us feedback or questions, if you like. [\textbf{Open-text question]} \\
}

\subsection{Results}
\label{appendix:survey:results}

\subsubsection{Recruiting and Demographics}
\label{appendix:survey:results:demographics}

We recruited our participants via the \textit{Prolific}\footnote{\url{https://www.prolific.co/}} platform.
Prolific collects basic demographic information\footnote{\url{https://researcher-help.prolific.co/hc/en-gb/articles/360009221093-How-do-I-use-Prolific-s-demographic-prescreening-}} about their participant pool, to which we added a pre-screening question to select only participants that stated that they use 2FA on at least two different websites.
We used Prolific to create a pre-screened participant pool whose demographics are representative of the US population and that has an approval rate of more than 90\% on Prolific. 
From this pre-screened participant pool, we selected a random sample of 30 participants for 3 rounds of pre-testing the survey to remove any ambiguity as much as possible.
After pre-testing, we opened the survey for 300 participants from this pool, where 309 in the end successfully finished the survey and we also paid the 9 participants that submitted their completion code too late.
For the basic survey (all questions except Q5), which we measured in pre-testing to take about 5~min, we paid \textsterling1.20.
This corresponds to an hourly wage of \textsterling14.40 ($\approx$\$16.50/hour).
For each additional website in Q5, we estimated 2~min work time and paid \textsterling~0.50 (i.e.,~\textsterling15.00 hourly wage).

\begin{table}[t]
    \centering
    \scriptsize
    \def\arraystretch{1.1} % height of row
    \begin{tabular}{l|r@{\hskip3pt}r}\toprule
    Number of participants & & 308 \\\hline
    \rowcolor[gray]{.9}
    \multicolumn{3}{l}{Gender}\\\hline
        Male            &  179 & (58.1\%) \\
        Female          &  128 & (41.6\%) \\
        No answer       &    1 & (0.3\%) \\
    \hline
    \rowcolor[gray]{.9}
    \multicolumn{3}{l}{Age group}\\\hline
        18-19 &   2 & (0.7\%) \\
        20-29 &  54 & (17.6\%) \\
        30-39 &  62 & (20.3\%) \\
        40-49 &  64 & (20.9\%) \\
        50-59 &  69 & (22.5\%) \\
        60-69 &  38 & (12.4\%) \\
        70-79 &  16 & (5.2\%) \\
        80-89 &   1 & (0.3\%) \\
        No answer & 2 & (0.01\%)\\
    \hline
    \rowcolor[gray]{.9}
    \multicolumn{3}{l}{Language}\\\hline
        English &  286 & (92.9\%) \\
        Other   &   22 & (7.1\%) \\
    \hline
    \rowcolor[gray]{.9}
    \multicolumn{3}{l}{Nationality}\\\hline
        United States &  291 & (94.5\%) \\
        Other         &   17 & (5.5\%) \\
    \bottomrule
   \end{tabular}
    \caption{Demographics by Prolific of our participants.}
    \label{appendix:tab:demographics}
\end{table}

One participant in our sample disagreed with our definition of 2FA (Q1), and the results were excluded from the final data set.
The demographics as collected by Prolific for the remaining 308 participants are summarized in Table~\ref{appendix:tab:demographics}.
The average age is 45.2$\pm$1.7 years.
Most participants are US citizens (94.5\%) and have English as a first language (92.9\%).
Of all participants, 58.1\% identified as male, which unfortunately differs from the US census 2021 estimation\footnote{\url{https://www.census.gov/quickfacts/fact/table/US/PST045221}} where 50.5\% identified as female.

\paragraph{Ethical considerations} For this survey we used the existing infrastructure by a fellow research group at our institution for a series of user studies on Prolific. Since our survey is more restricted in the collection of private data than the remaining user studies, the existing approval by our institutional ERB extended to our survey.

\subsubsection{Coding Open-Text Answers}
\label{appendix:survey:results:coding}

Our participants answered open-ended text questions about their 2FA user experiences on different websites.
Such qualitative data allows to capture individual perceptions, thoughts, and concerns of users.
We used inductive coding (see~\cite{Lazar:2017:IHI:3027063.3027102,doi:10.1177/1094428112452151,wilhelmy_2016,merriam2015qualitative}) to analyze their answers.
Two researchers jointly read a randomly sampled subset (25\%) of the open-text answers of our participants to Q5 and Q6, discussed them and developed an initial coding scheme for each question where each question was assigned one or more codes.
In the next step, the initial codes for each question were merged by axial coding to more abstract codes and the final code books for each question.
After this step, two researchers independently assigned a code from the final code books to each answer for Q5 and Q6.
We calculated Cohen's Kappa for the finally assigned codes to the answers of Q5 and Q6 to measure the inter-rater reliability (correspondence between the coders).
The two coders achieved a satisfactory to near-perfect mean inter-rater reliability, see Table~\ref{tab:appendix:irr}.

\begin{table}[t]
    \centering
    \scriptsize
    \def\arraystretch{1.3} % height of row
    \begin{tabular}{r|rrrrr|r}\toprule
            \textbf{Question}       & Q5a & Q5c & Q5d & Q6a & Q6b & \textbf{Mean} \\
            \textbf{Cohen's Kappa}  & 0.781 & 0.734 & 0.784 & 0.835 & 0.793 & 0.785 \\
            \bottomrule
    \end{tabular}
    \caption{Cohen's Kappa for inter-rater reliability for coding the open-text answers to Q5 and Q6.}
    \label{tab:appendix:irr}
\end{table}

\subsubsection{Qualitative and Quantitative Results}
\label{appendix:survey:results:qualitative}

Our participants named, on average 5.06$\pm$0.38 websites on which they currently use 2FA (question \textbf{Q2a}) and 0.61$\pm$0.12 websites on which they stopped using/setting up 2FA (\textbf{Q2b}).
However, these numbers have to be taken with a grain of salt since a few participants remarked that they could not recall all websites on which they use 2FA or that they could have listed more than the maximum of 20 we asked for.
Of all 308 participants, 150 participants (48.7\%) indicated a difference in the 2FA experience on at least one of the websites they listed (\textbf{Q3}), 158 did not note a difference.
Those 150 participants noticed, on average, differences on 1.51$\pm$0.31 websites of the previously named ones.
For the websites where the participants currently use 2FA, our participants noted differences on average on 1.25$\pm$0.32 sites.
For websites where they abandoned 2FA, they noticed differences on average on 0.26$\pm$0.07 sites.

\begin{figure}
    \centering
    \includegraphics[width=\linewidth]{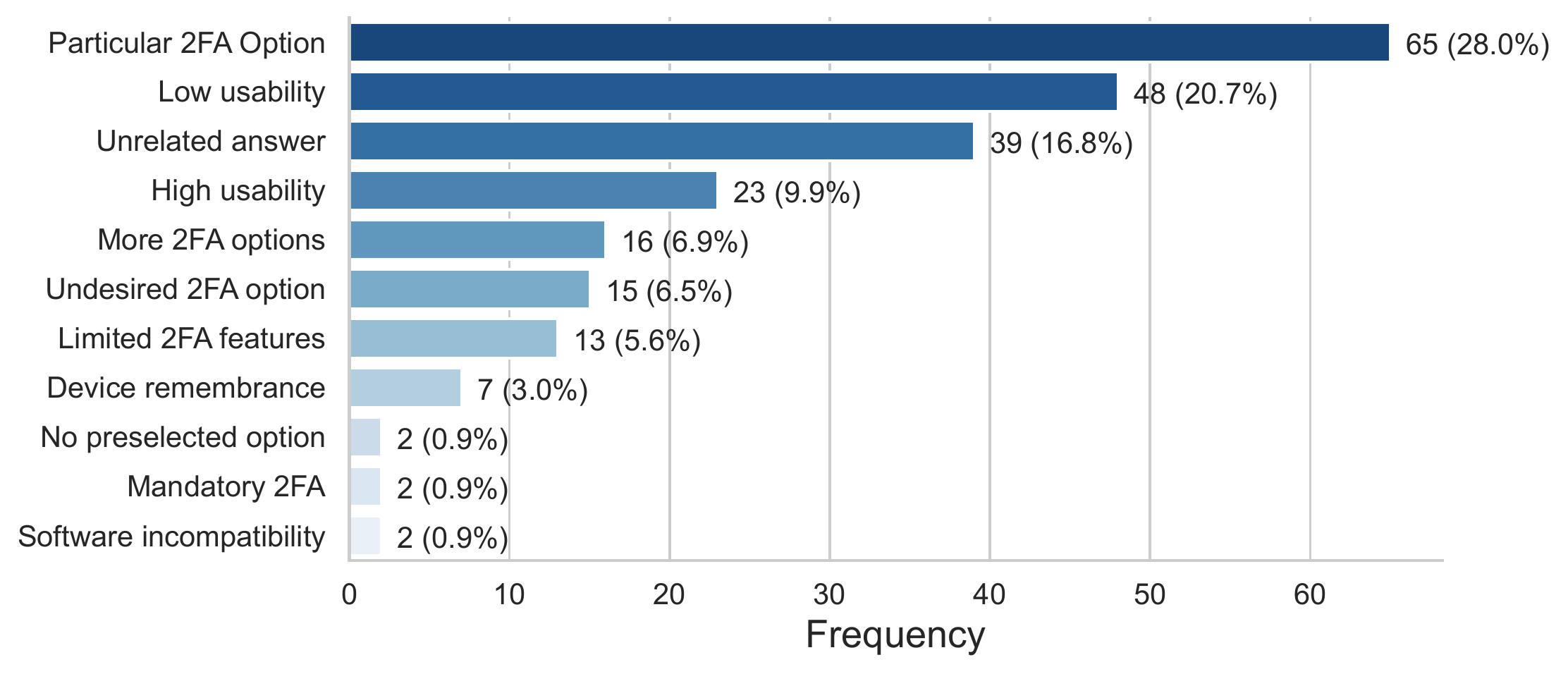}
    \caption{Codes for \textbf{Q5a} ordered by their frequency.}
    \label{fig:differences_hist}
\end{figure}

\begin{figure}
    \centering
    \includegraphics[width=\linewidth]{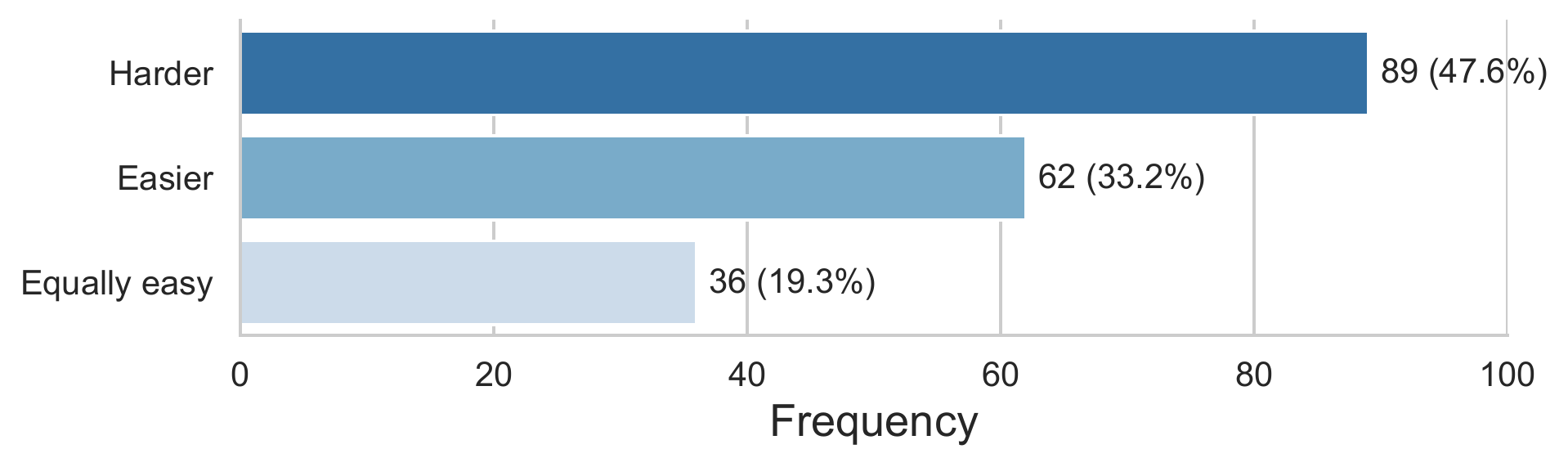}
    \caption{Quantitative answers for \textbf{Q5b} ordered by their frequency.}
    \label{fig:effect_hist}
\end{figure}

\begin{figure}[t]
    \centering
    \includegraphics[width=\linewidth]{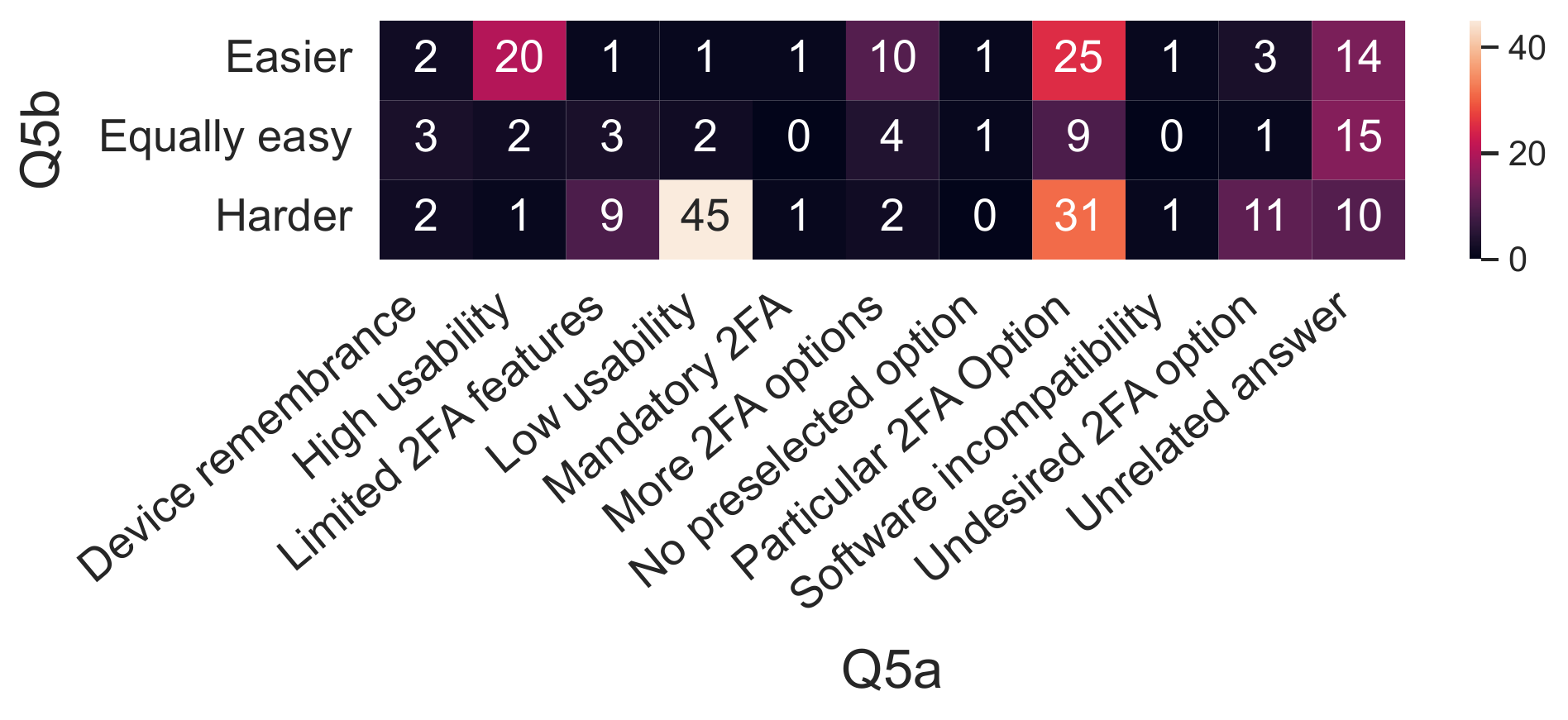}
    \caption{Contingency table between codes for \textbf{Q5a} and answers for \textbf{Q5b}.}
    \label{fig:confusion_matrix_differences_effect}
\end{figure}

\begin{figure}[t]
    \centering
    \includegraphics[width=\linewidth]{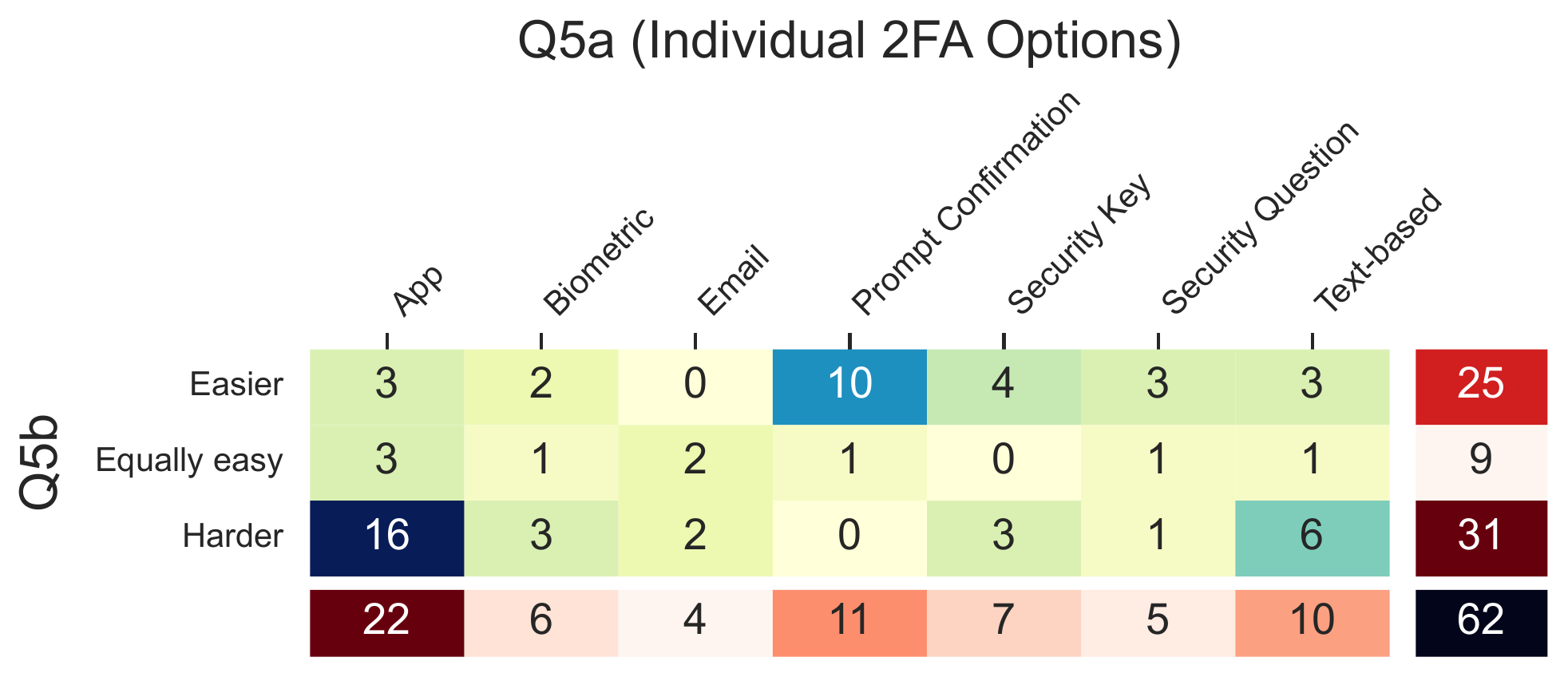}
    \caption{Contingency table between individual 2FA options mentioned in answers for \textbf{Q5a} with code "Particular 2FA option" and answers for \textbf{Q5b}.}
    \label{fig:confusion_matrix_options_effect}
\end{figure}

\begin{figure}[t]
    \centering
    \includegraphics[width=\linewidth]{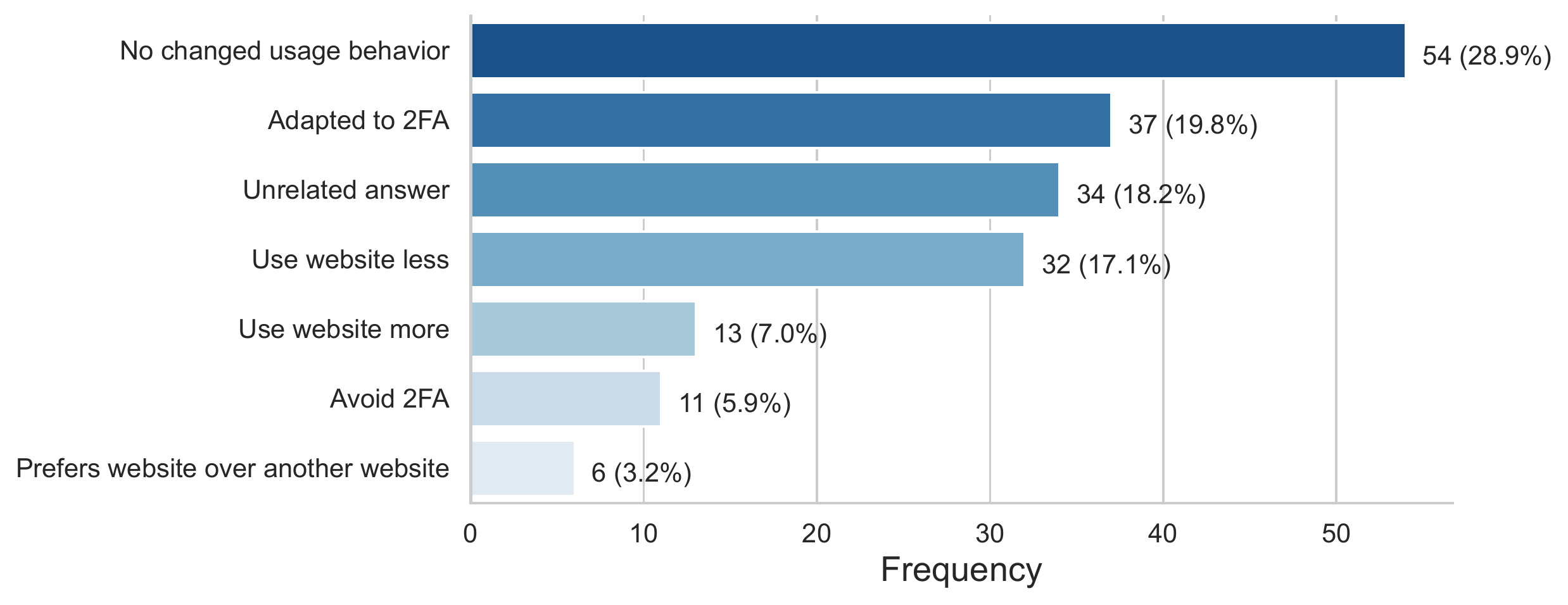}
    \caption{Codes for \textbf{Q5c} ordered by their frequency.}
    \label{fig:behavior_hist}
\end{figure}

\begin{figure}[t]
    \centering
    \includegraphics[width=\linewidth]{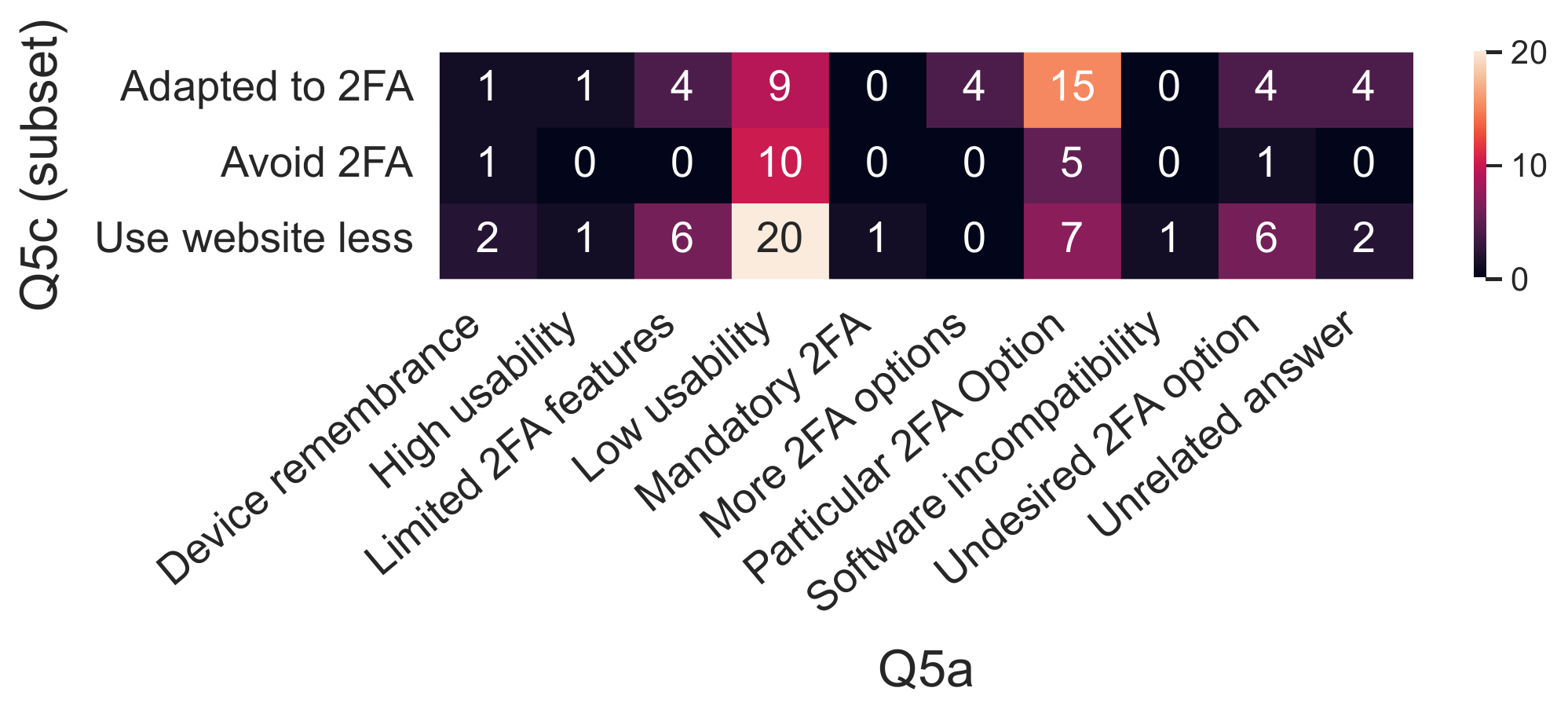}
    \caption{Contingency table between codes for \textbf{Q5a} and codes \textit{Adapted to 2FA}, \textit{Avoid 2FA}, and \textit{Use website less} for \textbf{Q5c}.}
    \label{fig:confusion_matrix_differences_behavior}
\end{figure}

\begin{figure}[t]
    \centering
    \includegraphics[width=\linewidth]{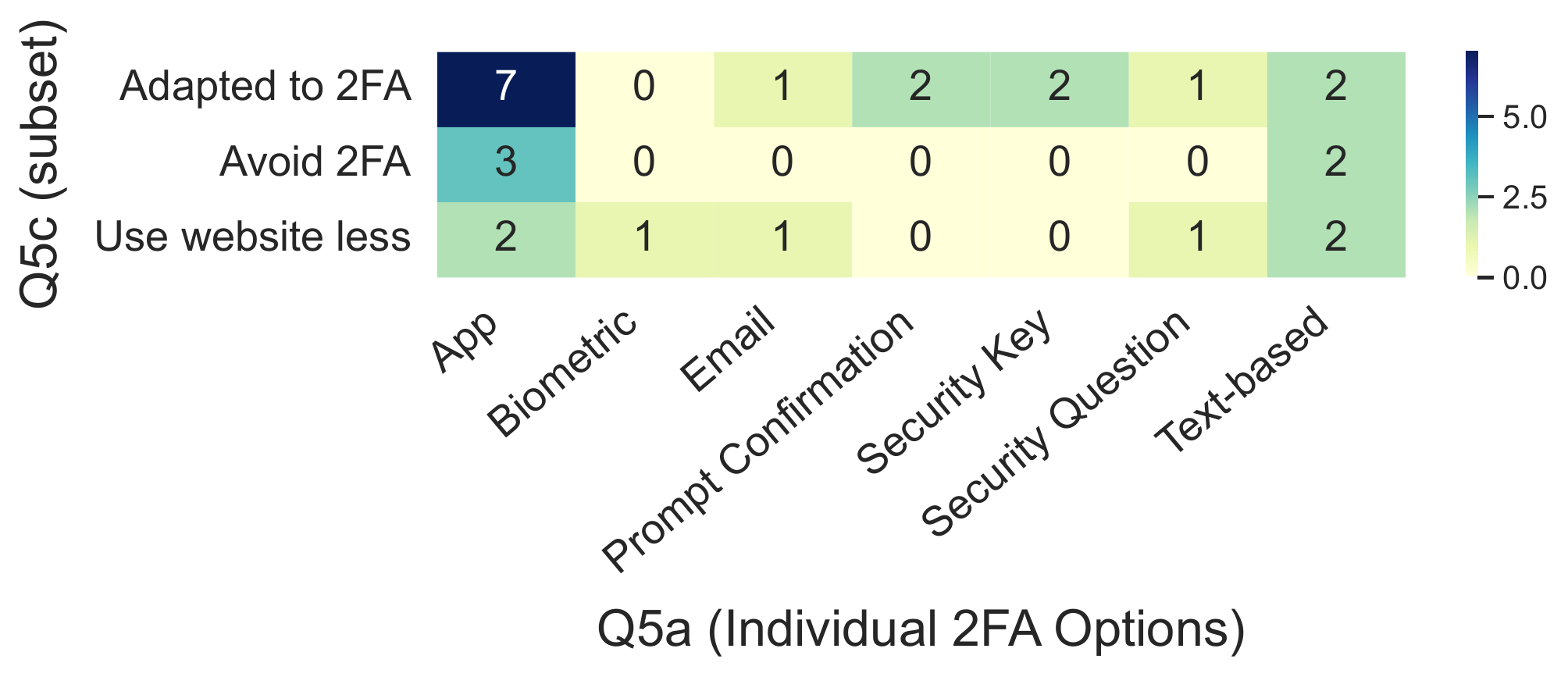}
    \caption{Contingency table between individual 2FA options mentioned in answers for \textbf{Q5a} with code "Particular 2FA option" and codes \textit{Adapted to 2FA}, \textit{Avoid 2FA}, and \textit{Use website less} for \textbf{Q5c}.}
    \label{fig:confusion_matrix_options_behavior}
\end{figure}

\begin{figure}[t]
    \centering
    \includegraphics[width=\linewidth]{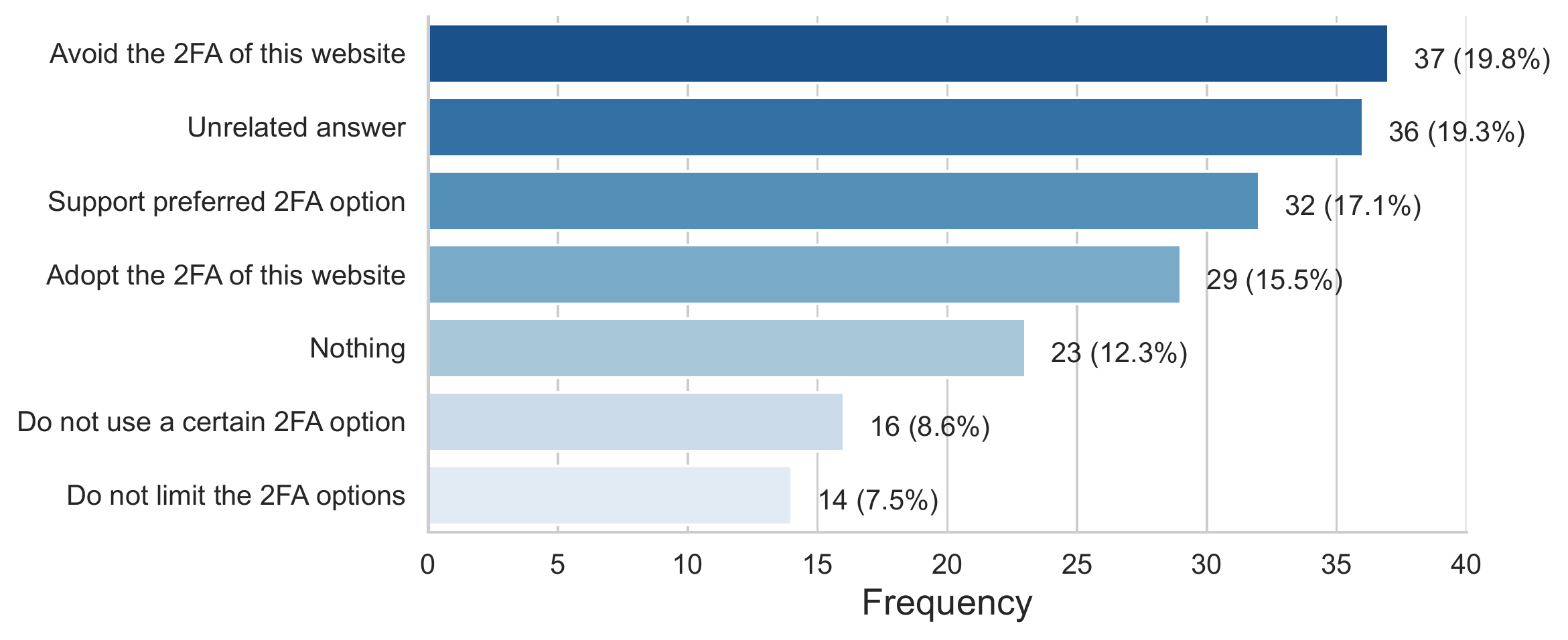}
    \caption{Codes for \textbf{Q5d} ordered by their frequency.}
    \label{fig:recommendation_hist}
\end{figure}

Of those 150 eligible participants, 112 opted-in (\textbf{Q4}) to answer the additional questions in \textbf{Q5}.
Figures~\ref{fig:differences_hist} to \ref{fig:recommendation_hist} depict the final codes for questions \textbf{Q5a} to \textbf{Q5d}, ordered by their frequency, which we elaborate on in the following.

\paragraph{Answers to \textbf{Q5a}}
When asked about the differences that they noticed on particular websites (\textbf{Q5a}), about a third (28\%) of the answers mentioned that the website offered a particular 2FA option (e.g., authenticator app, security keys, or confirmation prompts).
For example:
\begin{quote}
\textit{"it requires you to get a code from your email, I'm not sure if this is majorly different from other 2FA's but I have not seen it done this way before"} (P4, particular 2FA option )
\end{quote}
\begin{quote}
\textit{"This is the only site that is tied to Google Authenticator.  The other sites that I used, use simple text messages as the secondary method."} (P143, particular 2FA option)
\end{quote}
\begin{quote}
\textit{"Instead of sending you a verification code it sends you a push notification you have to click "yes it's me" or "decline". Others are always codes."} (P159, particular 2FA option)
\end{quote}
Almost a fifth of the answers (20.7\%) mentioned that this website differed by having comparatively lower usability of the 2FA experience, such as complex setups with too many steps, worse instructions than other sites, complex settings menu structure, or intransparent policies when 2FA is required during logins.
As examples:
\begin{quote}
\textit{"Veve sent you an email with a code instead of a text, and this made it harder to get the code. Plus, they had a short time limit that you had to get the code and put it in (Less than a minute or maybe even 30 seconds) and that was hard to make it sometimes."} (P208, low usability)
\end{quote}
In contrast, only 9.9\% of the answers mentioned that the website sticks out positively by providing higher usability of 2FA, for instance:
\begin{quote}
\textit{"Pay Pal explained how to use the 2 step system in more detail and in simpler more understandable terms than anyone else."} (P60, high usability) 
\end{quote}
\begin{quote}
\textit{"Google's 2FA is easy to set up and is very easy when using a hardware key because it supports U2F."} (P304, high usability)
\end{quote}
A few answers mentioned that the website differed in that it offered more 2FA options than other sites (6.9\%) or that it limits the 2FA features (i.e., 2FA and recovery options) too much (5.6\%), sometimes even to the extent that the participants desires another preferred option (6.5\%).
\begin{quote}
\textit{"They give me many options for how to verify myself."} (P21, more 2FA options)
\end{quote}
\begin{quote}
\textit{"Okta provides a multide (multiple) of options to enable 2FA on their site. Whenever I have an Okta prompt, I chose to setup an okta verify app that gives me a popup notification to click allow. They also offer text messaging with a code, emails, rotating numbers, and a few more that I don't recall right now as I only use the push notification to their app."} (P50, more 2FA options)
\end{quote}
\begin{quote}
\textit{"Facebook does not have an SMS part of their 2FA.  I prefer getting a quick message to my phone to authenticate my identity."} (P2, limited 2FA features)
\end{quote}
\begin{quote}
\textit{"I mention sofi because they do not have any backup for 2FA if I lose my phone. Most of the others sites have one-time-use codes we can print out or something like that. I set up two devices with google auth so that's my backup."} (P215, limited 2FA features)
\end{quote}
\begin{quote}
\textit{"All the other sites let me use an app to provide a code, but ally only allows a text based code. I don't feel it's nearly as secure."} (P108, undesired 2FA options)
\end{quote}
\begin{quote}
\textit{"I have to download a secondary app, Authenticator that has the 2FA code.  I'm generally fine with the quick SMS message, but I don't like to add additional apps to my phone when I don't think it is necessary."} (P148, undesired option)
\end{quote}
Lastly, a small number of answers mentioned very concrete technical differences regarding device remembrance support (7; 3.0\%), lacking support for selecting the 2FA during login (2; 0.9\%), mandating 2FA (2; 0.9\%), or incompatibilities with 3rd party software/services due to 2FA (2; 0.9\%).
Examples for those differences are:
\begin{quote}
\textit{"Most other websites seem to allow you to remember a device so you do not need to go through the 2FA process on a trusted device. For some reason Fidelity does not offer that and makes me go through 2FA every time I log in."} (P255, device remembrance)
\end{quote}
\begin{quote}
\textit{"Vk requires 2FA authentication to log in and doesn't give the option to not use it. A random number calls and the last 4 digits of that number are the authentication code. Sometimes the call takes up to a minute to appear, and you have to remember the number as it calls and type it in quickly."} (P302, mandatory 2FA)
\end{quote}
\begin{quote}
\textit{"53.com, the website for Fifth Third Bank, recently made 2FA mandatory. Despite my distaste in general for mandatory things, they made the process almost seamless. They seem to have planned very effectively for the change."} (P72, mandatory 2FA)
\end{quote}
\begin{quote}
\textit{"PAI offers several options for 2FA and offers them every time you log in. All other sites just offer you the choice once and every time after that you have to use the same option"} (P135, no preselected option)
\end{quote}
\begin{quote}
\textit{"They support FIDO2 which makes login a breeze. Their login is also simple with all your options presented to you to choose, except when FIDO2 is available to make login easier."} (P304, no preselected option)
\end{quote}
\begin{quote}
\textit{"The interface is very clumsy. The 2FA here also interferes with 3rd party apps like acorn."} (P62, software incompatibility)
\end{quote}
\begin{quote}
\textit{"Etrade requires an old-school RSA hardware key or special app, plus 2FA makes it hard to connect other services to it, so I turned it off. Their site is old and doesn't support the modern stuff."} (P215, software incompatibility)
\end{quote}
A little bit less than a fifth (16.8\%) of the answers were unrelated to the question, potentially because the participants misunderstood this or the previous question (\textbf{Q4}), e.g., they talked about general security benefits of 2FA or mentioned that the website does not differ from other websites (though we asked in \textbf{Q4} which websites \textit{differ} in their opinion).
For instance:
\begin{quote}
\textit{"First Citizen bank required 1 FA in person while online it required 2 FA via laptop."} (P191, unrelated answer)
\end{quote}
\begin{quote}
\textit{"enables security by providing 2FA but pretty much similar to other existing 2FA based apps or websites."} (P293, unrelated answer)
\end{quote}

\paragraph{Answers to \textbf{Q5b}}
Figure~\ref{fig:effect_hist} summarizes the quantitative data about how participants perceived those differences from \textbf{Q5a} in comparison to other websites.
Almost have of the answers (47.6\%) indicated that those differences made it harder to use the 2FA on that specific website, and only a third of the answers (33.2\%) indicated that the website's 2FA was easier to use.

Figures~\ref{fig:confusion_matrix_differences_effect} and \ref{fig:confusion_matrix_options_effect} provide a different view of this data.
Figure~\ref{fig:confusion_matrix_differences_effect} presents the contingency table between the codes for \textbf{Q5a} and the answers for \textbf{Q5b}.
We measured the statistical association the two variable with Cramer's V (with Bergsma bias-correction), and $V=0.43$ indicates that there is moderate to a strong association.
Not surprisingly, high usability makes the 2FA usage easier and low usability makes it more challenging.
However, the data also indicates that offering more 2FA options makes the 2FA usage easier while limiting 2FA options and/or forcing users to apply an undesired option makes the website harder to use.
Figure~\ref{fig:confusion_matrix_options_effect} splits the "particular 2FA option" code into the options identified in the participants' answers and shows the contingency table between individual options and the answers to \textbf{Q5b}.
The figure indicates that app-based two factor authentication is in particular correlated with a harder 2FA usage by our participants, while confirmation prompts are perceived as an easier to use 2FA option.

\paragraph{Answers to \textbf{Q5c}}
Figure~\ref{fig:behavior_hist} shows the final codes for the answers to question \textbf{Q5c}, ordered by frequency.
When asked how participants behaved in reaction to the differences in 2FA on a particular website, almost a third (28.9\%) responded that they did not change their behavior.
For example:
\begin{quote}
\textit{"I log into this website out the same as I normally would, I just find the process much more convenient than other options."} (P270, unchanged usage behavior)
\end{quote}
\begin{quote}
    \textit{"I've had no change in usage. But I do find it much easier to use 2FA instead of passwords."} (P220, unchanged usage behavior) 
\end{quote}
\begin{quote}
\textit{"No change to behaviour, but I do like it more."} (P305, unchanged usage behavior)
\end{quote}
Among the answers that described a changed usage behavior, most commonly, the participants indicated to have adapted to the particular 2FA of the website (19.8\% of all answers), e.g., learning how to use a prior unfamiliar 2FA option, or to use the website less frequently (17.1\%), in some cases even to the extent of abandoning the website or switching to a partner app for the service.
\begin{quote}
    \textit{"I went through all of the learning steps and I watched video tutorials." (P34, adapted to 2FA)}
\end{quote}
\begin{quote}
\textit{"I log in as much as I need to but prefer that they would let me pick the authenticator app or at least have more options available but since it has a time on it I try to get on and finish what I'm doing quickly without leaving my computer."} (P296, adapted to 2FA)
\end{quote}
\begin{quote}
\textit{"I use the website on my computer less and use the mobile app more often."} (P255, use website less)
\end{quote}
\begin{quote}
\textit{"I log on to the website less frequently than I typically would because the security questions get annoying after time."} (P296, use website less)
\end{quote}
A few participants described that they developed strategies to avoid having to use 2FA (5.9\%), e.g., trying to keep the logged in a session open as long as possible.
\begin{quote}
\textit{"I just try to remain logged in on Coinbase so I don't have to go through it again."} (P97, avoid 2FA)
\end{quote}
\begin{quote}
 \textit{"I try to not log out."} (P48, avoid 2FA)   
\end{quote}

\begin{quote}
\textit{"I try very hard not to have to like full login as in just leave my devices logged INTO google at ALL times so I can avoid the hassle of having all my devices go HOLD UP even for a moment."} (P201, avoid 2FA)
\end{quote}
Some participants with positive experiences with the 2FA on this particular site also described that they now use the website more often (7.0\%) or, prefer this website over another service (3.2\%).
For example:
\begin{quote}
\textit{"I log into this website more since it is an easy 2FA."} (P230, use website more often)
\end{quote}
\begin{quote}
\textit{"I login in more. It's quick and simple."} (P285, use website more often)
\end{quote}
\begin{quote}
\textit{"I login more because it is so quick and I just have to plugin or tap my yubikey."} (P304, use website more often)
\end{quote}
\begin{quote}
\textit{"I always choose it over the Barclays site"} (P167, prefer website over another)
\end{quote}
\begin{quote}
\textit{"I log into Chase more and bank with them because its easier rather than use a lot of the other banks. I like ease of use and I am a simple person."} (P299, prefer website over another)
\end{quote}
Lastly, about a fifth of the answers (18.2\%) were unrelated to the question, e.g., they referred to a companion app instead of the website, reported about an unrelated technical problem (e.g., a website was down), or related to problems with general security/privacy policies and risk-based authentication.
For example:
\begin{quote}
\textit{"I couldn't use Outlook outside of work for awhile"} (P37, unrelated answer)
\end{quote}
\begin{quote}
\textit{"I think I've started to think more about the data I share with other sites that feel less secure"} (P95, unrelated answer)
\end{quote}

Figure~\ref{fig:confusion_matrix_differences_behavior} relates the differences pointed out in \textbf{Q5a} with answers that mention that the participant had to adapt to 2FA on the specific website, tries to avoid the 2FA when possible or use the website less.
The data indicates that low usability, in contrast to other websites, caused the
participant to use the website in half of the cases less.
Limited 2FA features and undesired 2FA options almost equally caused participants to either adapt or to use the website less.

Figure~\ref{fig:confusion_matrix_options_behavior} depicts the contingency table for code \textit{Particular 2FA option} from Figure~\ref{fig:confusion_matrix_differences_behavior} broken down into individual 2FA options mentioned in answers to \textbf{Q5a}.
App-based 2FA has been mentioned most often in combination with \textit{Adapted to 2FA}, \textit{Avoid 2FA}, and \textit{Use website less}, where the majority of participants adapted to this option.
Only for app-based 2FA and text-based 2FA, our participants mentioned that they developed strategies to avoid two-factor authentication when possible.
Among our participants, all 2FA options except prompt confirmation and security keys have caused at least one participant to use a website less.

\paragraph{Answers to \textbf{Q5d}}
Figure~\ref{fig:recommendation_hist} shows the codes for answers to question \textbf{Q5d}, ordered by frequency.
When asked what participants would recommend other websites to adopt or avoid from 2FA on the website \textbf{Q5} is currently asking about, a fifth of the answers (19.8\%) suggested a general way to avoid the 2FA of that website.
This referred to general usability problems of 2FA, e.g., in the settings, setup, or usage.
For example:
\begin{quote}
\textit{"Avoid creating too many extra steps. Others like Yahoo do this seamlessly."} (P7, avoid 2FA of this website)
\end{quote} 
\begin{quote}
\textit{"Itunes should make their process simpler and and not force a re-login across different platforms like Google does, which is easier."} (P279, avoid 2FA of this website)
\end{quote}
\begin{quote}
\textit{"I do not think other websites should use the 2FA in a similar way to Udemy."} (P296, avoid 2FA of this website)
\end{quote}
\begin{quote}
\textit{"yes, use standard 2fa, don't try and reinvent the wheel."} (P43, avoid 2FA of this website)
\end{quote}
Fewer answers noted very particular things that other websites should avoid, such as a certain 2FA option (8.6\%) or limiting the 2FA options (7.5\%).
\begin{quote}
\textit{"Most other sites give multiple, more secure options for 2FA. Ally feels outdated and behind limiting you to either a text code or email code. Some sites allow multiple types of 2FA enabled at the same time (app, email, etc.) but Ally does not."} (P108, do not limit the 2FA options)
\end{quote}
\begin{quote}
\textit{"It's simple: Give users multiple different choices for 2FA!"} (P290, do not limit the 2FA options)
\end{quote}
\begin{quote}
\textit{"No other websites should use the text message code to authenticate an account."} (P270, do not use a certain 2FA option)
\end{quote}
\begin{quote}
\textit{"avoid SMS text messages."}(P308, do not use a certain 2FA option)
\end{quote}
Almost a fifth of the answers (17.1\%) concretely recommend that other websites should adopt their preferred 2FA option.
\begin{quote}
\textit{"I think all websites should have a click button or a push notification (whether app or to text) that is a one click verification."} (P50, support preferred 2FA option)
\end{quote}
\begin{quote}
\textit{"I think more sites should offer QR code login."} (P244, support preferred 2FA option)
\end{quote}
Of all answers, 15.5\% made general recommendations where other websites should adopt something from the two-factor authentication experience of this particular website.
For example:

\begin{quote}
\textit{"All sites should seek out an easier way like Yahoo where you only have to click OK rather than re-enter numbers."} (P7, adopt the 2FA of this website)
\end{quote}
\begin{quote}
\textit{"They should adopt the 2FA standards that Google support because it makes it much more easier."} (P304, adopt the 2FA of this website)
\end{quote}
Only 12.3\% of the answers did not make a recommendation about what to avoid or adopt on other websites.
About a fifth (19.3\%) of the answers were unrelated to the question, since they recommend general features unrelated to 2FA or made general statements about 2FA:
\begin{quote}
\textit{"I think, overall, that most sites dealing with personal/financial/health data should all be 2FA."} (P125, unrelated answer)
\end{quote}
\begin{quote}
\textit{"They should allow for a personally chosen user ID."} (309, unrelated answer)
\end{quote}

\begin{figure}
    \centering
    \includegraphics[width=\linewidth]{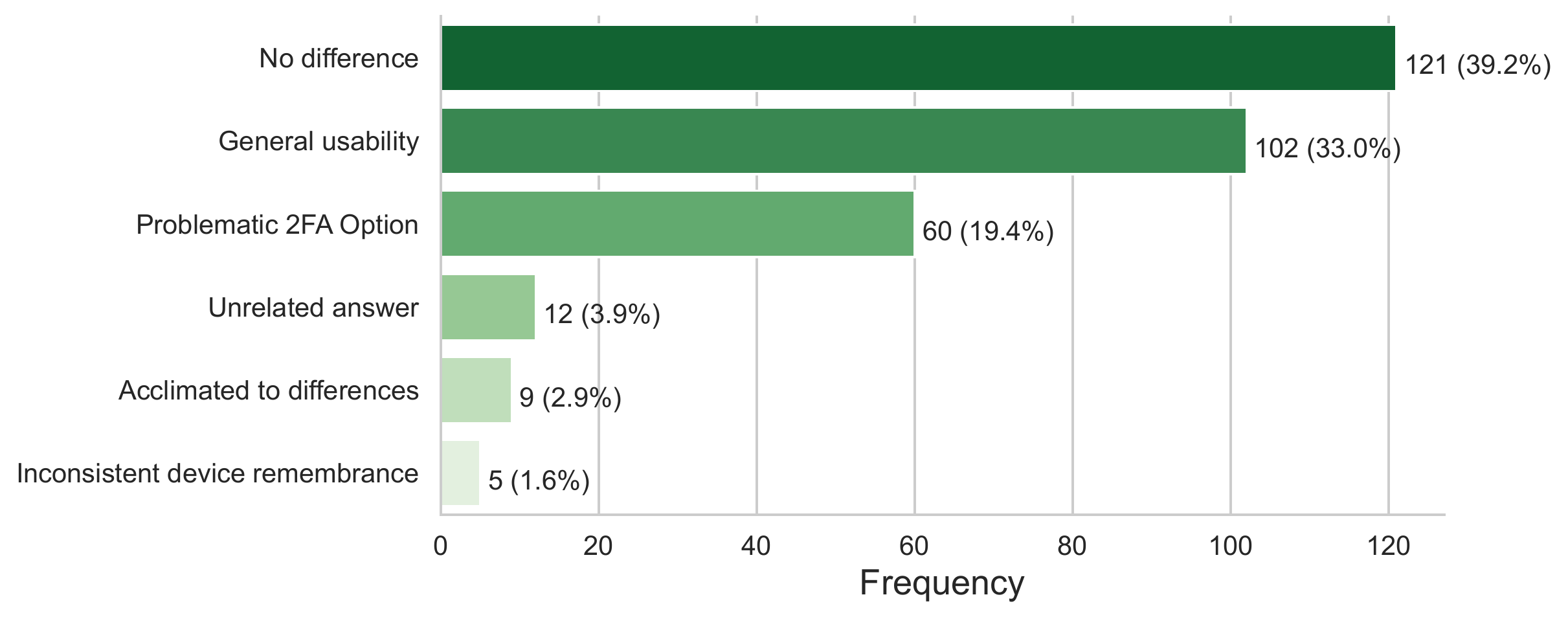}
    \caption{Codes for \textbf{Q6a} ordered by their frequency}
    \label{fig:merged_consq2_hist_df}
\end{figure}

\paragraph{Answers to \textbf{Q6a}}
In question \textbf{Q6a}, we asked every participant if they recall a specific situation in which 2FA was problematic due to differences in the 2FA user experience to what they were used to.
Figure~\ref{fig:merged_consq2_hist_df} lists the codes for the answers to \textbf{Q6a}.

About 40\% of the participants did not recall such a situation or indicated that they did not notice problematic differences.
\begin{quote}
\textit{"I cannot think of any situation where it was difficult, I have always found it pretty easy."} (P30, no difference)
\end{quote}
\begin{quote}
\textit{"No, they have all seemed about the same."} (257, no difference)
\end{quote}
And 3.9\% of the answers were unrelated to the question \textbf{Q6a}, e.g., referring to general problems in creating an account or logging in with passwords.

\begin{quote}
\textit{"I forgot one of the passwords, (I used a unique one only for this). I misentered something I guess."} (P42, unrelated answer)
\end{quote}
Of all 308 participants, the remaining 176 participants (57.1\%) recalled a problematic situation.
About half of them (102; 33\% of all answers) mentioned that a website had general usability problems in their 2FA experience, but the answer did not specify if and how this differs from the participants' usual 2FA experiences.
\begin{quote}
\textit{"My only thought is Paypal and that it isn't consistent. I probably need to check the settings. I have not had to do that with any of the other sites."} (P57, general usability)
\end{quote}
\begin{quote}
\textit{"Yes, see previous. Citi changed their method without informing me."} (P71, general usability)
\end{quote}
About a fifth of all answers (19.4\%) mentioned a problematic 2FA option in contrast to other websites, such as a custom 2FA option by that service, an unfamiliar 2FA option, or being required to use an undesired 2FA option.
For example:
\begin{quote}
\textit{"I usually don't continue a 2FA set-up if the 2FA involves a text. I prefer to use the authenticator app or my email."} (P164, problematic 2FA option)
\end{quote}
\begin{quote}
\textit{"Not really. Most use txt messages, one uses email.  Amazon also uses texts.  The Amazon one made you click on the link in the text, which I don't like."} (P131, problematic 2FA option)
\end{quote}
\begin{quote}
\textit{"I find it to be very annoying when websites have me download additional software to my phone such as an Authenticator app in order to go through the 2FA process. I can't stand downloading additional apps when a simple text or email will do."} (P200, problematic 2FA option)
\end{quote}
\begin{quote}
   \textit{"In the past, Verizon website had multiple additional steps of 2fa. In addition to 2fa itself, one also needed to remember a picture and code word combination that was set up earlier. I found both difficult to remember and was happy when Verizon switched to 2fa without any additional steps."} (P83, problematic 2FA option) 
\end{quote}
Five participants (1.6\%) recalled a situation in which the device remembrance logic of a website worked differently than they expected it or was used to it.
\begin{quote}
\textit{"2FA was more annoying on Runescape because you weren't able to select an option for your device to be remembered. So, each time I would log in I'd have to grab my phone, open the authentication app, and input the number sequence."} (P82, inconsistent device remembrance)
\end{quote}
\begin{quote}
\textit{"Some only use 2FA once unless it is a new device, Others use it everytime."} (P294, inconsistent device remembrance)
\end{quote}
Nine participants recalled a situation or even general inconsistencies in 2FA but directly stated that they got used to it and do not see a problem (anymore).
\begin{quote}
\textit{"I'm used to the differences; they seem kind of random: sometimes a texted code, sometimes a fingerprint; sometimes my use of LastPass Authenticater. All work well enough."} (P36, acclimated to differences)
\end{quote}
\begin{quote}
\textit{"Probably a 2FA experience I had with work. We use our own authenticator app because I work for a bank, and it is a little complex. For example you have to log into your computer, put in your password, then go to an app on your phone, enter the password again, it will throw up a generated code which you only have 20 seconds to type in. It took some getting used to for sure."} (P290, acclimated to differences)
\end{quote}

\begin{figure}[t]
    \centering
    \includegraphics[width=\linewidth]{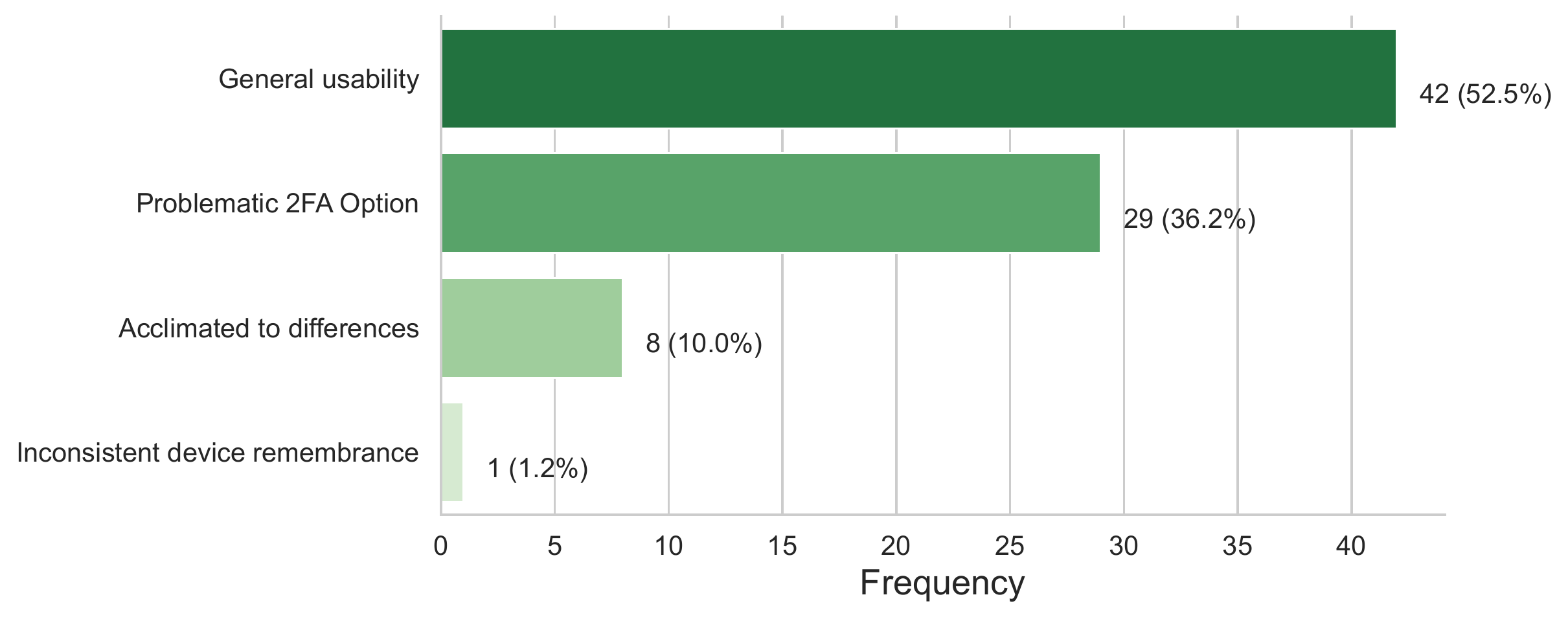}
    \caption{Codes for \textbf{Q6a} only for participants that did not selected websites in \textbf{Q3}.}
    \label{fig:add_situation_hist_df}
\end{figure}
Of those 176 participants that recalled a situation, 80 did not previously in \textbf{Q3} indicate differences in their 2FA experiences between websites (i.e., they did not go through \textbf{Q4} and \textbf{Q5}).
Figure~\ref{fig:add_situation_hist_df} summarizes their answers to \textbf{Q6a} separately.
Their answers show that although they did not initially indicated differences between 2FA experiences, they recalled in many cases, a situation in which a website had 2FA usability issues or a problematic 2FA option.

\begin{figure}[t]
    \centering
    \includegraphics[width=\linewidth]{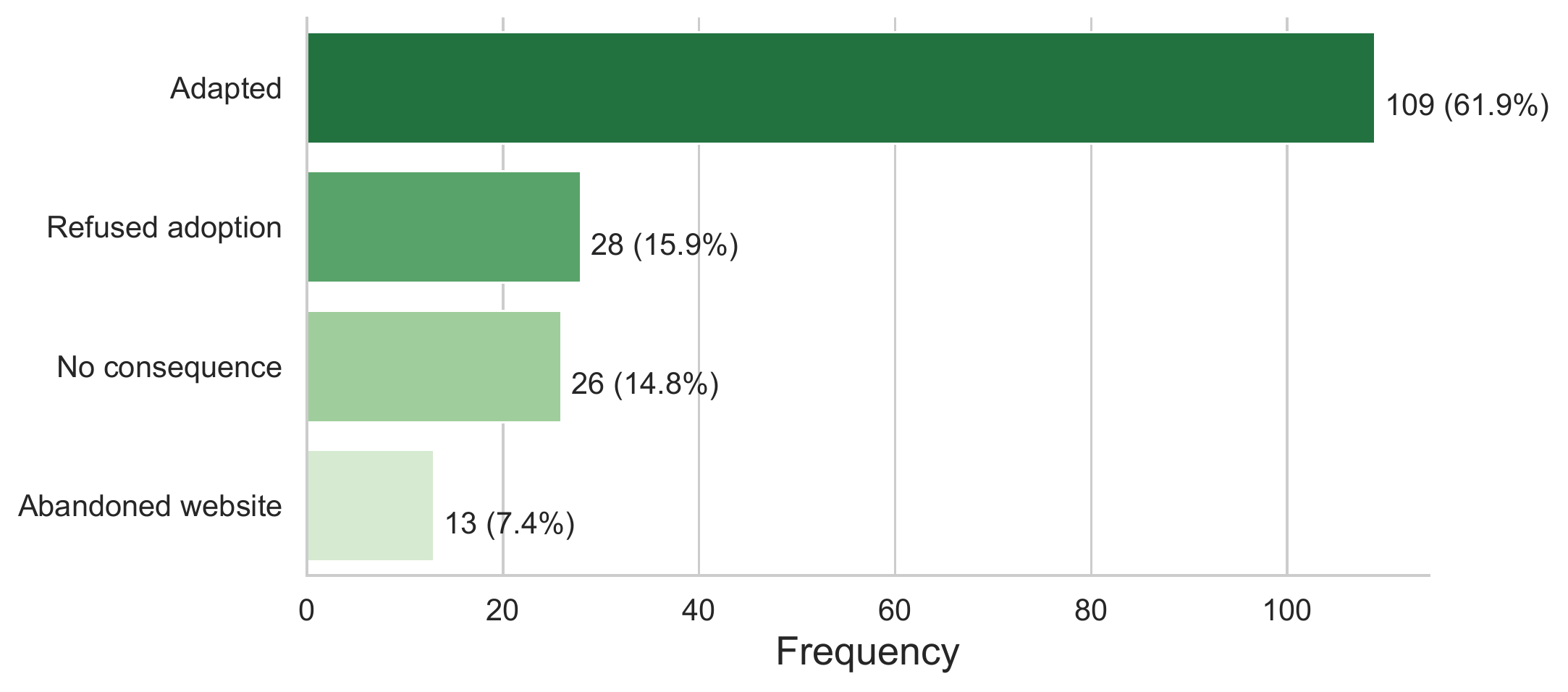}
    \caption{Codes for \textbf{Q6b} ordered by their frequency}
    \label{fig:merged_consq3_hist_df}
\end{figure}

\paragraph{Answers to \textbf{Q6b}}
Figure~\ref{fig:merged_consq3_hist_df} summarizes the answers for question \textbf{Q6b} from those 176 participants that recalled a problematic situation in \textbf{Q6a}.
The majority (61.9\%) of those participants explained that they had to adapt to the situation, e.g., by switching to a less preferred 2FA option, fighting their way through despite the problems, or learning how to use a new 2FA option.
\begin{quote}
\textit{"I generally like to opt-in for SMS messaging for 2FA, so if that is not offered, I will adjust."} (P193, adapted)
\end{quote}
\begin{quote}
\textit{"I still used 2FA because it was required, but I would have disabled it if I could. I also log into it a little less since I don't want to deal with getting the code."} (P208, adapted)
\end{quote}
However, several participants responded that they refused the adoption of 2FA in this situation (15.9\%) or abandoned the website due to the indicated problems (7.4\%).
\begin{quote}
\textit{"I chose not to sign up, and didn't use the site."} (P26, abandoned website)
\end{quote}
\begin{quote}
\textit{"I stopped using the site. If I can't trust their authentification info, I can't trust the site."} (P136, abandoned website)
\end{quote}
\begin{quote}
\textit{"Stopped banking with HSBC."} (P262, abandoned website)
\end{quote}
\begin{quote}
\textit{"I had to email them then once I got back on the account I shut 2FA off."} (P210, refused adoption of 2FA)
\end{quote}
\begin{quote}
\textit{"If I couldn't use VOIP I removed some 2FA."} (P69, refused adoption of 2FA)
\end{quote}
\begin{quote}
\textit{"I aborted setting up 2FA on those sites."} (P242, refused adoption of 2FA)
\end{quote}
\begin{figure}[t]
    \centering
    \includegraphics[width=\linewidth]{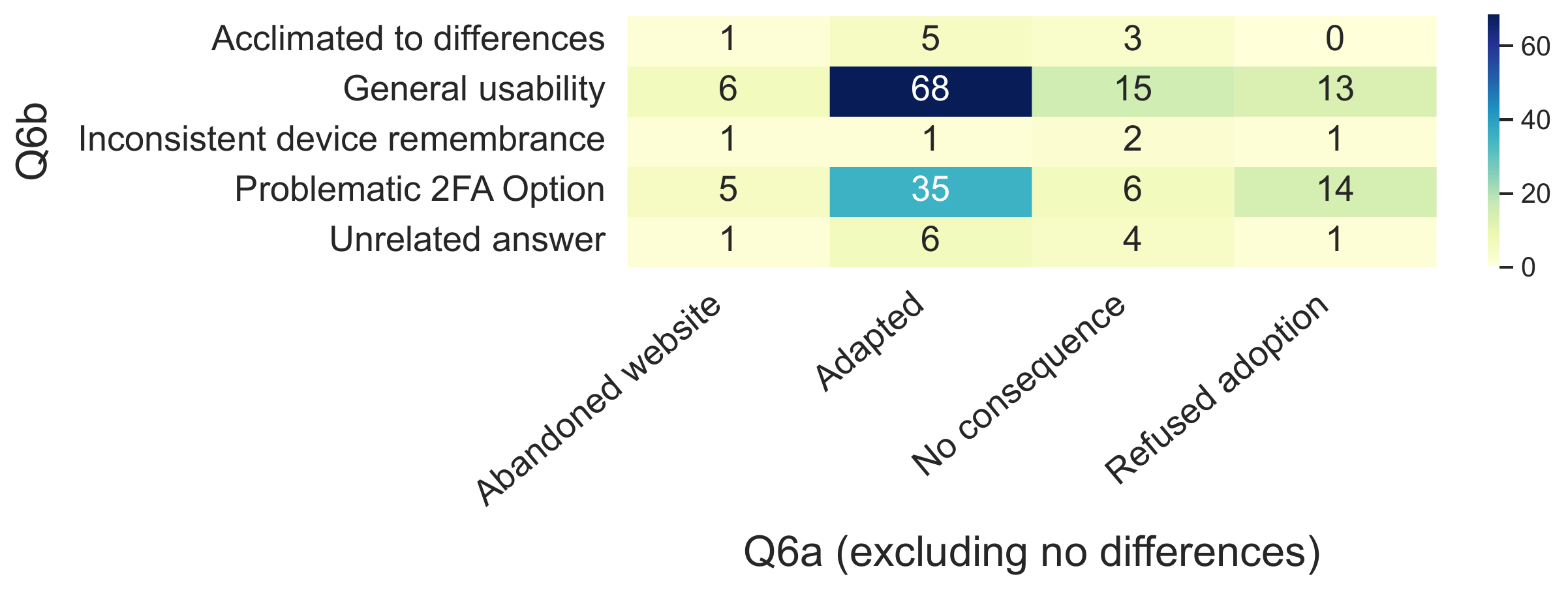}
    \caption{Contingency table between codes for \textbf{Q6a} (excluding \textit{No differences}) and \textbf{Q6b}}
    \label{fig:confusion_matrix_consq2_consq3_differences}
\end{figure}
Figure~\ref{fig:confusion_matrix_consq2_consq3_differences} depicts the contingency table between codes for \textbf{Q6a} (excluding \textit{No differences}) and \textbf{Q6b}.
While most participants adapted to problematic 2FA options or general usability issues, 14 and 13 participants, respectively, refused to adopt 2FA.
One participant refused adoption of 2FA because of friction with the device remembrance policy that differed from other websites. 

Of all 176 participants, 26 (14.8\%) answered that there were no consequences for them.

\subsection{Conclusion of our survey}
\label{appendix:survey:conclusion}

In regards to the question of whether users had negative experiences transferring their 2FA experiences between websites and whether this has stopped them from enabling or continuing to use 2FA, 28 (9.1\%) of our 308 participants mentioned for at least one website, which differed in their opinion from others in its 2FA experience (\textbf{Q5}), that they use this website less due to this 2FA experience.
Additionally, 41 (15.9\%) of the participants recalled a concrete situation with 2FA that was challenging because the 2FA experience differed from what they were used to (\textbf{Q6}) and, as a result, they abandoned the website or refused adoption of (a specific) 2FA option.
Taken together, 60 (19.5\%) of our participants reported using a website less, abandoning a website, or refusing adoption of (a specific) 2FA option.
Of those, 28 (9.1\% of all participants) refused adoption due to differences in the usability of 2FA in contrast to other websites, undesired/unfamiliar/custom 2FA options, or in one case due to an inconsistent device remembrance policy.

\section{Codebook for Device Remembrance}
\label{appendix:codebook_dr}

\begin{table}[t]
    \caption{Codebook for device remembrance}
    \label{tab:device_remembrance_codebook}
    \centering

    \def\arraystretch{1.2} % height of row
    \scriptsize
    
\begin{tabular}{@{}l|l}
\toprule
                \textbf{Code} &                             \textbf{Examples} \\
\midrule
\multirow{9}{1.5cm}{\textbf{Remember}} &       Remember verification for this computer \\
                              &           Recognize this device in the future \\
                              &            Do not require OTP on this browser \\
                              & Skip two-factor authentication on this device \\
                              &                                  Save browser \\
                              &               Do not ask again on this device \\
                              &                          Remember this device \\
                              &         Remember this computer for \{duration\} \\
                              &            Do not ask for code on this device \\\hline
   \multirow{6}{1.5cm}{\textbf{Trust}} &                    Trust this device (opt-in) \\
                              &                   Trust this device (opt-out) \\
                              &            Do not trust this device (opt-out) \\
                              &             Do not trust this device (opt-in) \\
                              &              Trust this device for \{duration\} \\
                              &                           Untrust this device \\\hline
    \multirow{3}{1.5cm}{\textbf{Skip}} &          Require code to login for \{duration\} \\
                              &          We won't ask for the next \{duration\} \\
                              &                      Skip this for \{duration\} \\\hline
                        \textbf{Other} &                         Stay signed/logged in \\
\bottomrule
\multicolumn{2}{c}{\textit{opt-out}: checkbox is pre-checked; \textit{opt-in}: checkbox is not pre-checked}\\
\multicolumn{2}{c}{\textit{duration}: a number of days, weeks, or logins}
\end{tabular}
\end{table}

Table~\ref{tab:device_remembrance_codebook} provides the codebook for the device remembrance descriptions. 

\section{Website Details}
\label{appendix:codebook}

\begin{figure}[t]
    \centering
    \includegraphics[width=\linewidth]{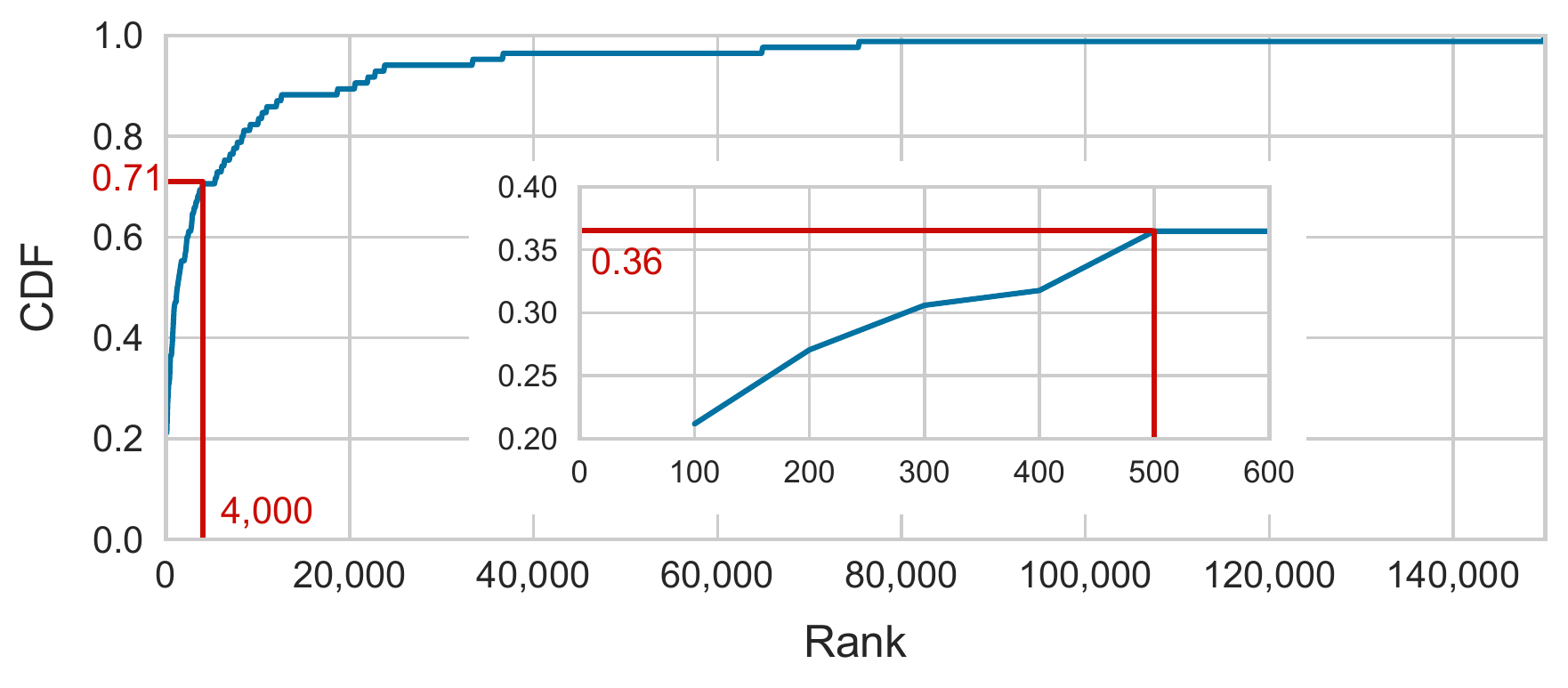}
    \caption{Cumulative frequency distribution of Tranco~\cite{tranco_paper, tranco_website} rankings of our 85 websites. Percentiles for 0.36 (31~websites) and 0.71 (60~websites) are marked, which correspond to websites in the top-500 and in the top-4000 in Tranco.}
    \label{fig:cfd_ranking}
\end{figure}

Figure~\ref{fig:cfd_ranking} shows the CFD of the Tranco website rankings in our data set. We split our 85 websites into three roughly equal sized groups through the 36th and 71st percentile of the websites' Tranco ranks.

\begin{figure}[t]
    \centering
    \includegraphics[width=\linewidth]{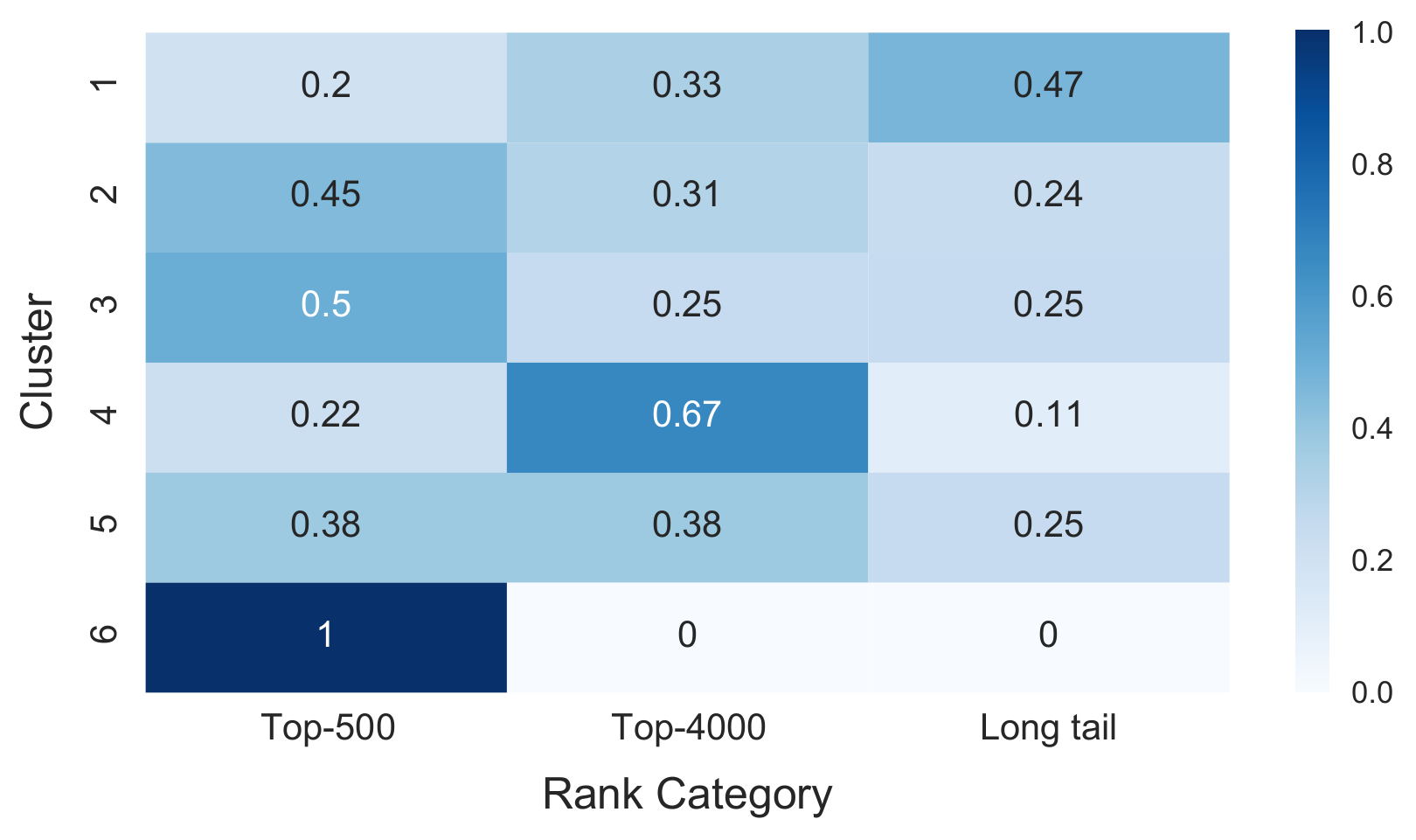}
    \caption{Heatmap of the normalized contingency table (see Table~\ref{tab:contingency}) for website cluster vs.~website rank category}
    \label{fig:heatmap_rank_cluster}
\end{figure}

Table~\ref{tab:contingency} and Figure~\ref{fig:heatmap_rank_cluster} depict the contingency table between website rank group vs.~\textit{inter}-cluster, and illustrate the normalized contingency table, respectively.

\begin{table}[t]
    \caption{Contingency table for cluster vs.~rank category}
    \label{tab:contingency}
    \centering
    \def\arraystretch{1.1} % height of row
    
\begin{tabular}{l|rrr|r}
\multicolumn{1}{l}{} & \multicolumn{3}{c}{\textbf{Rank category}} &\multicolumn{1}{l}{}\\\cmidrule(lr){2-4}
\multicolumn{1}{l}{\textbf{Cluster}}        &   Top-500 &   Top-4000 &   \multicolumn{1}{r|}{Long tail} & \multicolumn{1}{l}{$\sum$}\\
\midrule
1             &   6 &  10 &  14 & 30\\
\rowcolor{gray!25} 2             &  13 &   9 &   7 & 29\\
3             &   2 &   1 &   1 &  4\\
\rowcolor{gray!25} 4             &   2 &   6 &   1 &  9\\
5             &   3 &   3 &   2 &  8\\
\rowcolor{gray!25} 6             &   5 &   0 &   0 &  5\\ \midrule
$\sum$        &  31 &  29 &  25 & 85\\
\bottomrule
\end{tabular}
\end{table}

Table~\ref{tab:websitedetails} provides additional details about the websites in our data set, including their Tranco rank, rank category (see Section~\ref{sec:results:clustervsrank}), naming and location of 2FA settings, description of 2FA, and form of device remembrance.

\begin{table*}%[!htbp]
  \caption{Details about websites in our data set.}\label{tab:websitedetails}
  \centering
  
%   \scriptsize
    % \tiny
    \footnotesize
  \setlength{\tabcolsep}{2 pt} % width of column
  \def\arraystretch{1.3} % height of row
  
\begin{tabular}{@{}p{1.7 cm}|p{2.5cm}|>{\raggedleft\arraybackslash}p{1cm}|>{\raggedleft\arraybackslash}p{2cm}|r|l|l|l}
\toprule
                 \multicolumn{1}{c|}{\textbf{Category}} &                \multicolumn{1}{c|}{\textbf{Website}} &   \multicolumn{1}{c|}{\textbf{Rank}$^{\dagger}$} &  \textbf{Rank Category} &                   \textbf{Naming} &                \multicolumn{1}{c|}{\textbf{Location}} &        \textbf{Description} &             \textbf{Remembrance} \\
\midrule

    \multirow{7}{1.5cm}{Backup and Sync} &      backblaze.com &   6973 &             Long tail &      other & Security / Account &                      --- &                          Trust \\
                                         &        dropbox.com &     56 &             Top-500 &        2SV & Security / Account &               Security &                          Trust \\
                                         &       evernote.com &    412 &             Top-500 &        2SV & Security / Account &               Security &                       Remember \\
                                         &         icloud.com &    111 &             Top-500 &        2FA & Security / Account &                      --- &                          Trust \\
                                         &     jottacloud.com & 149805 &             Long tail &        2FA & Security / Account &               Security &                              --- \\
                                         &            mega.io &  21990 &             Long tail &        2FA & Security / Account &               Security &                              --- \\
                                         &       synology.com &   3267 &             Top-4000 &        2SA & Security / Account &               Security &                       Remember \\\hline
    \multirow{2}{1.5cm}{Cloud Computing} &   digitalocean.com &   1573 &             Top-4000 &        2FA & Security / Account &               Security &                              --- \\
                                         &        laravel.com &   3424 &             Top-4000 &        2FA & Security / Account &                      --- &                              --- \\\hline
      \multirow{4}{1.5cm}{Communi\-ca\-tion} &       basecamp.com &   1483 &             Top-4000 &        2FA & Security / Account &               Security &                              --- \\
                                         &        discord.com &    460 &             Top-500 &        2FA & Security / Account &               Security &                              --- \\
                                         &      mailchimp.com &    222 &             Top-500 &        2FA & Security / Account &               Security &                           Skip \\
                                         &            zoom.us &     34 &             Top-500 &        2FA & Security / Account &                      --- &                              --- \\\hline
                            Crowdfunding &    kickstarter.com &    292 &             Top-500 &        2FA & Security / Account &               Security &                       Remember \\\hline
   \multirow{5}{1.5cm}{Crypto\-currencies} &        binance.com &    466 &             Top-500 &        2FA & Security / Account &               Security &                              --- \\
                                         &       bitfinex.com &   2870 &             Top-4000 &        2FA & Security / Account &               Security &                              --- \\
                                         &    blockchain.info &   8244 &             Long tail &        2SV & Security / Account &                 Device &                              --- \\
                                         &          bybit.com &   9189 &             Long tail &        2FA & Security / Account &                      --- &                              --- \\
                                         &         kraken.com &   3950 &             Top-4000 &        2FA & Security / Account &               Security &                              --- \\\hline
          \multirow{4}{1.5cm}{Developer} &      atlassian.com &    671 &             Top-4000 &        2SV & Security / Account &                 Device &                              --- \\
                                         &         github.com &     35 &             Top-500 &        2FA & Security / Account &               Security &                              --- \\
                                         &         gitlab.com &    712 &             Top-4000 &        2FA & Security / Account &               Security &                              --- \\
                                         &          unity.com &   2480 &             Top-4000 &        2FA & Security / Account &                      --- &                              --- \\\hline
            \multirow{4}{1.5cm}{Domains} &        easydns.com &  36665 &             Long tail &        2FA & Security / Account &                      --- &                              --- \\
                                         &          gandi.net &   3644 &             Top-4000 &        2FA & Security / Account &               Security &                              --- \\
                                         &          hover.com &   5585 &             Long tail &      other & Security / Account &               Security &                              --- \\
                                         &      namecheap.com &    852 &             Top-4000 &        2FA & Security / Account &               Security &                              --- \\\hline
              \multirow{3}{1.5cm}{Email} &         google.com &      1 &             Top-500 &        2SV & Security / Account &               Security &                       Remember \\
                                         &          yahoo.com &     18 &             Top-500 &        2SV & Security / Account &               Security &                       Remember \\
                                         &           zoho.com &    365 &             Top-500 &        MFA & Security / Account &               Security &                           Skip \\\hline
                           Entertainment &          twitch.tv &     76 &             Top-500 &        2FA & Security / Account &               Security &                       Remember \\\hline
            \multirow{2}{1.5cm}{Finance} &           xero.com &   1238 &             Top-4000 &        MFA & Security / Account &               Security &                           Skip \\
                                         & youneedabudget.com &   8478 &             Long tail &        2SV & Security / Account &               Security &                              --- \\\hline
             \multirow{5}{1.5cm}{Gaming} &       blizzard.com &   2258 &             Top-4000 &      other & Security / Account &                 Device &                              --- \\
                                         &             ea.com &    789 &             Top-4000 &      other & Security / Account &               Security &                              --- \\
                                         &      epicgames.com &    920 &             Top-4000 &        2FA & Security / Account &               Security &                       Remember \\
                                         &    playstation.com &    707 &             Top-4000 &        2SV & Security / Account &               Security &                              --- \\
                                         &         roblox.com &    248 &             Top-500 &        2SV & Security / Account &               Security &                          Trust \\\hline
                              Government &             va.gov &   1102 &             Top-4000 &        MFA & Security / Account &               Security &                              --- \\\hline
             \multirow{2}{1.5cm}{Health} &        23andme.com &   6345 &             Long tail &        2SV & Security / Account &                      --- &                              --- \\
                                         &      runsignup.com &   7326 &             Long tail &        MFA &              Other &               Security &                              --- \\\hline
               Hotels/Acom. &        booking.com &    139 &             Top-500 &        2FA & Security / Account &                      --- &                              --- \\

\bottomrule
\multicolumn{8}{r}{\textit{continued on next page}}
\end{tabular}
\end{table*}

\begin{table*}%[!htbp]
  \centering
  
%   \scriptsize
    % \tiny
    \footnotesize
  \setlength{\tabcolsep}{2 pt} % width of column
  \def\arraystretch{1.3} % height of row
  
\begin{tabular}{@{}p{1.7 cm}|p{2.5cm}|>{\raggedleft\arraybackslash}p{1cm}|>{\raggedleft\arraybackslash}p{2cm}|r|l|l|l}
\toprule
                 \multicolumn{1}{c|}{\textbf{Category}} &                \multicolumn{1}{c|}{\textbf{Website}} &   \multicolumn{1}{c|}{\textbf{Rank}$^{\dagger}$} &  \textbf{Rank Category} &                   \textbf{Naming} &                \multicolumn{1}{c|}{\textbf{Location}} &        \textbf{Description} &             \textbf{Remembrance} \\
\midrule

\multirow{7}{1.5cm}{Identity Management} &      1password.com &   3012 &             Top-4000 &        2FA &              Other &               Security &                              --- \\
                                         &      bitwarden.com &  10002 &             Long tail &      other & Security / Account &               Security &                       Remember \\
                                         &              id.me &   6012 &             Top-4000 &      other & Security / Account &               Security &                              --- \\
                                         & keepersecurity.com &  12065 &             Long tail &        2FA & Security / Account &                      --- &                           Skip \\
                                         &       lastpass.com &   1333 &             Top-4000 &      other & Security / Account &               Security &                          Trust \\
                                         &          orcid.org &   1697 &             Top-4000 &        2FA & Security / Account &               Security &                              --- \\
                                         &       roboform.com &   5312 &             Long tail &      other & Security / Account &               Security &                              --- \\\hline
                \multirow{2}{1.5cm}{IoT} &           arlo.com &  10953 &             Long tail &        2SV &              Other &               Security &                              --- \\
                                         &          ifttt.com &   2847 &             Top-4000 &        2SV & Security / Account &               Security &                              --- \\\hline
              \multirow{2}{1.5cm}{Legal} &           clio.com &  20552 &             Long tail &      other & Security / Account &               Security &                              --- \\
                                         &       docusign.com &    815 &             Top-4000 &        2SV & Security / Account &               Security &                       Remember \\\hline
              \multirow{2}{1.5cm}{Other} &          adobe.com &     23 &             Top-500 &        2SV & Security / Account &               Security &                          Other \\
                                         &          opera.com &    133 &             Top-500 &        2FA &              Other &               Security &                              --- \\\hline
           \multirow{2}{1.5cm}{Payments} &         paypal.com &     75 &             Top-500 &        2SV & Security / Account &               Security &                          Trust \\
                                         &         stripe.com &    630 &             Top-4000 &        2SA &              Other &               Security &                              --- \\\hline
      \multirow{4}{1.5cm}{Remote Access} &            join.me &  12578 &             Long tail &        2SV & Security / Account &               Security &                              --- \\
                                         &        logmein.com &   2714 &             Top-4000 &        2SV & Security / Account &               Security &                          Trust \\
                                         &        realvnc.com &  10442 &             Long tail &        2SV & Security / Account &               Security &                              --- \\
                                         &     teamviewer.com &    455 &             Top-500 &        2FA &              Other &               Security &                              --- \\\hline
             \multirow{4}{1.5cm}{Retail} &         amazon.com &     17 &             Top-500 &        2SV & Security / Account &               Security &                       Remember \\
                                         &           ebay.com &     70 &             Top-500 &        2SV & Security / Account &               Security &                              --- \\
                                         &           etsy.com &    107 &             Top-500 &        2FA & Security / Account &                 Device &                          Other \\
                                         &         newegg.com &   1141 &             Top-4000 &        2SV & Security / Account &               Security &                       Remember \\\hline
           \multirow{5}{1.5cm}{Security} &    bitdefender.com &   2198 &             Top-4000 &        2FA & Security / Account &               Security &                          Trust \\
                                         &     cloudflare.com &     79 &             Top-500 &        2FA & Security / Account &               Security &                              --- \\
                                         &       digicert.com &    158 &             Top-500 &        2FA & Security / Account &               Security &                       Remember \\
                                         &         norton.com &    872 &             Top-4000 &        2FA & Security / Account &               Security &                          Trust \\
                                         &     virustotal.com &   2295 &             Top-4000 &        2FA & Security / Account &               Security &                       Remember \\\hline
             \multirow{7}{1.5cm}{Social} &       facebook.com &      4 &             Top-500 &        2FA & Security / Account &                 Device &                              --- \\
                                         &      instagram.com &      7 &             Top-500 &        2FA & Security / Account &                 Device &                              --- \\
                                         &       linkedin.com &      9 &             Top-500 &        2SV & Security / Account &               Security &                       Remember \\
                                         &         reddit.com &     32 &             Top-500 &        2FA & Security / Account &               Security &                              --- \\
                                         &         tumblr.com &     45 &             Top-500 &        2FA & Security / Account &               Security &                              --- \\
                                         &        twitter.com &      6 &             Top-500 &        2FA & Security / Account &               Security &                              --- \\
                                         &             vk.com &     50 &             Top-500 &        2SV & Security / Account &               Security &                       Remember \\\hline
    \multirow{5}{1.5cm}{Task Management} &       airtable.com &   2052 &             Top-4000 &        2SA & Security / Account &               Security &                              --- \\
                                         &        clickup.com &   7723 &             Long tail &        2FA & Security / Account &               Security &                              --- \\
                                         &       clubhouse.io &  23778 &             Long tail &        2FA & Security / Account &               Security &                              --- \\
                                         &    meistertask.com &  18630 &             Long tail &        2FA & Security / Account &               Security &                              --- \\
                                         &       toodledo.com &  33344 &             Long tail &      other & Security / Account &               Security &                              --- \\\hline
          \multirow{2}{1.5cm}{Utilities} &    callcentric.com &  75387 &             Long tail &      other &              Other &               Security &                       Remember \\
                                         &          coned.com &  22715 &             Long tail &        2SV & Security / Account &               Security &                              --- \\\hline
                           VPN           &         airvpn.org &  64852 &             Long tail &      other & Security / Account &               Security &                              --- \\

\bottomrule

\multicolumn{8}{c}{---\ = not applicable;\quad $^{\dagger}$ based on Tranco~\cite{tranco_paper,tranco_website}}
\end{tabular}

\end{table*}

\section{Details on Pairwise Hamming Distances}
\label{appendix:websitesimilarity}

\begin{figure}[h]
    \centering
    \begin{subfigure}[b]{\linewidth}
        \centering
        \includegraphics[width=\linewidth]{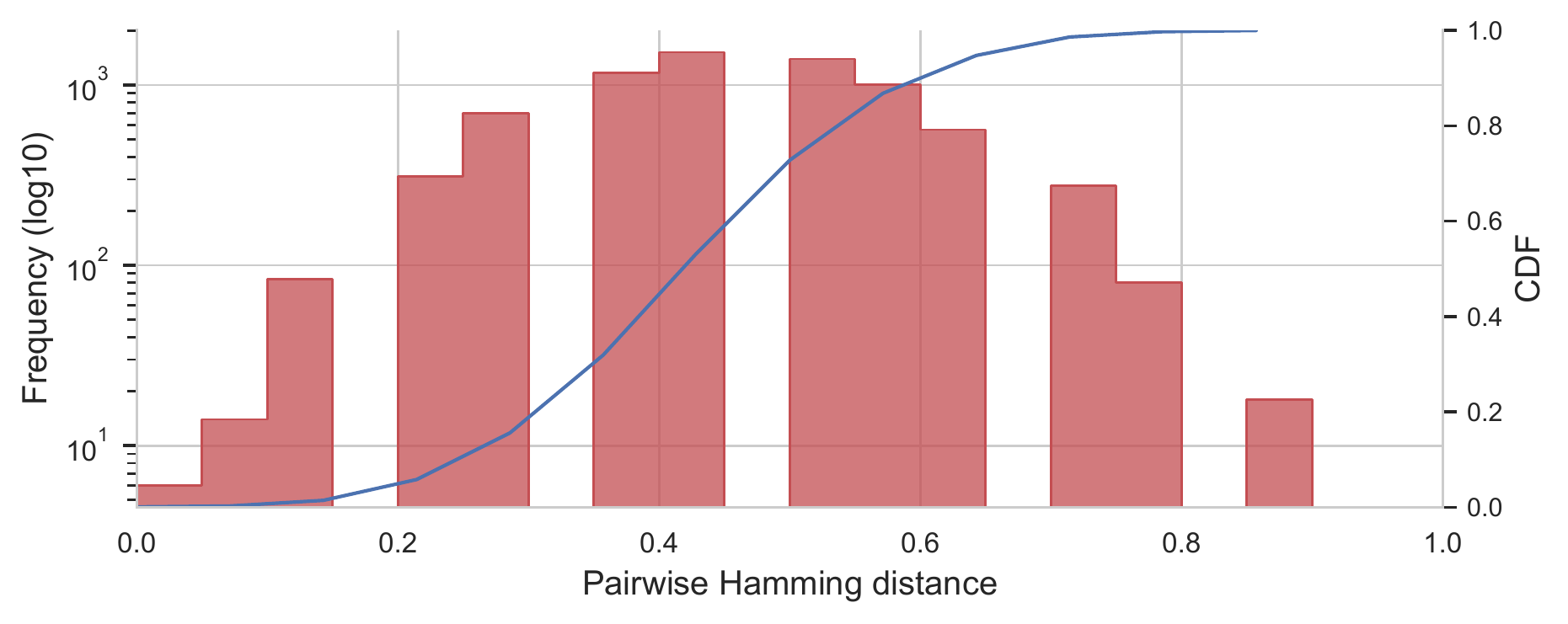}
        \caption{Overall distribution of distances.}
        \label{fig:similarity_all}
    \end{subfigure}
      \vspace{0.1cm}
    
    \begin{subfigure}[b]{\linewidth}
        \centering
        \includegraphics[width=\linewidth]{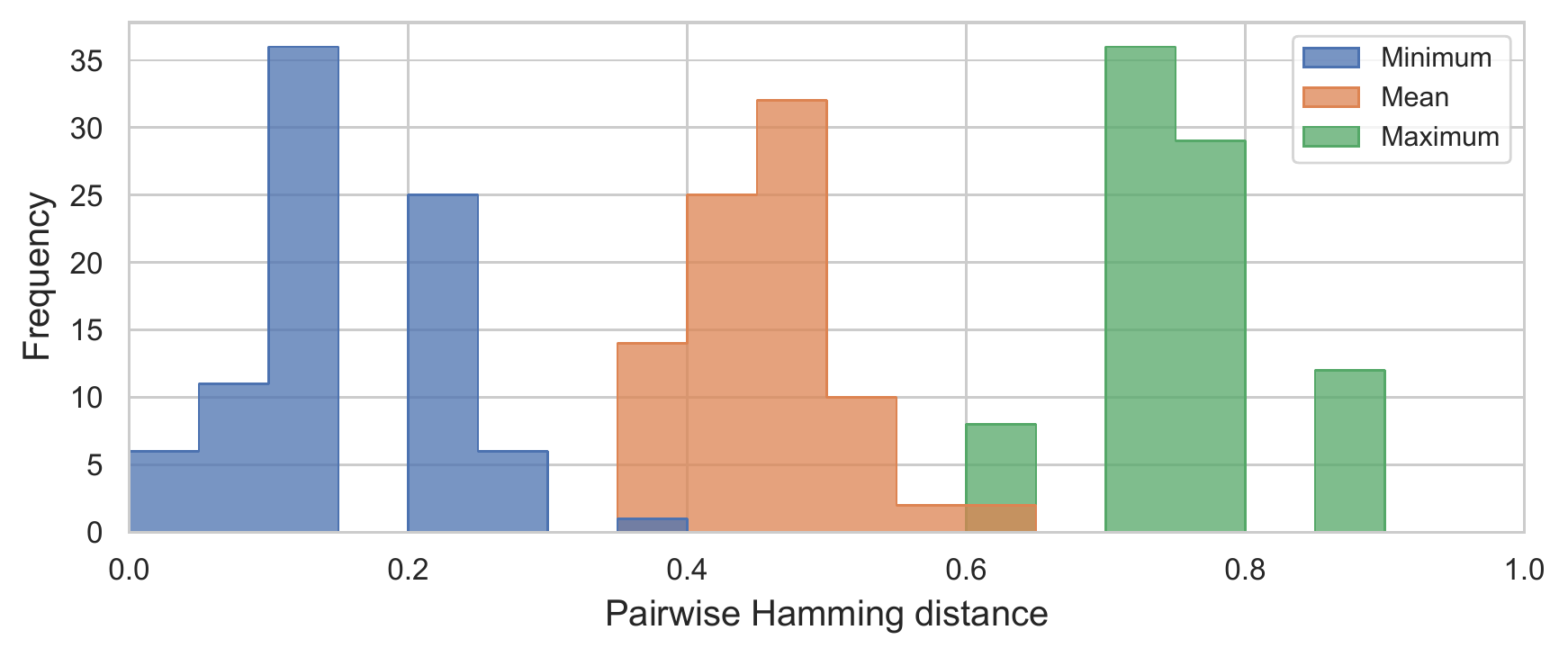}
        \caption{Min, max, and median distances.}
        \label{fig:similarity_mmm}
    \end{subfigure}
    
    \caption{Frequency distributions of pairwise Hamming distances of non-conditional comparison factors between all websites in our dataset.}
    \label{fig:similarity}
\end{figure}

As explained in Section~\ref{sec:websitesimilarity}, we compare the 85 websites in our dataset using pairwise Hamming distance between the 14 non-conditional factors of each website. Figure~\ref{fig:similarity_all} depicts the (cumulative) frequency distribution of the pairwise Hamming distances between all websites in our data set, where a distance of $0$ means equality in all factors and a distance of $1$ means complete inequality of all 14 factors. This distribution is very symmetrically (skew=$0.062$) and only slightly heavy-tailed (kurtosis=$-0.112$), but an omnibus test of normality~\cite{scipy_normaltest} ($p>.05$) indicates that it is not Gaussian.
Further, Figure~\ref{fig:similarity_mmm} shows each website's frequency distribution for minimum, mean, and maximum distance. The average website in our data has a minimum Hamming distance of $0.16\pm0.02$ (for a confidence interval of 95\%), a mean distance of $0.46\pm0.01$, and a maximum distance of $0.75\pm0.01$. In other words, the average website in our data set differs on average in 6--7 out of 14 factors from the other websites and differs on average in at least 2--3 factors. Nevertheless, no pair of websites has a distance larger than 0.86, i.e., there are always two factors identical for each pair of websites.

\section{High-level View of Clusters}
\label{sec:appendix:clusters}

\begin{figure}[h]
    \centering
    \includegraphics[width=\linewidth]{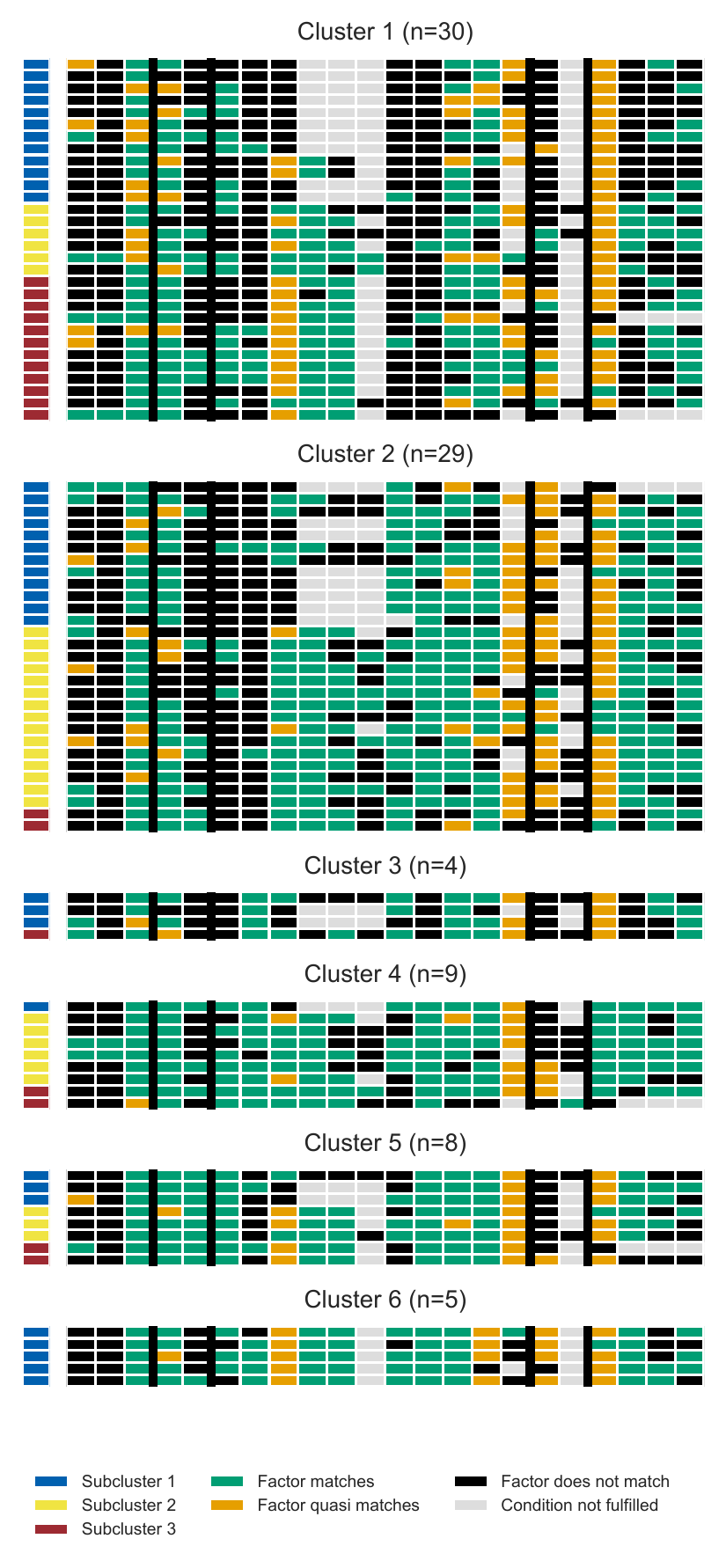}
    \caption{Clusters of websites based on comparison factors. Subclusters based on the conditional factors are indicated in the first column. Thick lines separate factors from different steps in the user journey (see also Table~\ref{tab:comparisonfactors}).}
    \label{fig:clusters}
\end{figure}

\begin{figure}[h]
    \centering
    \includegraphics[width=.8\linewidth]{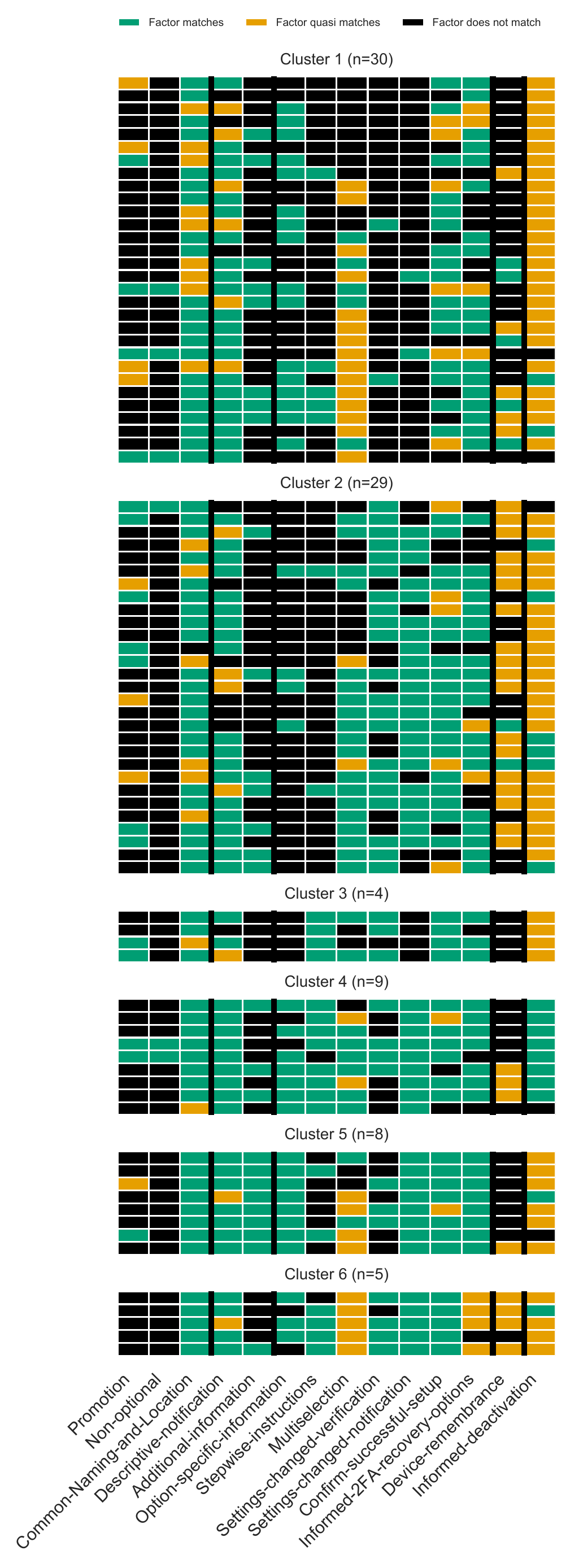}
    \caption{Clusters of websites based on non-conditional factors. Only non-conditional factors are shown.}
    \label{fig:noncond_clusters}
\end{figure}

\begin{figure}[h]
    \centering
    \includegraphics[width=.8\linewidth]{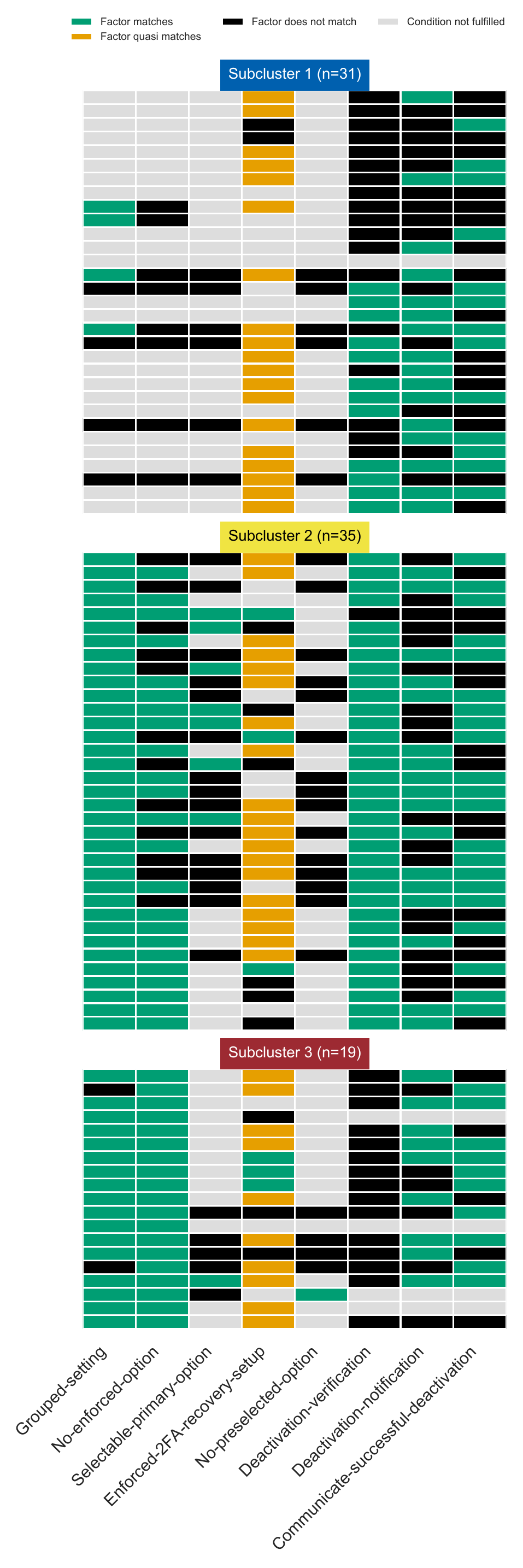}
    \caption{Subclusters of websites based on conditional factors. Only conditional factors are shown.}
    \label{fig:cond_clusters}
\end{figure}

Figure~\ref{fig:clusters} provides a less noisy view of the clusters depicted in Table~\ref{tab:comparisonfactors} to see the clusters' structure easily. Similarly, Figure~\ref{fig:noncond_clusters} depicts only the \textit{non-conditional} factors for which we identified six \textit{inter-clusters}, and Figure~\ref{fig:cond_clusters} depicts only the \textit{conditional} factors for which we found three \textit{subclusters} or \textit{intra-clusters}.

\section{Opinionated Separation of Comparison Factors}
\label{appendix:separation_factors}

We conducted an expert assessment of our comparison factors with local experts to create an opinionated separation of our factors based on their impact on \textit{user experience}, \textit{security}, \textit{both UX and security}, or \textit{neither}.
In addition to the authors, we invited four usable security researchers from two other research groups at our institution to participate as experts in this assessment.
The invited experts had extensive experience in usability studies and were familiar with web authentication.
We presented each expert with our comparison factors with their definition.
In an online survey, each expert separately decided which of the four categories each factor falls and provided a short rationale for their categorization.
The authors discussed the individual assessments and assigned the final category to each factor.
This decision was based on a clear majority among the experts, evidence from the literature and best practices, rationales, and personal opinion by the authors.
Table~\ref{tab:separation_factors} lists the categories for each factor, and in the following Section~\ref{appendix:separation_factors:list} we briefly summarize the rationales that led to this separation.
A point worth noting from our discussion is the definition of \textit{"impact on security."}
Improvements in usability and user experience are very often entangled with the users' security---for example, nudging more users to adopt 2FA will increase the overall security; streamlining the setup of 2FA options lowers the friction and can cause a higher adoption; or highlighting certain options, like security keys, might persuade users to adopt stronger options over, e.g., text-based OTP.
We decided to limit "security" to the \textit{direct} impact on the security of 2FA and the user's account, and to assign the "security" or "user experience" categories when either category clearly outweighs the other (in our opinion).
Afterward, in Section~\ref{appendix:separation_factors:clusters}, we present additional views on the factor clusters with separated comparison factors (see also Section~\ref{sec:results:separated_factors}).

\begin{table}[t]
    \centering
    \scriptsize
    \def\arraystretch{1.2} % height of row
    \begin{tabular}{l|r|l}
        \multicolumn{1}{c|}{\textbf{Category}}        & \multicolumn{1}{c|}{\textbf{Conditional}} & \multicolumn{1}{c}{\textbf{Factors}} \\\toprule
        User Experience & No & Promotion \\
                        & No & Common-Naming-and-Location \\
                        & No & Descriptive-Notification \\
                        & No & Option-Specific-Information \\
                        & Yes & Grouped-Setting \\
                        & Yes & Selectable-Primary-Option \\
                        & No & Confirm-Successful-Setup \\
                        & No & Informed-2FA-Recovery-Options \\
                        & Yes & Enforced-2FA-Recovery-Setup \\
                        & No & Device-Remembrance \\
                        & Yes & No-Preselected-Option \\
                        & Yes & Communicate-Successful-Deactivation\\ \hline
        Security        & No & Settings-Changed-Verification\\
                        & Yes & Deactivation-Verification \\ \hline
        Both Security and            & No & Non-optional \\
        User Experience                & No & Step-Wise-Instructions \\
                        & No & Multiselection \\
                        & Yes & No-Enforced-Options \\
                        & No & Settings-Changed-Notification \\
                        & No & Informed-Deactivation \\
                        & Yes & Deactivation-Notification \\ \hline
        Neither         & No & Additional-Information \\ \bottomrule
    \end{tabular}
    \caption{Separation of comparison factors}
    \label{tab:separation_factors}
\end{table}

\subsubsection{List of Comparison Factors}
\label{appendix:separation_factors:list} \hfill

\parahighlightnoindent{Promotion [UX]} By making 2FA support more visible, users might perceive the website as more secure. The promotion dialog can be the entry point for the 2FA journey. Golla et al.~\cite{driving_2fa} recently showed that a certain kind of messaging and UX design patterns can effectively improve 2FA adoption. However, the promotion can only convince users to start their 2FA journey, but its impact on a security depends on other factors that determine the success of this journey. Thus, we think the impact on UX outweighs the impact on security.

\parahighlightnoindent{Non-Optional [Both]} Abbott and Patil~\cite{10.1145/3313831.3376457} have shown that mandatory 2FA can increase users' frustration, affecting the UX. 
However, when every user account is additionally secured with 2FA has been shown to be an effective defense against account takeovers~\cite{10.1145/3308558.3313481}.

\parahighlightnoindent{Common-Naming-and-Location [UX]} By providing users with a familiar interface/workflow, the website facilitates the discovery of 2FA and reduces user frustration. Following a common naming and location also fits the usability heuristic ``Consistency and standards''~\cite{nielsen_10heuristics}.

\parahighlightnoindent{Descriptive-Notification [UX]} Similarly to Promotion, this can help users to make more informed decisions and might nudge them to start their 2FA journey at this point but does not directly affect the security of the user's 2FA.

\parahighlightnoindent{Additional-Information [Neither]} While more information can help users better understand 2FA, this particular additional information is not presented within context (like descriptive notification or option-specific information) and requires users to take a detour to find it. Thus, the impact on the user experience is considered only marginal, and it does not directly affect the security of 2FA.

\parahighlightnoindent{Option-Specific-Information [UX]} This factor implements UX guidelines to provide adequate contextual help~\cite{9781118766576} and timely guidance~\cite{9780321965516}.
A better understanding of the 2FA options definitely affects user experience.
If users have a free choice of 2FA options, this understanding \textit{might} also lead them to select a more secure 2FA choice.
However, However, as shown in the literature, we consider the impact on UX to outweigh these
potential security benefits since such information alone do not guarantee that users \textit{actually} choose (\textit{exclusively}) stronger 2FA options.

\parahighlightnoindent{Step-Wise-Instructions [Both]} Similar to option-specific information, this implements UX guidelines to provide adequate contextual help, break down complex tasks, and offer assistive interfaces without the need to break the user's flow~\cite{9780321965516,9781118766576,9781492055310,nngroup_10th_heuristic} and, hence, improve the UX of the 2FA setup. Additionally, clear instructions should reduce the risk of mistakes and errors that might affect security. Fewer errors should also increase the success rate of users in finishing the 2FA setup.

\parahighlightnoindent{Multiselection [Both]} Offering users multiple options allows them to choose the one that fits them best, affecting their 2FA experience. This is particularly clear from the answers to our survey (see Appendix~\ref{appendix:survey}).
In contrast to option-specific information, multiselection affects security directly since it reflects whether users can actually choose between options with potentially different security levels.

\parahighlightnoindent{Grouped-Setting [UX]} A single location for the settings decreases the burden on the user. 
This is also summarized in UX guidelines, such as the ``gestalt principles'' for the common region and proximity~\cite{gestalt_principles,lawsofux_website}.
Not grouping the 2FA settings \textit{might} cause users to miss stronger 2FA options.
However, similarly to option-specific information, we consider the already demonstrated impact on UX to outweigh this potential security implication.

\parahighlightnoindent{No-Enforced-Options [Both]} Prior work~\cite{238317} has shown that enforcing a certain 2FA option while not communicating that additional 2FA options become available afterward has confused users that were explicitly looking for a specific 2FA option, which differed from the enforced one.
We also consider this an illustration of Norman's Gulf of Execution~\cite{norman_user_1986}.
Thus, this factor can have an impact on the user experience.
Further, forcing users to set up a ``weak'' 2FA option first, e.g., SMS-based OTP, directly affects security by preventing by-design that users can achieve the strongest possible account security.

\parahighlightnoindent{Selectable-Primary-Option [UX]} Allowing users to personalize their 2FA login experience is in accordance with UX guidelines, increasing the usability of the login process and, hence, reducing friction.
Setting a primary option does not affect the choices of 2FA options, just their order during login. Hence, the factor does not affect the 2FA security.

\parahighlightnoindent{Settings-Changed-Verification [Security]} The verification authorizes a security-critical change in the settings, hence, this factor has an impact on the 2FA security of the website.
Assuming that the authorization is done with the default authentication, e.g., password plus any second factors, this authorization does not differ significantly from a regular login.
We assume that this is the usual scenario and, hence, the impact on security outweighs the impact on user experience.

\parahighlightnoindent{Settings-Changed-Notification [Both]} Notification about the successful change of the settings reassures users, and we consider it part of the UX best practice for visibility of the system status~\cite{nngroup_visibility}.
A notification also informs the user, even if they are not interacting with the system currently, about potentially unauthorized actions and allows them to take remediation steps.

\parahighlightnoindent{Confirm-Successful-Setup [UX]} This factor relates to the UX since confirming a successful setup provides visibility of the system status~\cite{nngroup_visibility}, illustrates Norman's Gulf of Evaluation~\cite{norman_user_1986}, and also addresses the peak-end rule for UX~\cite{lawsofux_website}.
It also enhances the experience by reducing the chance that the user accidentally locks themselves out of their account.
This feedback could also enhance security if the user notices failures and mistakes easier, e.g., incomplete or failed TOTP setup that leads the user to incorrectly believe that their account is secured.
However, we considered the impact on UX to outweigh this potential impact on security.

\parahighlightnoindent{Informed-2FA-Recovery-Options [UX]} Offering users an (easy) way to reduce the chance of being locked out of their account increases the usability and the experience.
The availability of recovery options as a "fail-safe" might encourage users to test 2FA.
Recovery options can affect the account security if they are weaker than the 2FA option and provide an easier attack surface.
However, in the end, considering the answers to our survey in Appendix~\ref{appendix:survey}, the impact on UX when users are locked out of their accounts seems to outweigh (for now) the potential risks that a weak recovery option poses.

\parahighlightnoindent{Enforced-2FA-Recovery-Setup [UX]} Mandating a recovery option can help users to reduce the risk of account lock-out but also might increase friction and annoy users.
The argument for the impact on security remains the same as for Informed-2FA-Recovery-Options.

\parahighlightnoindent{Device-Remembrance [UX]} Ciolino et al.~\cite{238317} reported that an unexpected device remembrance policy frustrated their participants. Reynolds et al.~\cite{tale_of_two_studies} as well as our survey results corroborate this issue.
On the other hand, device remembrance reduces the frequency of 2FA, which some of our survey participants appreciated and considered an enhancement of their UX.
Device remembrance can create an attack surface if the attacker has access to a trusted device (e.g., a private device or an accidentally trusted shared/public computer).
However, we consider the impact on UX for all users to outweigh the security drawbacks in this narrow threat model.
 
\parahighlightnoindent{No-Preselected-Option [UX]} Participants of Ciolino et al.~\cite{238317} expressed a desire to personalize the 2FA experience and also UX guidelines recommend ways to allow users to customize the system to their preferences.
Like Selectable-Primary-Option, offering users to choose between their available 2FA options during login does not affect which "weaker" or "stronger" options of 2FA the user has set up, just their order during login. Hence, the factor does not affect the 2FA security.

\parahighlightnoindent{Informed-Deactivation [Both]} Allowing users to deactivate 2FA is in accordance with UX guidelines, like the heuristic for user control and freedom~\cite{nngroup_freedom}, while not allowing them to deactivate can frustrate users.
On the other hand, allowing users to deactivate 2FA also directly allows them to make their accounts less secure.

\parahighlightnoindent{Deactivation-Verification [Security]} Like Settings-Changed-Verification, a security-critical change in the settings has to be authorized. We assume that also here, the authorization takes place via a default authentication (e.g., password plus any configured 2FA option) that does not add additional friction for the user.

\parahighlightnoindent{Deactivation-Notification [Both]} Notification about the deactivation of 2FA can reassure users, and we consider it part of the UX best practice for visibility of the system status~\cite{nngroup_visibility}.
Like Settings-Changed-Notification, an out-of-band notification also informs the user about potentially unauthorized actions and allows them to take remediation steps.

\parahighlightnoindent{Communicate-Successful-Deactivation [UX]} Clearly communicating the deactivation of 2FA follows UX best practice for communicating the system status~\cite{nngroup_visibility}.
While this feedback could warn a user in case they accidentally deactivated 2FA and put their account at higher risk, we consider this rather unlikely and the UX impact to outweigh the potential security benefits.

\subsubsection{Pairwise Hamming distance and Factor Clusters}
\label{appendix:separation_factors:clusters}

We repeat the pairwise Hamming distance measurement and the factor clustering from Section~\ref{sec:results:comparison} with separated sets of comparison factors.
We split the factors into those purely UX-related (i.e., [UX]) and those with security-relevance (i.e., [Security] and [Both]) from Appendix~\ref{appendix:separation_factors:list}.
We further differentiate conditional versus non-conditional factors.
This results in a set of four disjoint sets of comparison factors for clustering: \textit{Non-conditional-UX}, \textit{Non-conditional-Security}, \textit{Conditional-UX} and \textit{Conditional-Security}.

\paragraph{Pairwise Hamming distance}
We repeated measuring the pairwise Hamming distances from Appendix~\ref{appendix:websitesimilarity} for the sets of separated factors.
Considering only \textit{non-conditional-UX} factors, the average website has a minimum Hamming distance of $0.1\pm0.02$, a mean distance of $0.47\pm0.02$, and a maximum distance of $0.86\pm0.01$.
For the set of \textit{non-conditional-security} factors, the average website has a minimum distance of $0.04\pm0.02$, a mean distance of $0.44\pm0.02$, and a maximum distance of $0.90\pm0.01$.
Thus, compared to all factors, the general impression is that these separate sets of factors do not exhibit a better overall consistency across the websites in our data set.

\begin{figure}[h]
    \centering
    \includegraphics[width=.8\linewidth]{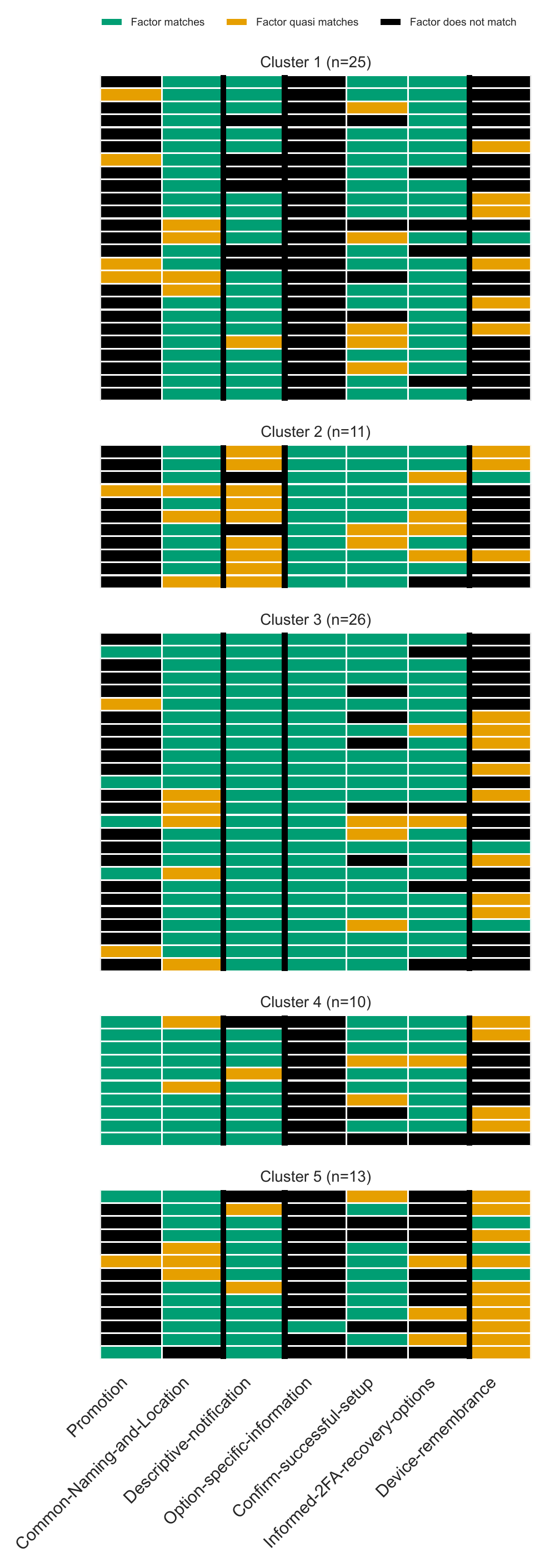}
    \caption{Clusters based on \textit{Non-conditional-UX} comparison factors}
    
    \label{fig:cluster_noncond_ux}
\end{figure}

\begin{figure}[h]
    \centering
    \includegraphics[width=.8\linewidth]{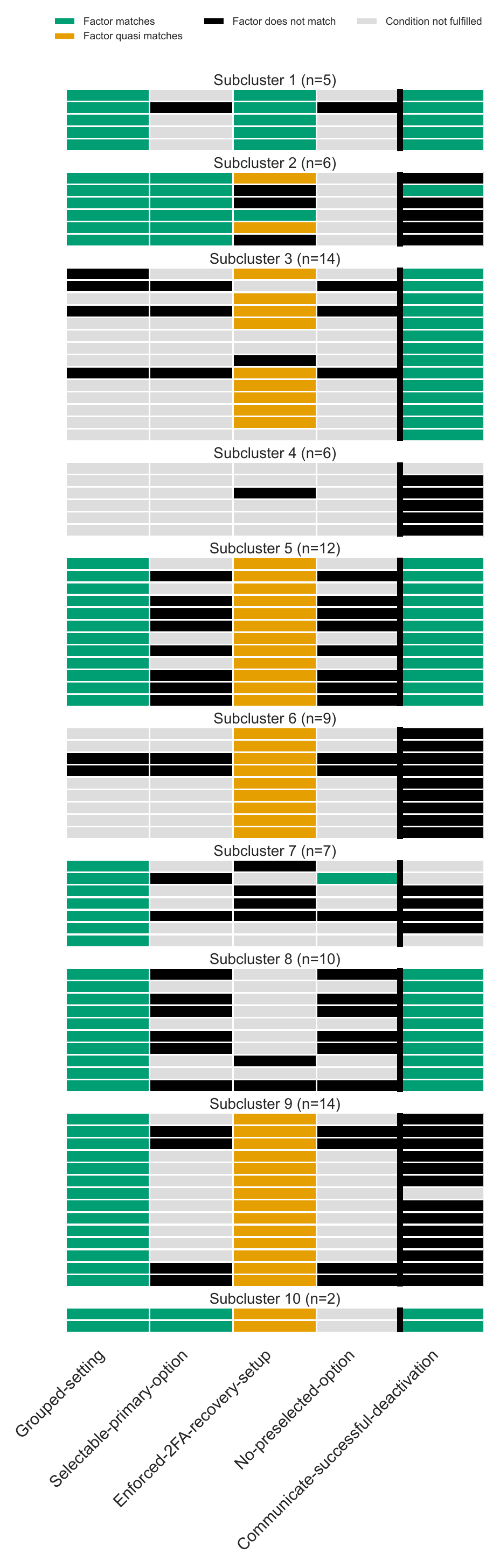}
    \caption{Clusters based on \textit{Conditional-UX} comparison factors}
    \label{fig:cluster_cond_ux}
\end{figure}

\begin{figure}[h]
    \centering
    \includegraphics[width=.8\linewidth]{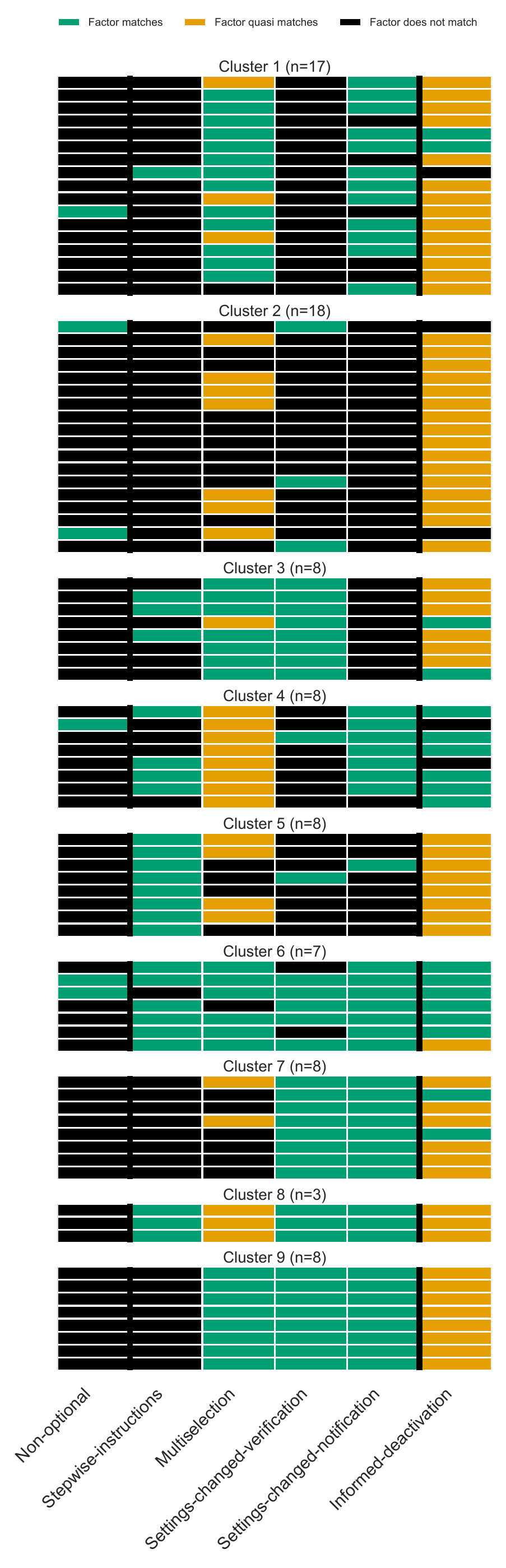}
    \caption{Clusters based on \textit{Non-conditional-security} comparison factors}
    \label{fig:cluster_noncond_security}
\end{figure}

\begin{figure}[h]
    \centering
    \includegraphics[width=.8\linewidth]{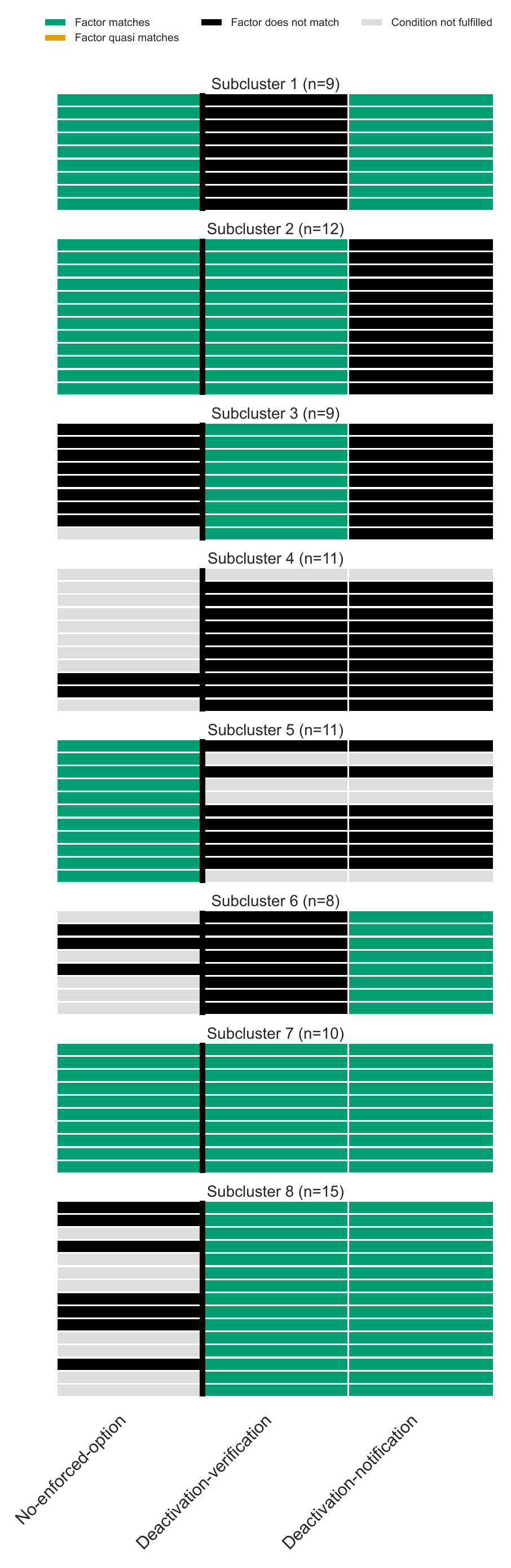}
    \caption{Clusters based on \textit{Conditional-security} comparison factors}
    \label{fig:cluster_cond_security}
\end{figure}

\begin{figure}[h]
    \centering
    \includegraphics[width=.8\linewidth]{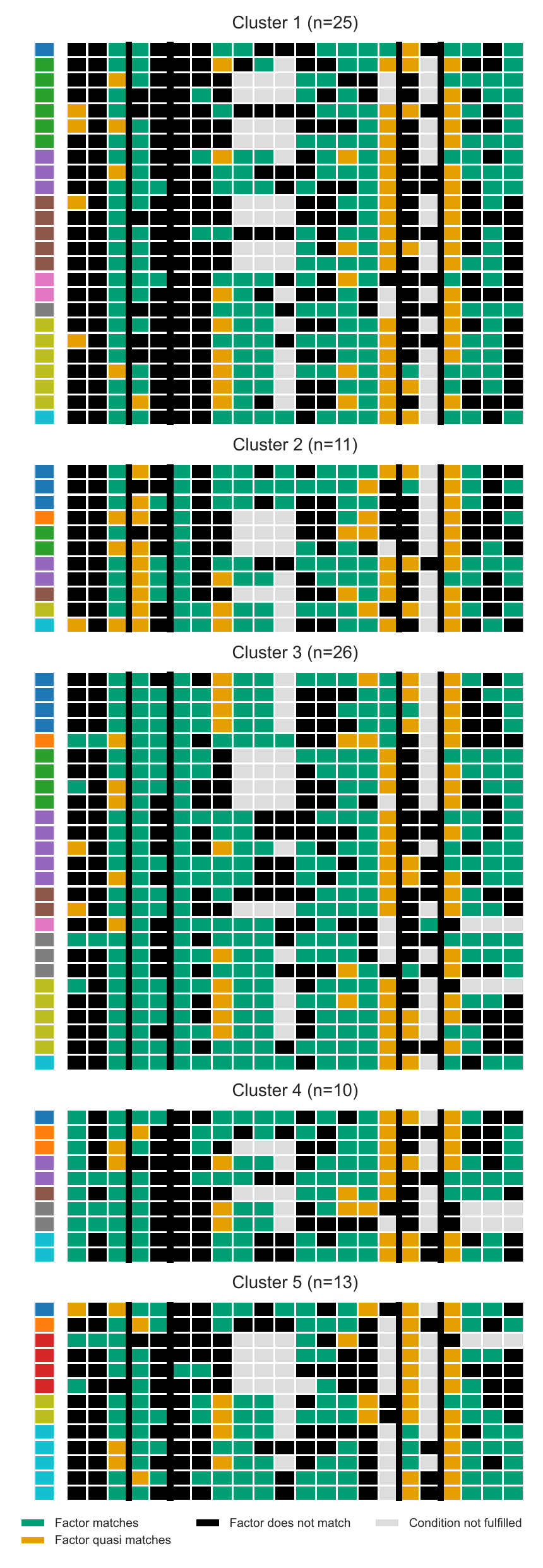}
    \caption{Clusters based on \textit{all purely UX-related comparison factors}, using non-conditional factors for \textit{inter}-clustering and conditional factors for \textit{intra}-clustering (left column).}
    \label{fig:cluster_all_ux}
\end{figure}

\begin{figure}[h]
    \centering
    \includegraphics[width=.8\linewidth]{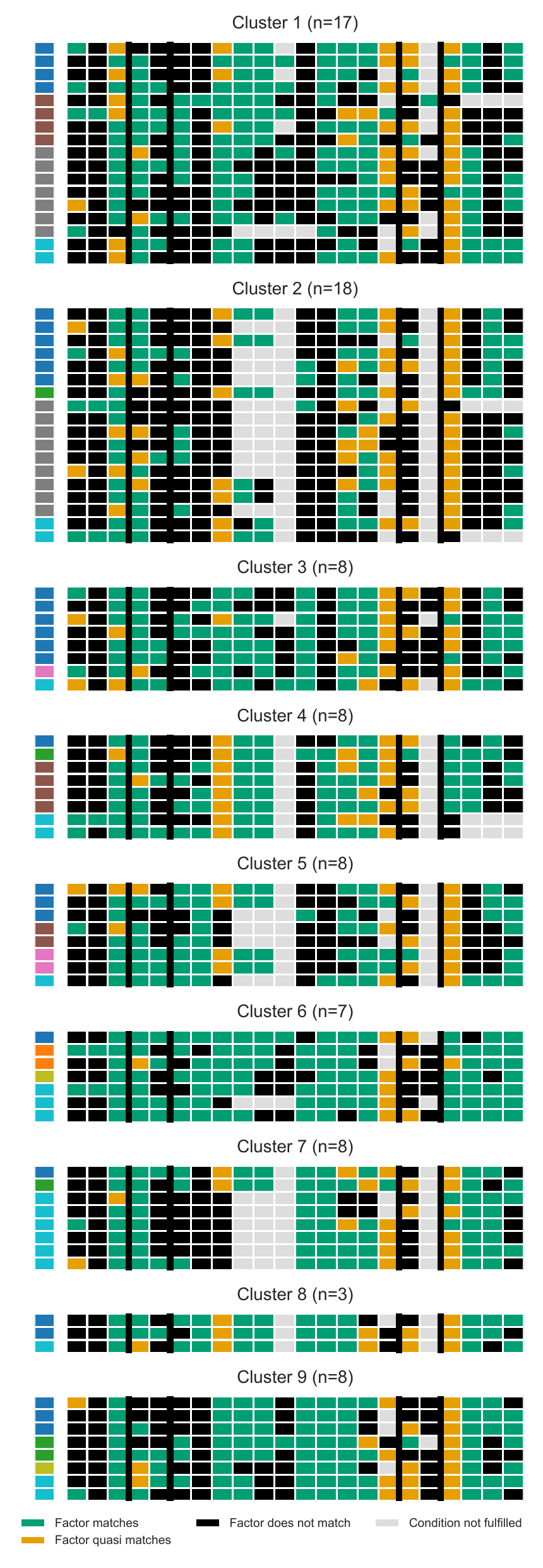}
    \caption{Clusters based on \textit{all security-related comparison factors}, using non-conditional factors for \textit{inter}-clustering and conditional factors for \textit{intra}-clustering (left column).}
    \label{fig:cluster_all_security}
\end{figure}

\paragraph{Factor clusters}
Similar to the high-level view of our clusters with all comparison factors in Appendix~\ref{sec:appendix:clusters}, we provide different views on clusters based on the separated sets of factors.
We again computed the mean Silhouette Coefficient for each set of factors for different clusters with KModes to determine the best number of clusters to describe our data set.
For \textit{Non-conditional-UX} comparison factors, we computed 2 and 5 to be the best number of clusters. Since, with 2 clusters, the websites were only clustered by their strategy for \textit{Option-specific-information}, we chose 5 as the more expressive number of clusters. Figure~\ref{fig:cluster_noncond_ux} illustrates the result of the clustering.
We computed 10 as the best number of clusters for \textit{Conditional-UX} comparison factors, illustrated in Figure~\ref{fig:cluster_cond_ux}.
For \textit{Non-conditional-security} comparison factors, we computed 9 clusters as the best number of clusters. Figure~\ref{fig:cluster_noncond_security} illustrates the resulting clusters.
Moreover, for \textit{Conditional-security} comparison factors, Silhouette testing showed 8 as the best number of clusters, see Figure~\ref{fig:cluster_cond_security}.
We also combined the non-conditional and conditional clusters to get the complete picture of the purely UX-related and security-related clusters.
The results are depicted in Figures~\ref{fig:cluster_all_ux} and \ref{fig:cluster_all_security}, however, the high number of intra-clusters and, for security-related factors, inter-clusters make a meaningful interpretation of this noisy view hard.

\clearpage
\section{Part of 2FA user journey on icloud.com}
\label{appendix:icloudcom}

\begin{figure}[h]
    \centering
    \begin{subfigure}[b]{\linewidth}
    \centering
        \includegraphics[width=.7\textwidth]{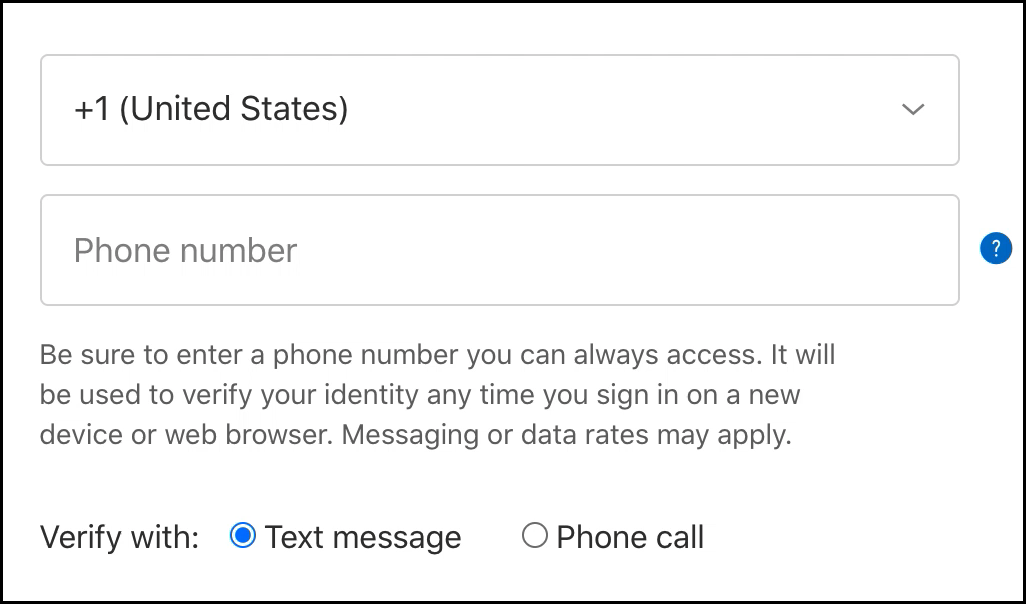}
        \caption{Mandatory phone number for account creation but only brief description of the additional 2FA purpose.}
        \label{fig:icloud1}
    \end{subfigure}
   
      \vspace{0.1cm}
      
    \begin{subfigure}[b]{\linewidth}
        \centering
        \includegraphics[width=.6\textwidth]{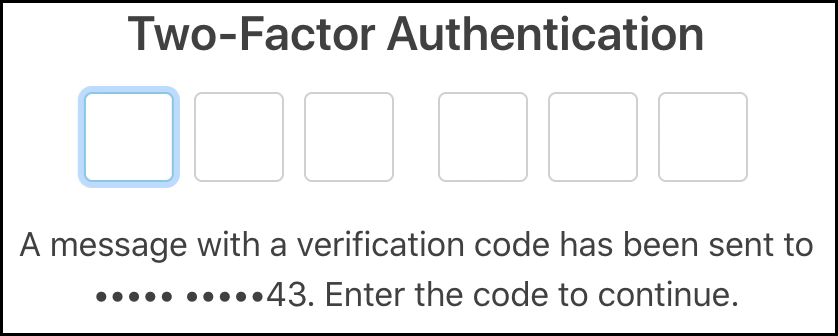}
        \caption{Phone number denoted clearly as 2FA on login.}
        \label{fig:icloud2}
    \end{subfigure}
    \vspace{0.1cm}
    
    \begin{subfigure}[b]{\linewidth}
        \centering
        \includegraphics[width=.9\textwidth]{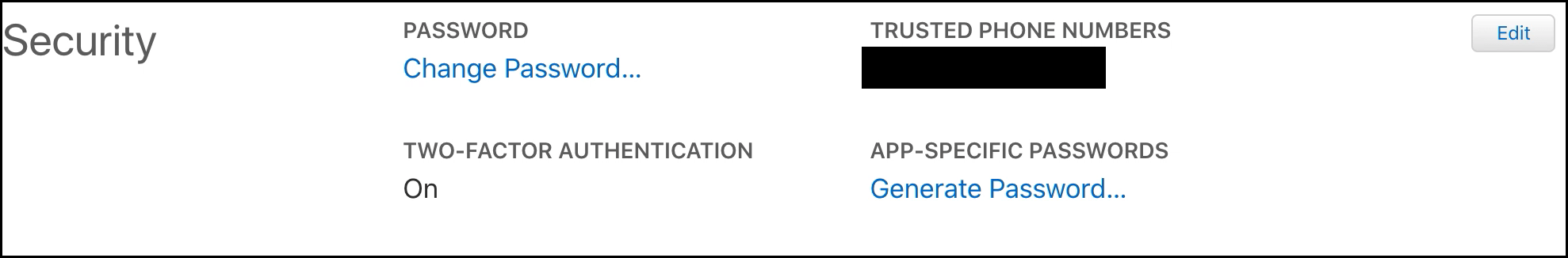}
        \caption{Settings do not allow change or deactivation of 2FA.}
        \label{fig:iclud3}
    \end{subfigure}
    \caption{Part of 2FA user journey of \url{icloud.com}.}

    \label{fig:icloud}
\end{figure}

Figure~\ref{fig:icloud} depicts parts of \url{icloud.com}'s user journey that are inconsistent with the journey's on most other websites.

\newpage
\section{Examples from Websites}
\label{appendix:examples}

We provide examples of different aspects of the 2FA user journey from different websites. For each example, we note how it matches certain comparison factors that we identified in our study of 2FA user journeys (see Section~\ref{sec:comparisonfactors}).

\begin{figure}[h]
    \centering
    \begin{subfigure}[b]{\linewidth}
        \centering
        \includegraphics[width=0.6\textwidth]{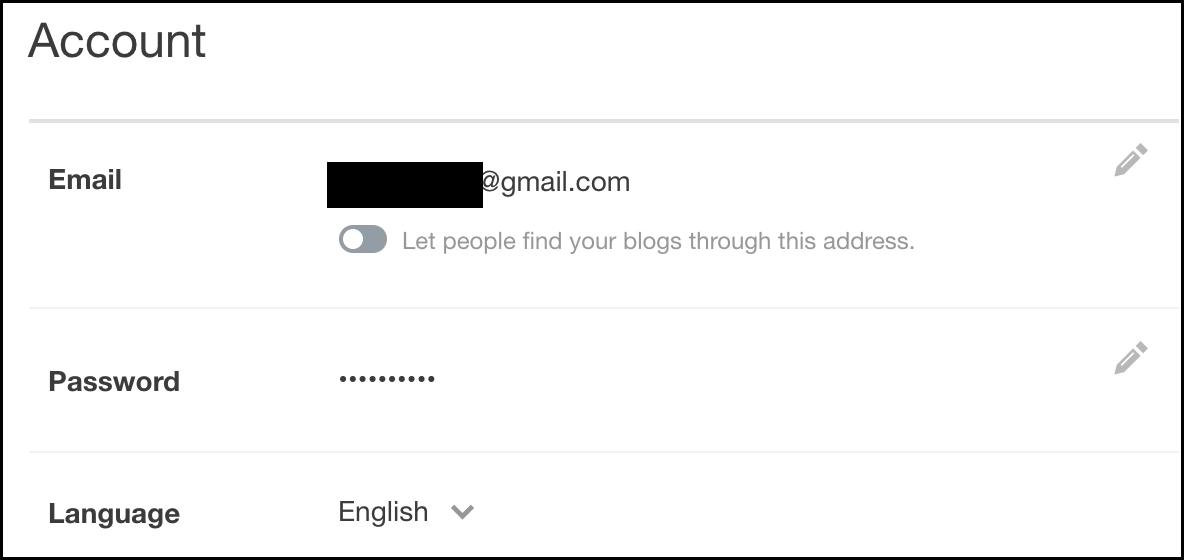}
        \caption{\url{tumblr.com} before email verification}
        \label{fig:temblr1}
    \end{subfigure}
   
      \vspace{0.1cm}
      
    \begin{subfigure}[b]{\linewidth}
        \centering
        \includegraphics[width=0.6\textwidth]{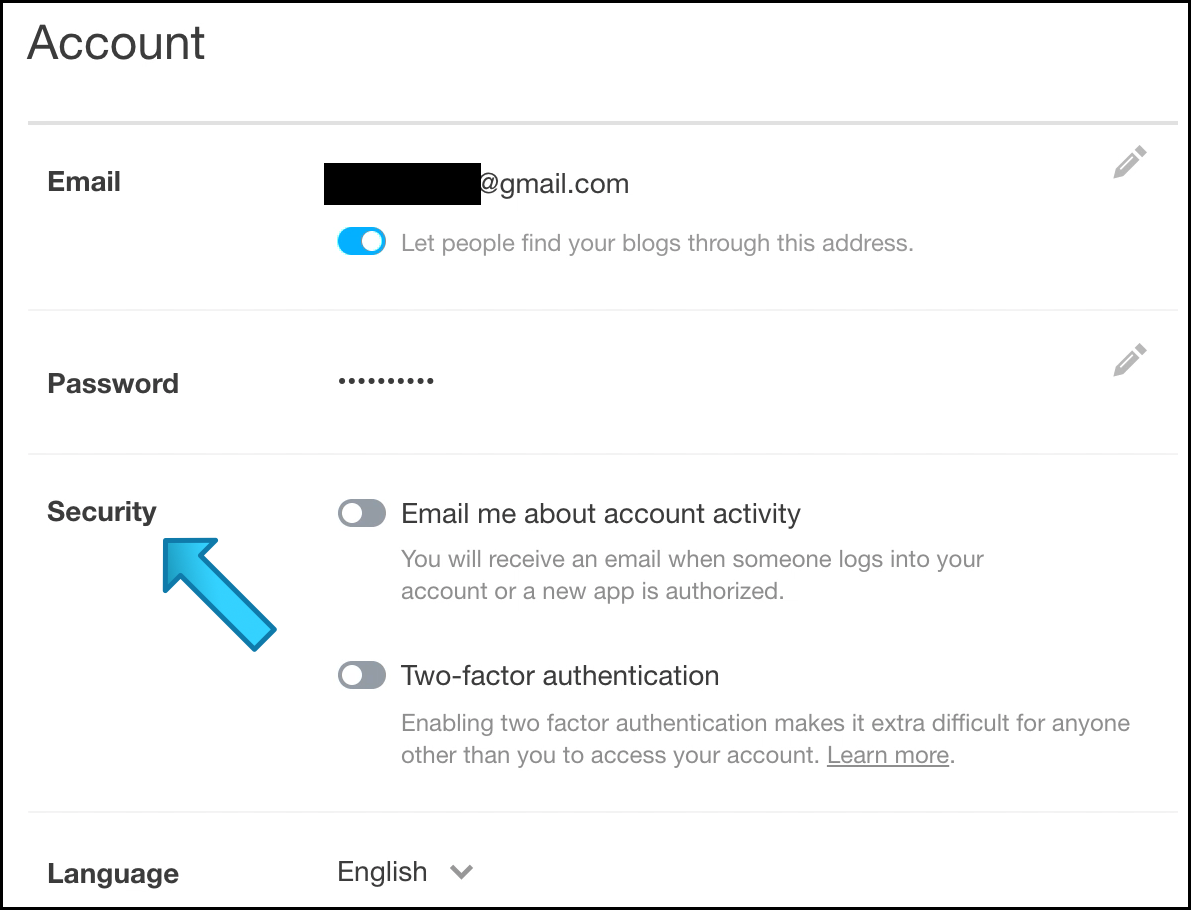}
        \caption{\url{tumblr.com} after email verification}
        \label{fig:tubmlr2}
    \end{subfigure}
    \caption{\url{tumblr.com} has a \textit{Common-Naming-and-Location}~(\yes), but the security settings are initially hidden until the user verifies their email address.}

    \label{fig:Tumblr}
\end{figure}

\begin{figure}[h]
    \centering
    \includegraphics[width=.6\linewidth]{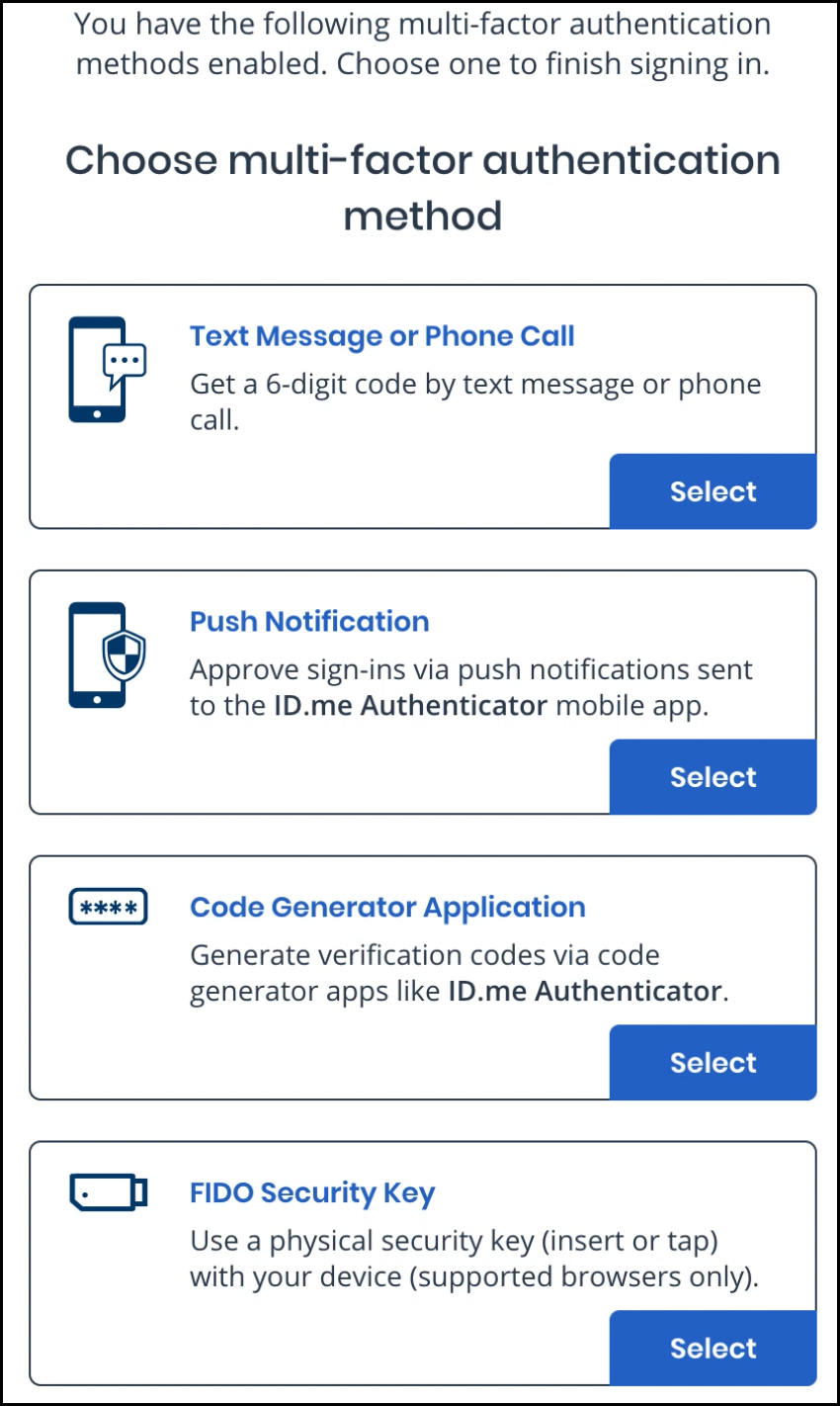}
    \caption{\url{id.me} allows users to choose their 2FA option upfront during login (\textit{No-preselected-option}: \yes) and supports multiple activated 2FA options (\textit{Multiselection}: \yes).}
    \label{fig:idme}
\end{figure}

\begin{figure}[h]
    \centering
    
     \begin{subfigure}[b]{\linewidth}
        \centering
        \includegraphics[width=.8\textwidth,frame]{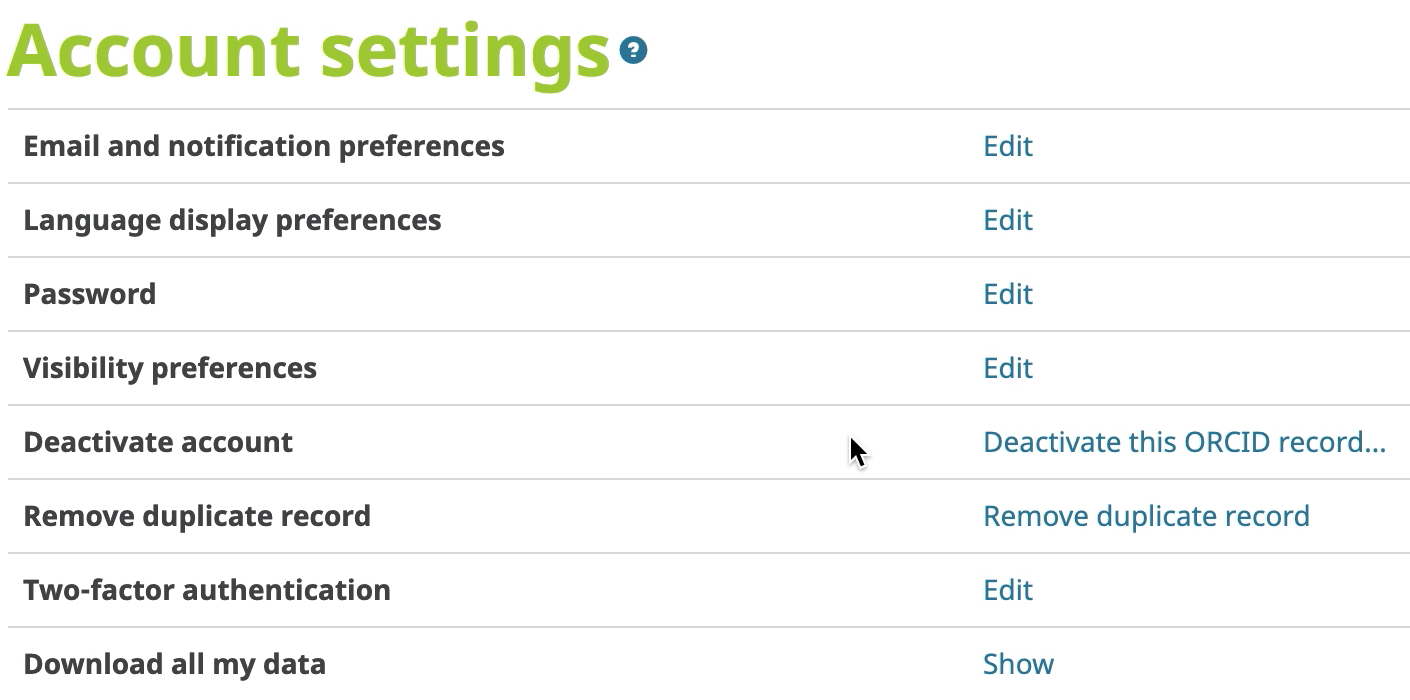}
        \caption{\url{orcid.org}~(\yes) has common name ("Two Factor Authentication") that places the 2FA settings at a common location ("Account settings" tab).}
        \label{fig:orcid}
    \end{subfigure}  
     \vspace{0.1cm}
     
    \begin{subfigure}[b]{\linewidth}
    \centering
        \includegraphics[width=\textwidth]{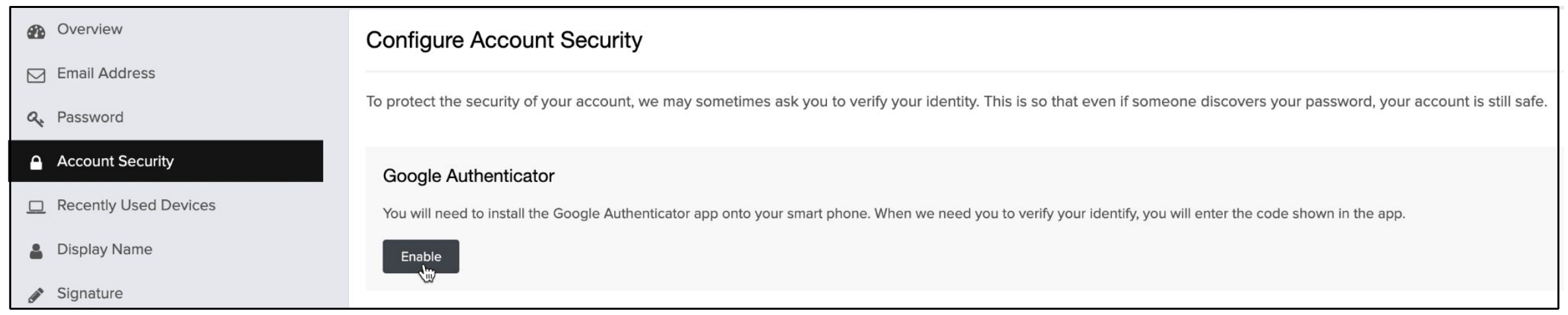}
        \caption{\url{airvpn.org}~(\kinda) has an uncommon name ("Configure account security") that places the 2FA settings at the common location ("Account security" tab).}
        \label{fig:airvpn}
    \end{subfigure}
   
      \vspace{0.1cm}
      
         \begin{subfigure}[b]{\linewidth}
        \centering
        \includegraphics[width=.8\linewidth,frame]{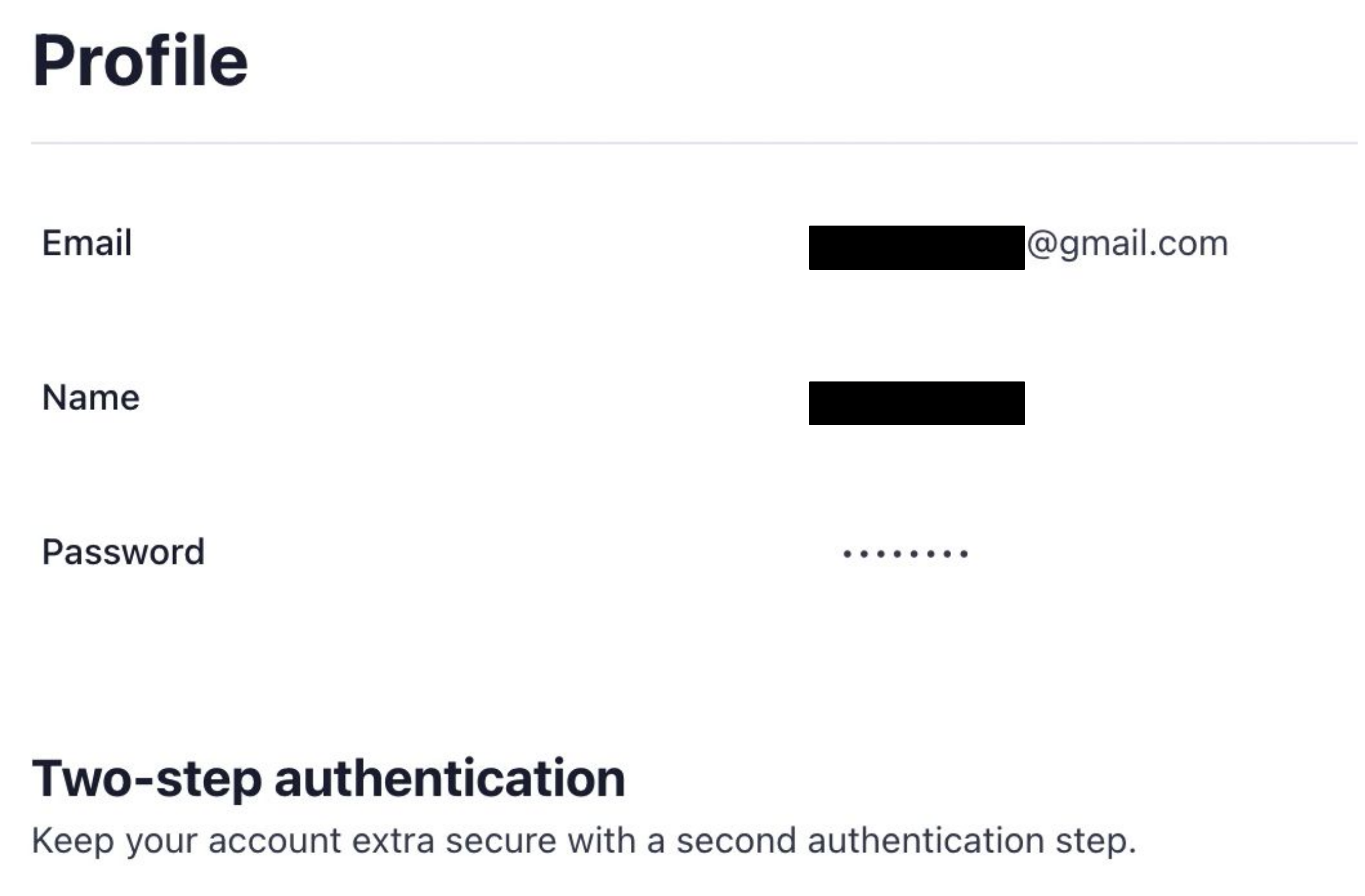}
       \caption{\url{stripe.com}~(\kindaright) has common name that places the 2FA settings at an uncommon location ("Profile" tab).}
        \label{fig:stripe}
    \end{subfigure}
       \vspace{0.1cm}
      
    \begin{subfigure}[b]{\linewidth}
        \centering
        \includegraphics[width=\linewidth,frame]{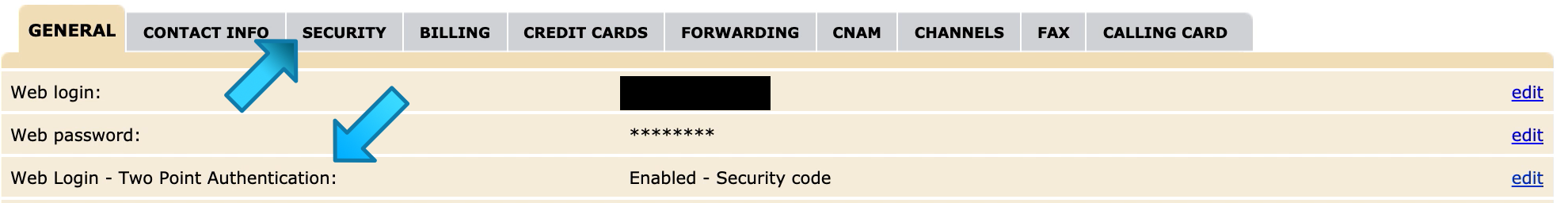}
       \caption{\url{callcentric.com}~(\no) has uncommon name ("Two Point Authentication") for 2FA that places the 2FA settings at an uncommon location ("General" tab) despite the dedicated "Security" tab.}
        \label{fig:callcentric}
    \end{subfigure}

    \caption{Examples of websites that (quasi/not) matched the \textit{Common-Naming-and-Location} factor.}

    \label{fig:commonanduncommon}
\end{figure}

\begin{figure}[h]
    \centering
    
    \begin{subfigure}[b]{\linewidth}
    \centering
        \includegraphics[width=0.7\textwidth,frame]{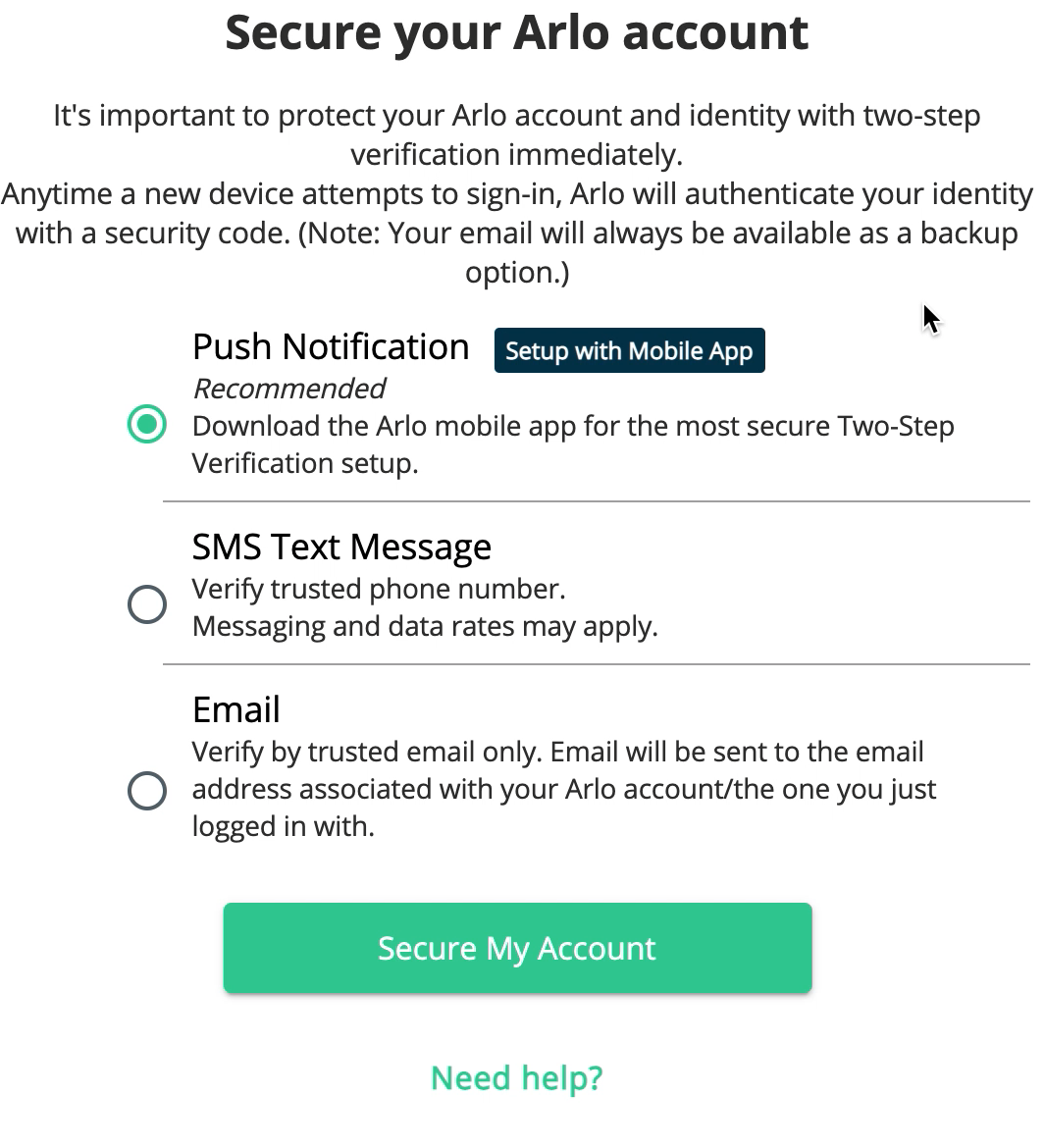}
        \caption{\url{arlo.com} promotes its 2FA as a pop-up immediately after an account creation~(\textit{Promotion}: \yes) and it is mandatory to setup 2FA for their user account~(\textit{Non-optional}: \yes). Further, the pop-up provides \textit{Descriptive-notification}~(\yes), \textit{Additional-information}~(\yes), and \textit{Option-specific-information}~(\yes).}
        \label{fig:arlo}
    \end{subfigure}
    
      \vspace{0.1cm}
      
    \begin{subfigure}[b]{\linewidth}
        \centering
        \includegraphics[width=0.7\linewidth,frame]{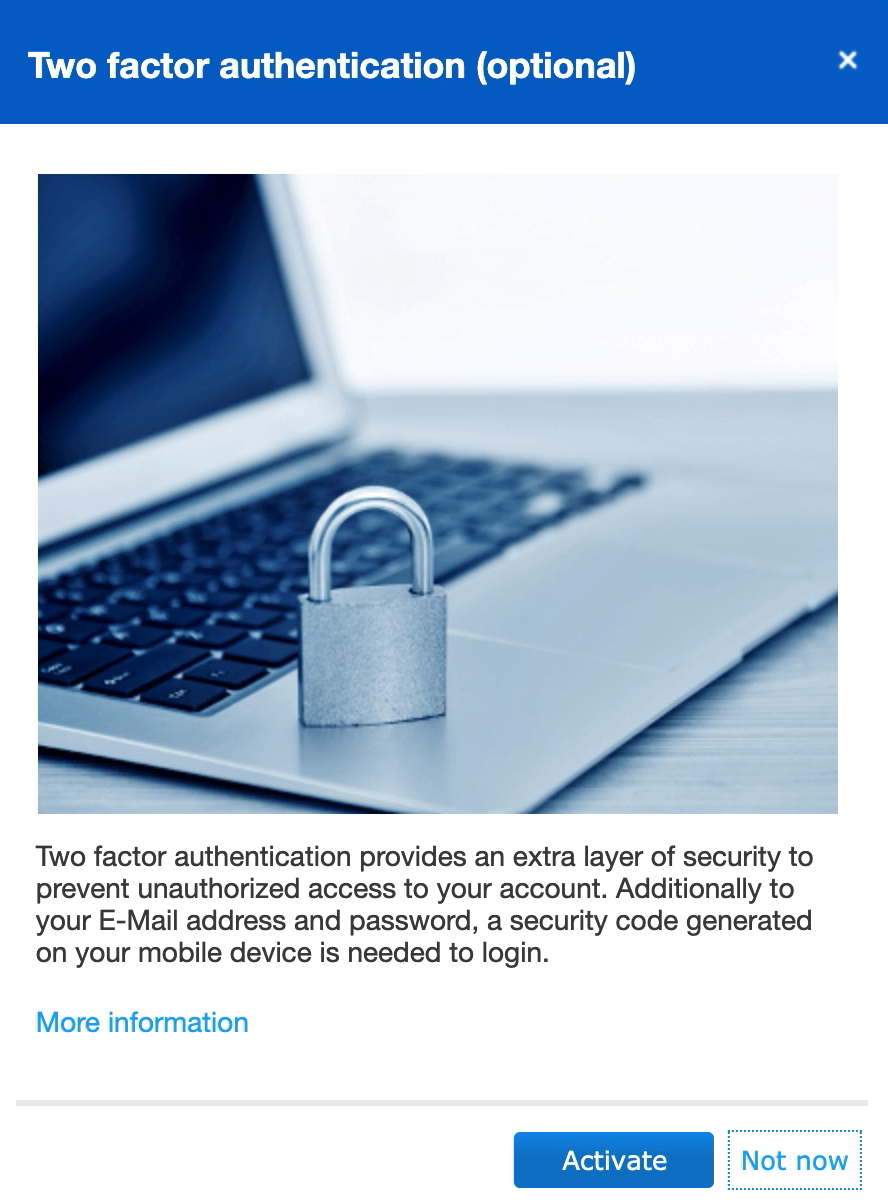}
       \caption{\url{teamviewer.com} promotes its 2FA as a pop-up immediately after an account creation~(\textit{Promotion}: \yes) but it is optional to setup 2FA~(\textit{Non-optional}: \no). Further, the pop-up provides \textit{Descriptive-notification}~(\yes) and \textit{Additional-information}~(\yes).}
        \label{fig:teamviewer}
    \end{subfigure}

    \caption{Examples of websites that match the \textit{Promotion} factor and (not) matched the \textit{Non-Optional} factor.}

    \label{fig:promandnonoptional}
\end{figure}

\begin{figure}[h]
    \centering
    
    \begin{subfigure}[b]{\linewidth}
    \centering
        \includegraphics[width=\textwidth,frame]{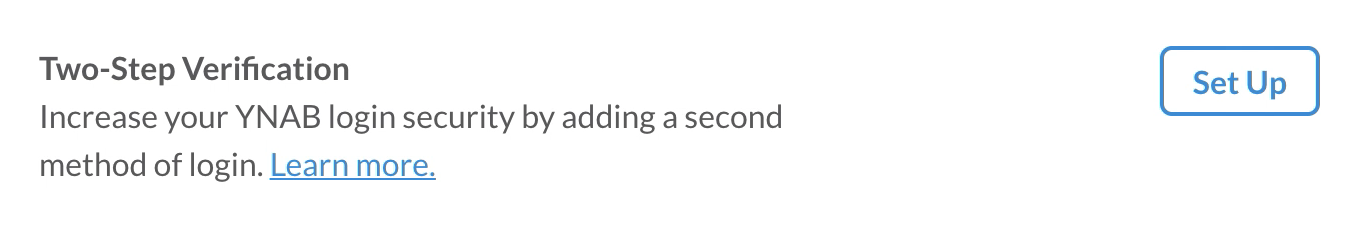}
        \caption{\url{youneedabudget.com} provides their user a description of the 2FA in advanced before user clicks to "set up"~(\textit{Descriptive-notification:} \yes) and offers additional information via "learn more" link to their users~(\textit{Additional-information}: \yes).}
        \label{fig:youneedabudget}
    \end{subfigure}
   
      \vspace{0.1cm}
      
       \begin{subfigure}[b]{\linewidth}
    \centering
        \includegraphics[width=\textwidth]{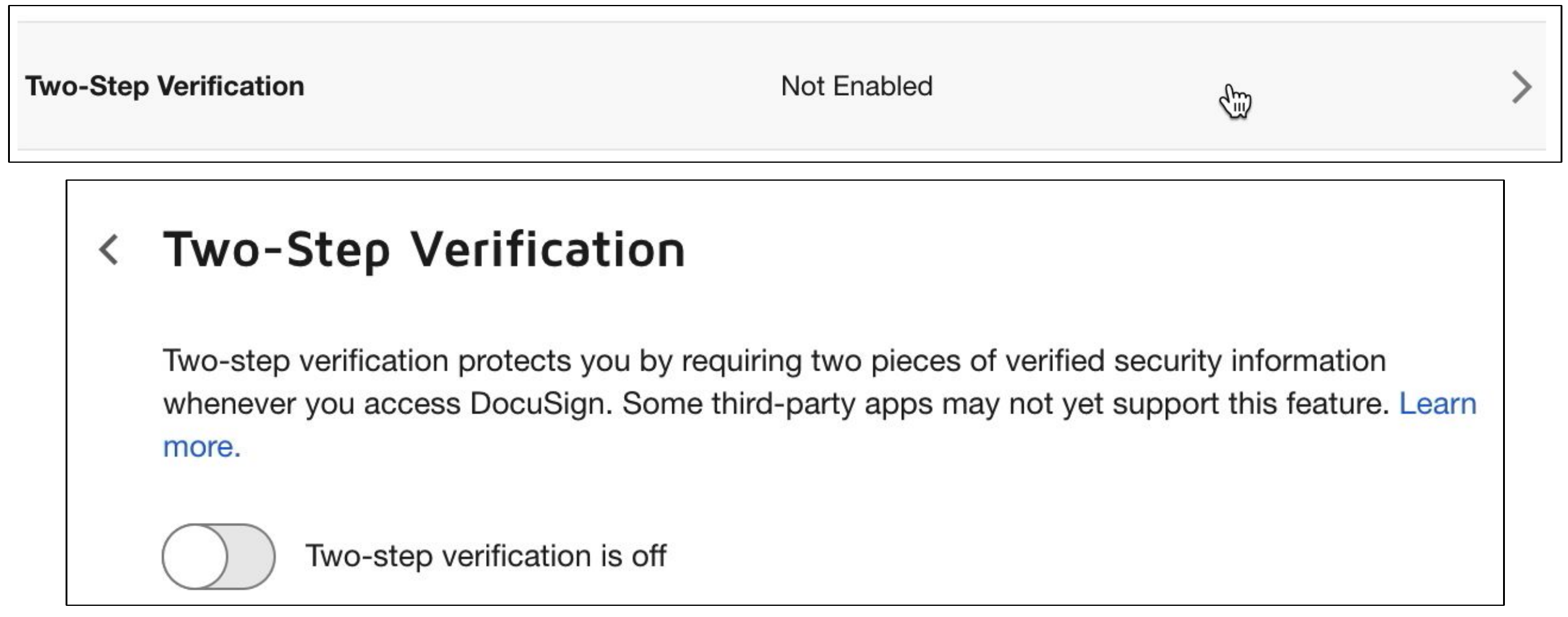}
        \caption{\url{docusign.com} provides their user a description of the 2FA after user "clicks to enable"~(\textit{Descriptive-notification}: \kinda) and offers additional information via "learn more" link to their users~(\textit{Additional-information}: \yes).}
        \label{fig:kickstar}
    \end{subfigure}
    
      \vspace{0.1cm}

    \begin{subfigure}[b]{\linewidth}
        \centering
        \includegraphics[width=\linewidth,frame]{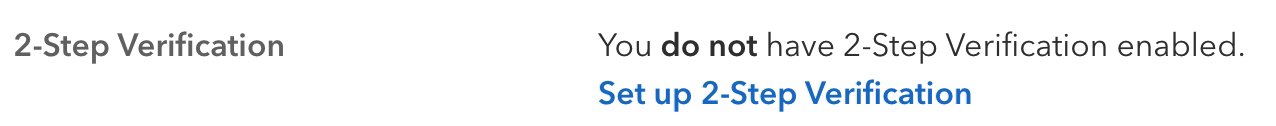}
       \caption{\url{23andme.com} does not present a description of 2FA~(\textit{Descriptive-notification}: \no) and does not offer any additional information to their end users~(\textit{Additional-information}: \no).}
        \label{fig:23andme}
    \end{subfigure}

    \caption{Examples of websites that (quasi/not) matched \textit{Descriptive-notification} and (not) matched \textit{Additional-information}.}

    \label{fig:descriptionandadditionalinfo}
\end{figure}

\begin{figure}[h]
    \centering
    
    \begin{subfigure}[b]{\linewidth}
    \centering
      \includegraphics[width=\linewidth,frame]{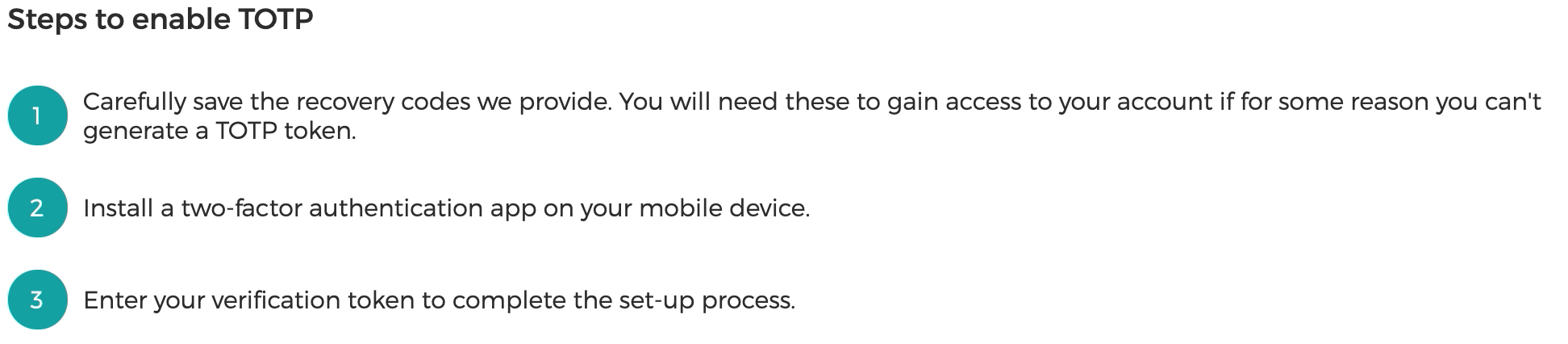}
    \caption{\url{gandi.net}~(\yes) gives an overview of the involved steps of setting 2FA by introducing them in advance.}
    \label{fig:gandi}
    \end{subfigure}

      \vspace{0.1cm}
      
    \begin{subfigure}[b]{\linewidth}
        \centering
        \includegraphics[width=\linewidth,frame]{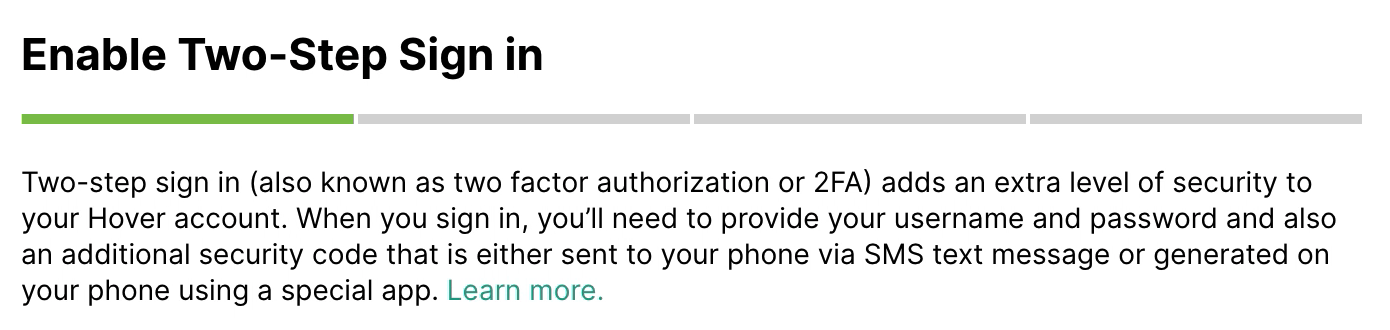}
       \caption{\url{hover.com}~(\yes) gives an overview of the involved steps of setting 2FA as a progress bar.}
        \label{fig:hover}
    \end{subfigure}

    \caption{Examples of websites that matched the \textit{Stepwise-Instruction} factor.}

    \label{fig:stepwise}
\end{figure}

\begin{figure}[h]
    \centering
    
    \begin{subfigure}[b]{\linewidth}
    \centering
      \includegraphics[width=\linewidth,frame]{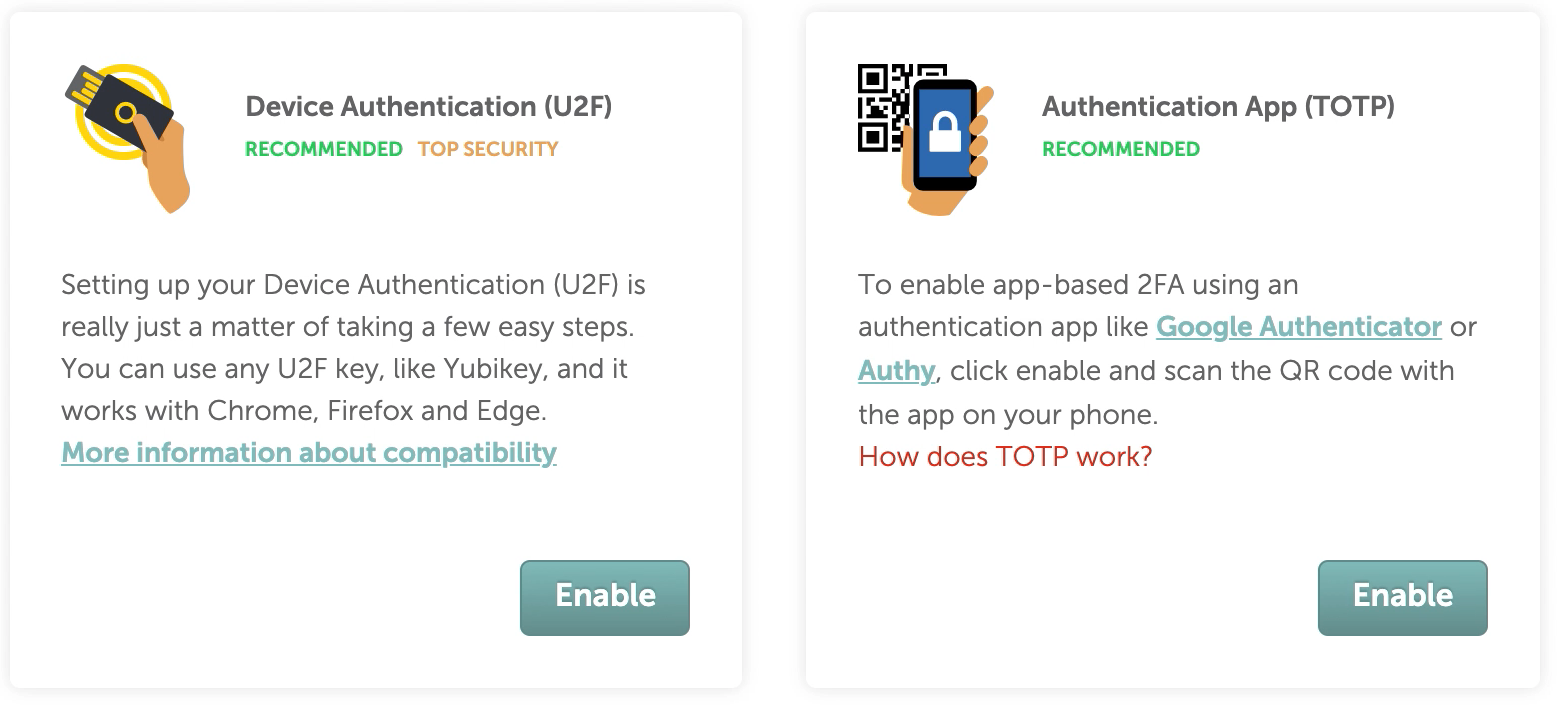}
    \caption{\url{namecheap.com}~(\yes) is an example of a website that provides specific information about the 2FA options that it supports. }
    \label{fig:namecheap}
    \end{subfigure}
   
      \vspace{0.1cm}
      
      \begin{subfigure}[b]{\linewidth}
    \centering
      \includegraphics[width=.7\linewidth,frame]{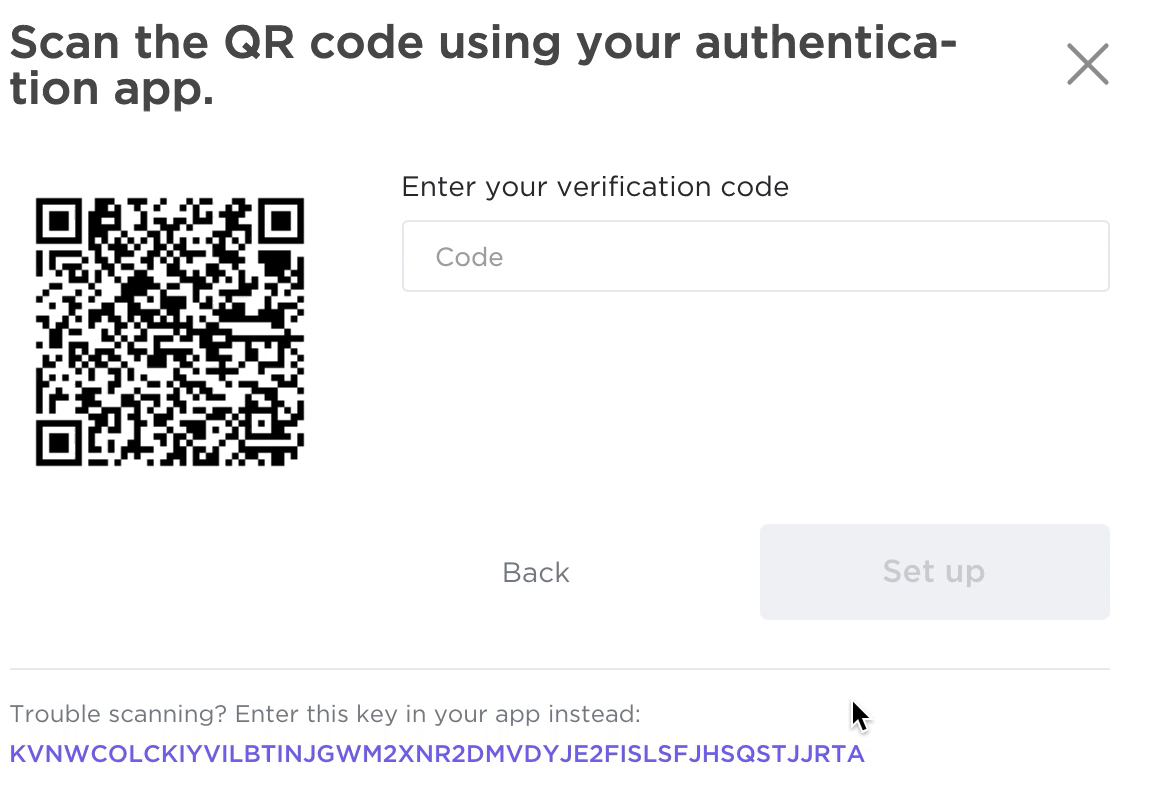}
    \caption{\url{clickup.com}~(\no) does not explain the setup of TOTP to its users but immediately prompts the user to make use of the authenticator app. }
    \label{fig:namecheap}
    \end{subfigure}
    
      \vspace{0.1cm}
      
    \begin{subfigure}[b]{\linewidth}
        \centering
        \includegraphics[width=.7\linewidth,frame]{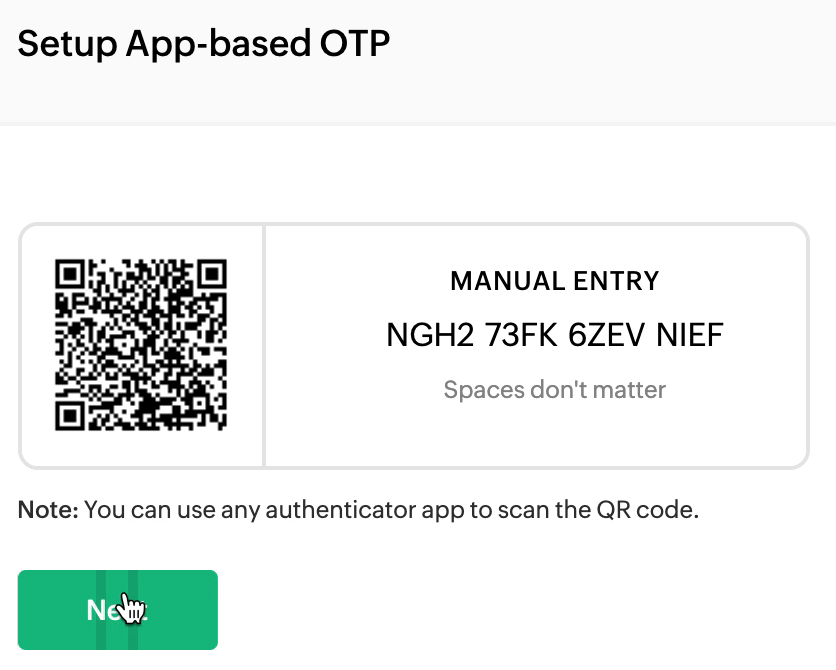}
       \caption{\url{zoho.com}~(\no) does not explain the setup of TOTP to its users but immediately prompts the user to make use of the authenticator app.}
        \label{fig:roboform}
    \end{subfigure}

    \caption{Examples of websites that (not) matched the \textit{Option-Specific-Information} factor.}

    \label{fig:educate}
\end{figure}

\begin{figure}[h]
    \centering
    
    \begin{subfigure}[b]{\linewidth}
    \centering
      \includegraphics[width=.8\linewidth,frame]{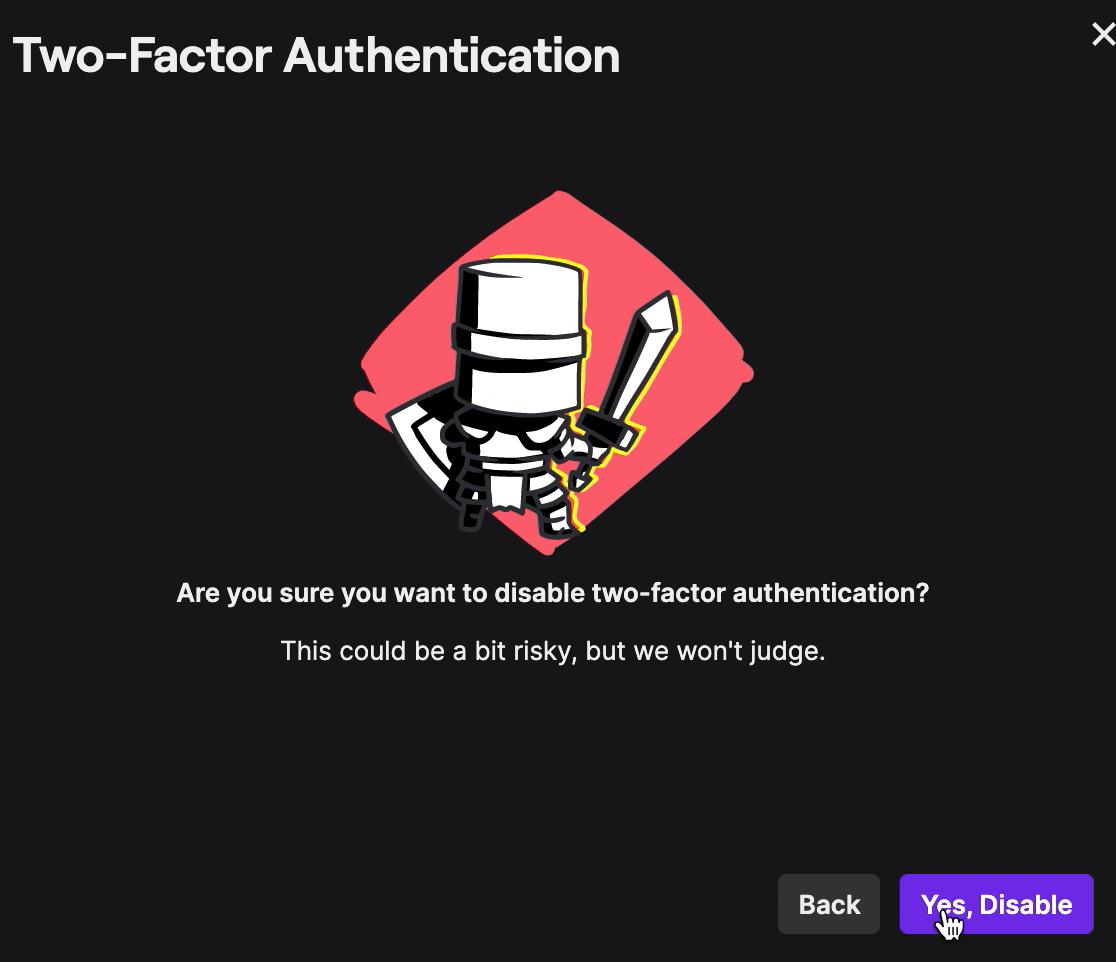}
    \caption{\url{twitch.tv}~(\yes) allows the user to deactivate 2FA and warns about the potential risk associated with that.}
    \label{fig:twitch-deactivation}
    \end{subfigure}
      \vspace{0.1cm}
      
    \begin{subfigure}[b]{\linewidth}
        \centering
        \includegraphics[width=.8\linewidth,frame]{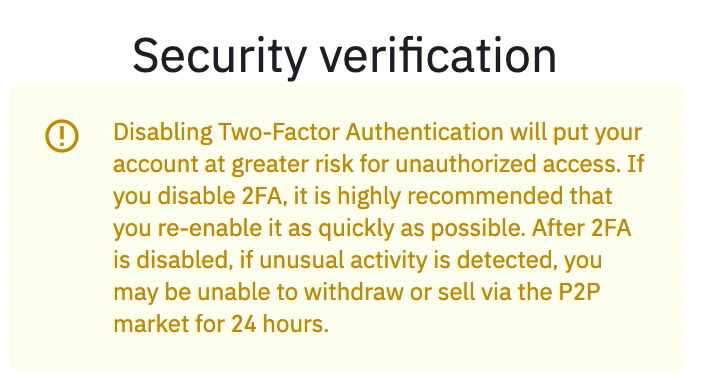}
       \caption{\url{binance.com}~(\yes) allows the user to deactivate 2FA and warns about the potential risk associated with that.}
        \label{fig:binance-deactivation}
    \end{subfigure}
     \vspace{0.1cm}

        \begin{subfigure}[b]{\linewidth}
        \centering
        \includegraphics[width=0.8\linewidth,frame]{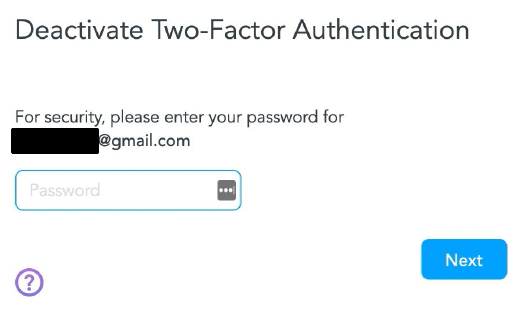}
      \caption{\url{meistertask.com}~(\kinda) allows the user to deactivate the 2FA but does not communicate the risk associated with it.}
        \label{fig:meistertask-deactivation}
    \end{subfigure}
    \vspace{0.1cm}
    
    \begin{subfigure}[b]{\linewidth}
        \centering
        \includegraphics[width=\linewidth,frame]{Image/icloud3line.png}
      \caption{\url{icloud.com}~(\no) does not allow the user to deactivate the 2FA.}
        \label{fig:apple-deactivation}
    \end{subfigure}
    
    \caption{Examples of websites that (quasi/not) matched the \textit{Informed-deactivation} factor.}

    \label{fig:deactivation-feedback}
\end{figure}

\begin{figure}[h]
    \centering
    
    \begin{subfigure}[b]{\linewidth}
    \centering
      \includegraphics[width=.8\linewidth,frame]{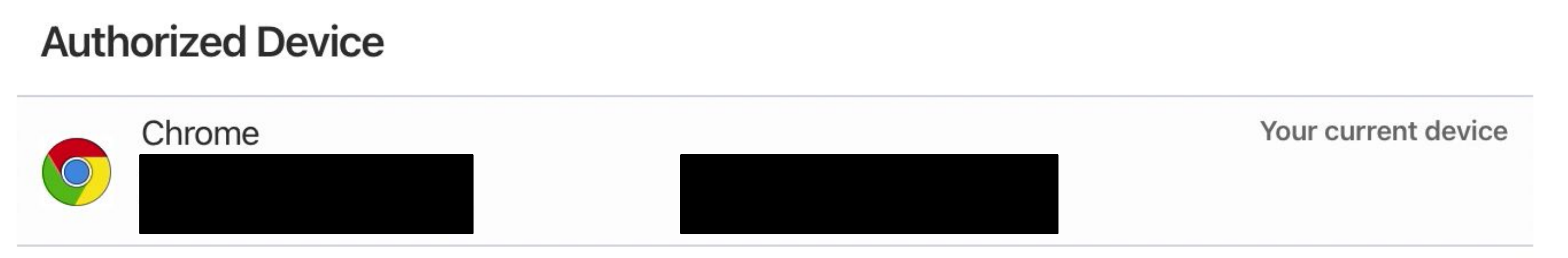}
    \caption{\url{1password.com}~(\yes) automatically sets device remembrance without involving the user.}
    \label{fig:1password-remember}
    \end{subfigure}
   
      \vspace{0.1cm}
      
    \begin{subfigure}[b]{\linewidth}
        \centering
        \includegraphics[width=.8\linewidth,frame]{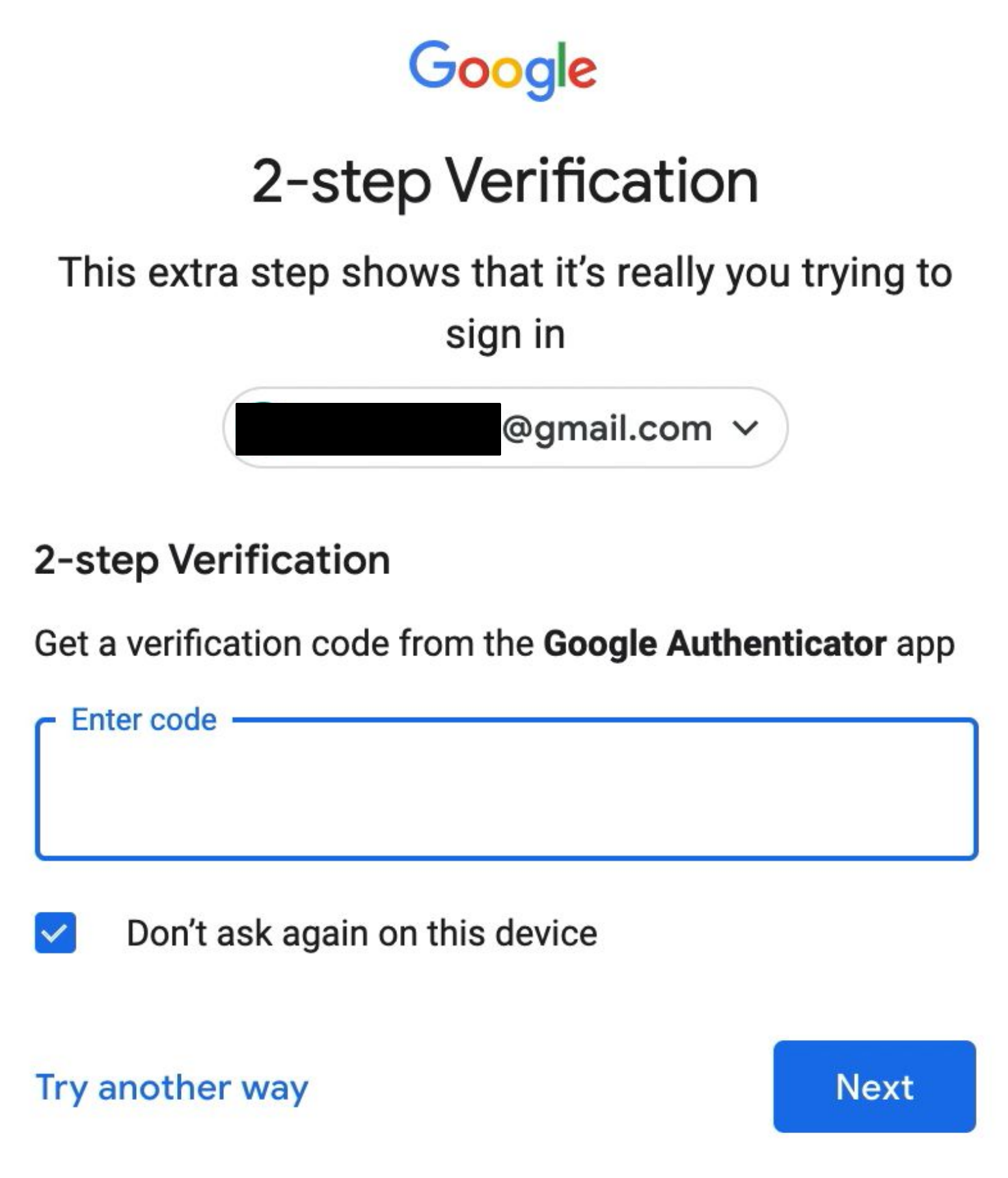}
       \caption{\url{google.com}~(\kindaright) puts the device remembrance at the discretion of the user and states it as opt-\textit{out}.}
        \label{fig:google-remember}
    \end{subfigure}
     \vspace{0.1cm}

        \begin{subfigure}[b]{\linewidth}
        \centering
        \includegraphics[width=0.8\linewidth,frame]{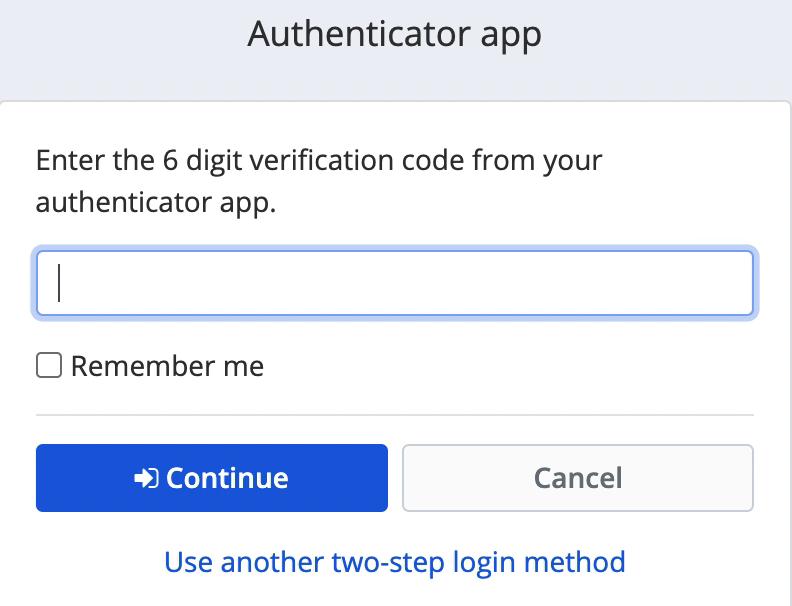}
      \caption{\url{meistertask.com}~(\kinda) states the device remembrance as opt-\textit{in}.}
        \label{fig:bitwarden-remember}
    \end{subfigure}

    \caption{Examples of websites that (quasi) matched the \textit{Device-remembrance} factor.}

    \label{fig:deviceremembrance-feedback}
\end{figure}

\begin{figure}[h]
    \centering
    
    \begin{subfigure}[b]{\linewidth}
    \centering
      \includegraphics[width=.8\linewidth,frame]{Image/clickup-confirm-setup.png}
    \caption{\url{clickup.com} asks users to confirm the successful setup of TOTP by entering the current code.}
    \end{subfigure}
   
      \vspace{0.1cm}
 
   \begin{subfigure}[b]{\linewidth}
    \centering
      \includegraphics[width=.8\linewidth,frame]{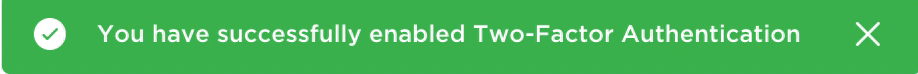}
    \caption{\url{clickup.com} shows a highlighted message to inform the user about the successful activation of 2FA.}
    \end{subfigure}

    \caption{\url{clickup.com} as an example for websites that matched the \textit{Confirm-Successful-Setup}~(\yes) factor.}

    \label{fig:basecamp-confirm-setup}
\end{figure}

\begin{figure}[h]
    \centering
    
   \includegraphics[width=\linewidth,frame]{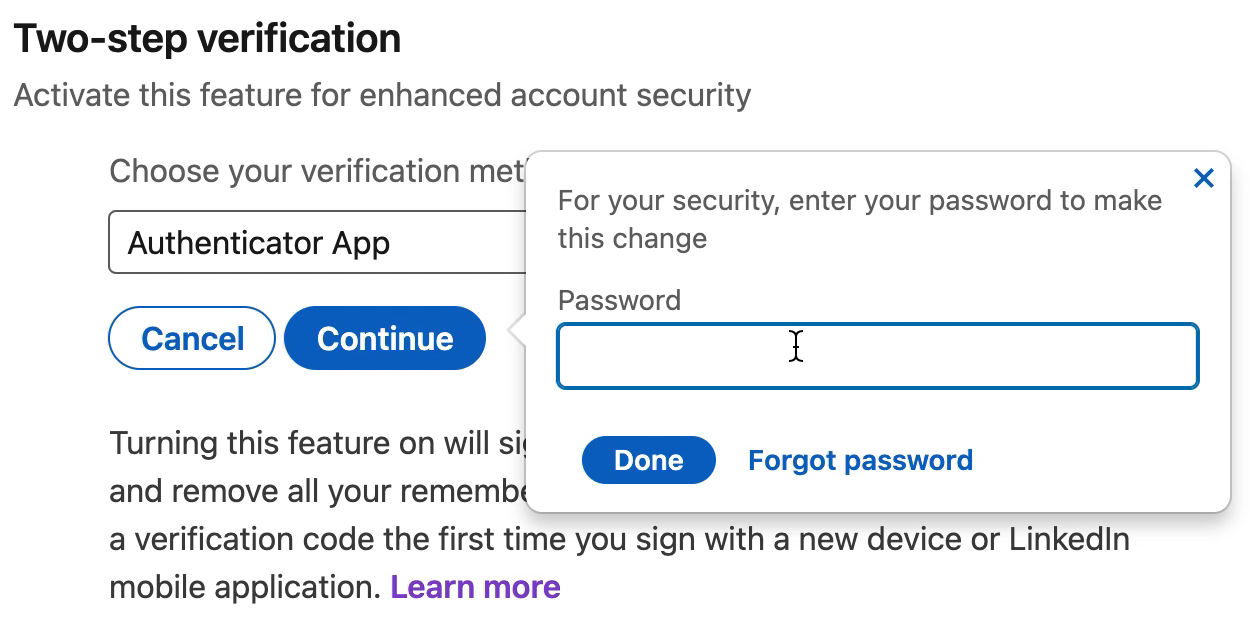}
    \caption{\url{linkedin.com} is an example for a website that asks users to verify their identity before being able to setup 2FA~(\textit{Settings-Changed-Verification}: \yes).}

    \label{fig:linkedin-setting-setup-verification}
\end{figure}

\begin{figure}[h]
    \centering
    
    \begin{subfigure}[b]{\linewidth}
    \centering
      \includegraphics[width=.8\linewidth,frame]{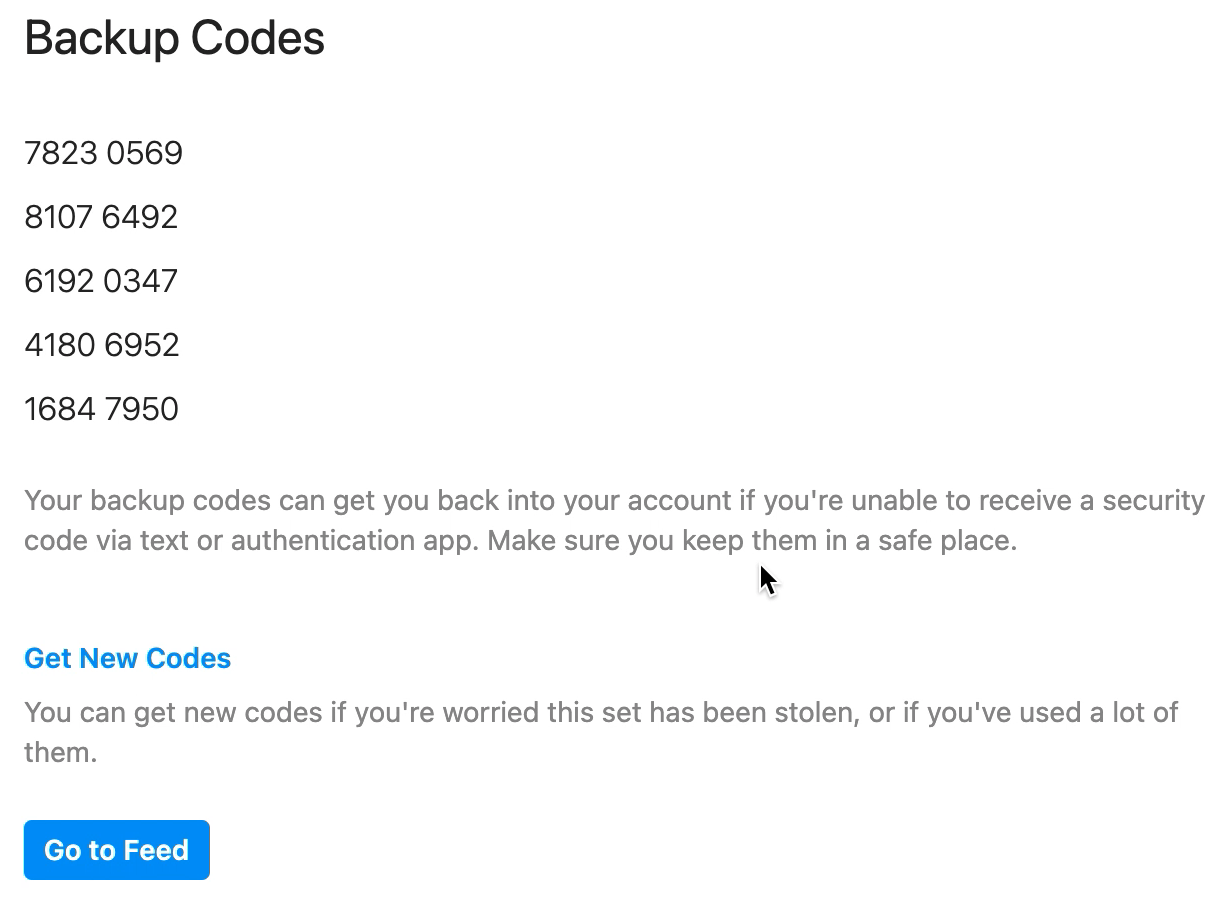}
    \caption{\url{instagram.com}~(\yes) offers a dedicated recovery option and explains to the user why it is important to enable recovery options.}
    \label{fig:basecamp}
    \end{subfigure}
   
      \vspace{0.1cm}
 
  \begin{subfigure}[b]{\linewidth}
    \centering
      \includegraphics[width=.8\linewidth,frame]{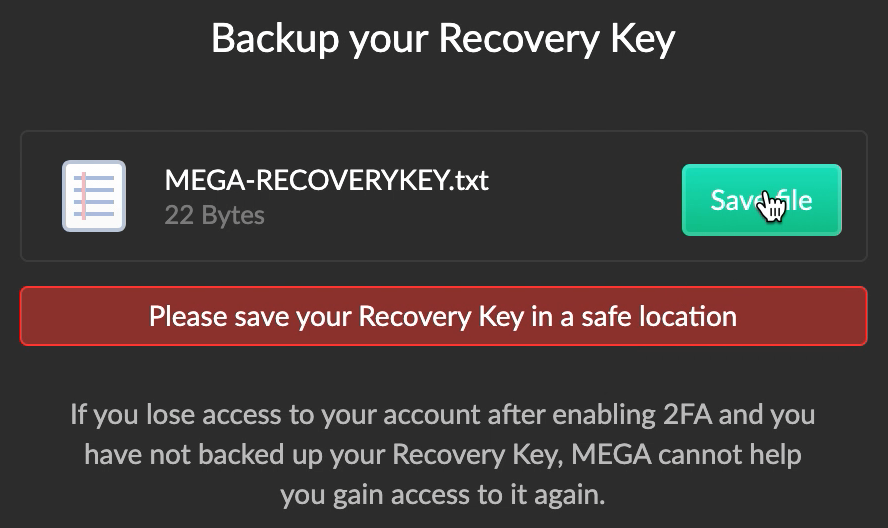}
    \caption{\url{instagram.com}~(\yes) offers a dedicated recovery option and explains to the user why it is important to enable recovery options.}
    \label{fig:basecamp}
    \end{subfigure}
   
      \vspace{0.1cm}
 
   \begin{subfigure}[b]{\linewidth}
    \centering
      \includegraphics[width=.8\linewidth,frame]{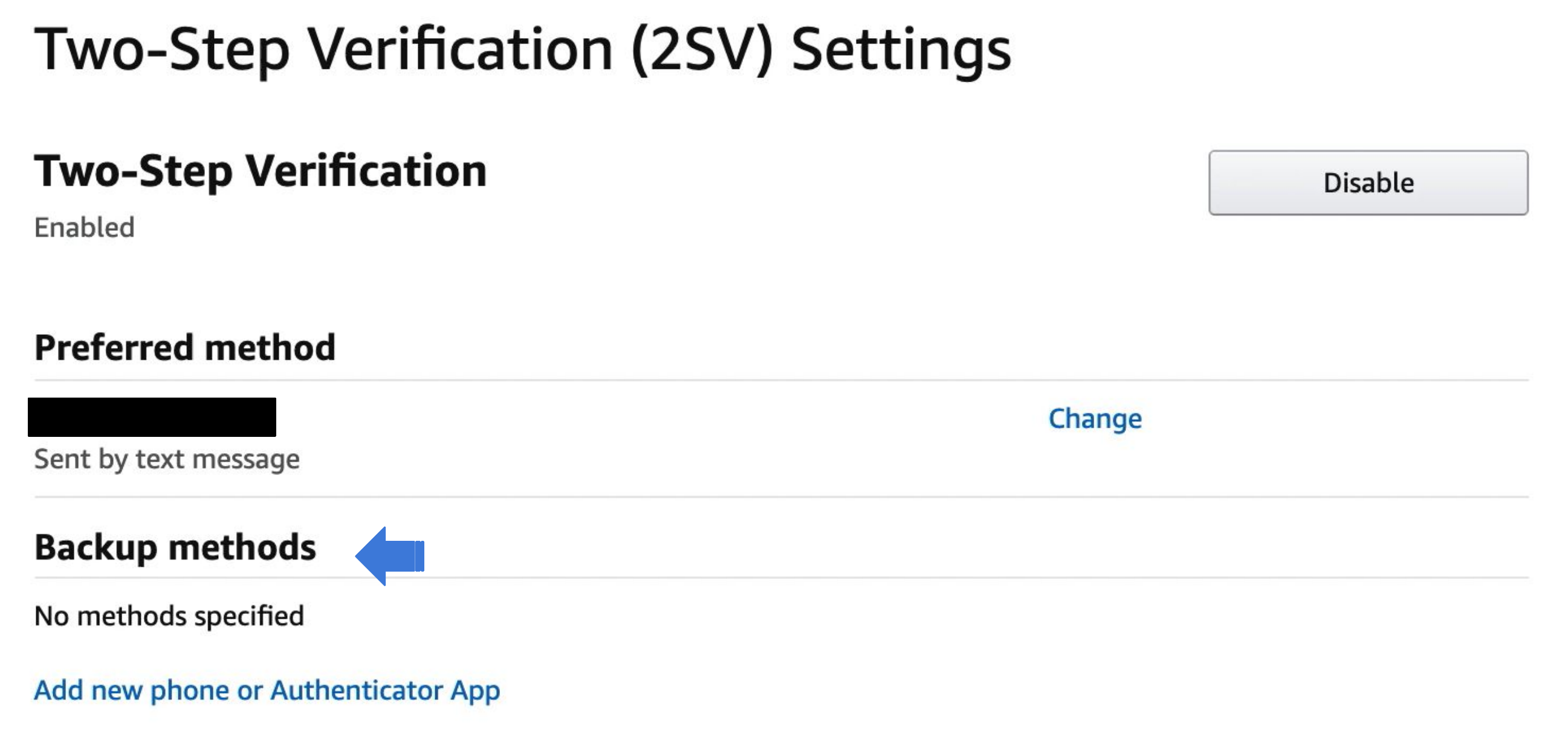}
    \caption{\url{clickup.com}~(\kinda) offers dedicated recovery options but does not provide any explanation to the user why it is important to enable recovery options.}
    \label{fig:basecamp}
    \end{subfigure}

    \caption{Examples of websites that (quasi) matched the \textit{Informed-2FA-Recovery-Options} factor.}

    \label{fig:Informed-recovery}
\end{figure}

\begin{figure}[h]
    \centering
    
    \begin{subfigure}[b]{\linewidth}
    \centering
      \includegraphics[width=.6\linewidth,frame]{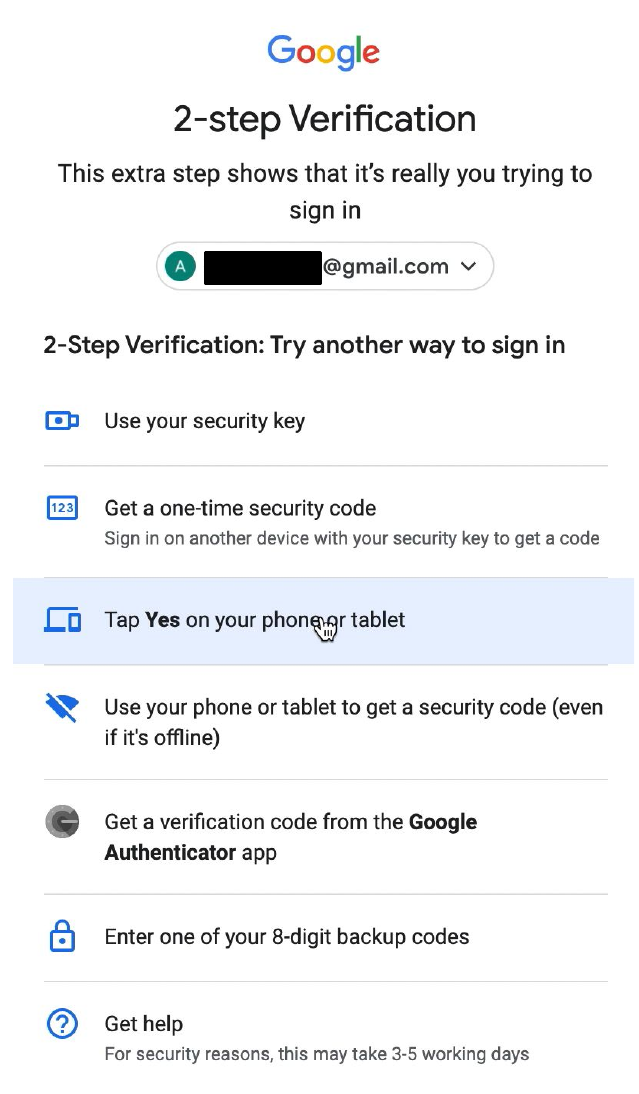}
    \caption{\url{google.com}~(\yes) offers multiple 2FA options and allows the user to setup and activate multiple 2FA options at the same time.}
    \label{fig:google-multiselection}
    \end{subfigure}
   
      \vspace{0.1cm}
 
  \begin{subfigure}[b]{\linewidth}
    \centering
      \includegraphics[width=.8\linewidth,frame]{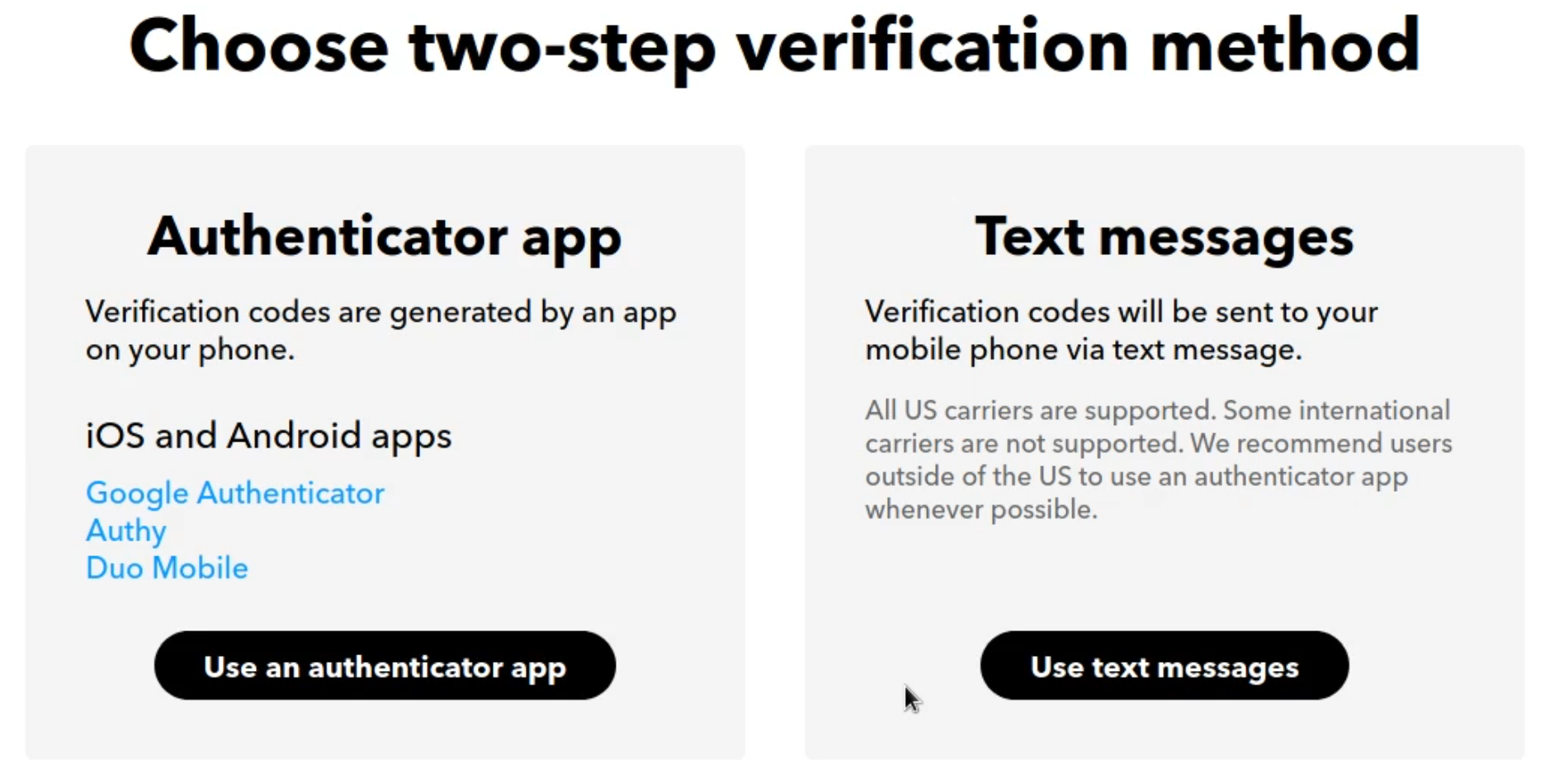}
    \caption{\url{ifttt.com}~(\kinda) offers multiple 2FA options but allows the user to only select one active option at the same time.}
    \label{fig:iftttmultiselection}
    \end{subfigure}
    
      \vspace{0.1cm}
     
      \begin{subfigure}[b]{\linewidth}
    \centering
      \includegraphics[width=.6\linewidth,frame]{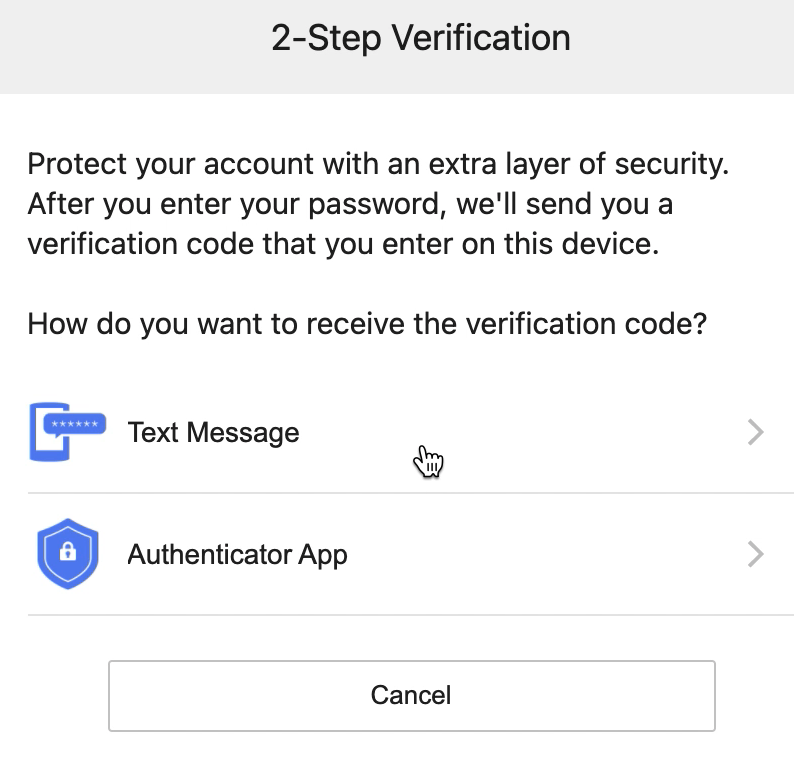}
    \caption{\url{playstation.com}~(\kinda) offers multiple 2FA options but allows the user to only select one active option at the same time.}
    \label{fig:playstationtmultiselection}
    \end{subfigure}
    
      \vspace{0.1cm}

    \caption{Examples of websites that (quasi) matched the \textit{Multiselection} factor.}

    \label{fig:multiselection}
\end{figure}

\end{document}